\documentclass[12pt]{article} 
\usepackage[hyperfootnotes=false]{hyperref}
     \usepackage{amsmath}
      \usepackage{amssymb}
  \usepackage{graphicx}
  \usepackage{xcolor}
  \newlength{\abstractwidth}
 \usepackage{mciteplus} 
\usepackage{subcaption}
 \usepackage{bbold}

  \newcommand{\be}{\begin{equation}}
  \newcommand{\bea}{\begin{eqnarray}}
  \newcommand{\eea}{\end{eqnarray}}
  \newcommand{\beq}{\begin{equation}}
  \newcommand{\ee}{\end{equation}}
  \newcommand{\eeq}{\end{equation}}

  \newcommand{\half}{{1\over 2}}

\def\la{\label}

\def\32{{3 \over 2 } }

  \def\ba{\begin{eqnarray}}
  \def\ea{\end{eqnarray}}

 \def\simleq{\; \raise0.3ex\hbox{$<$\kern-0.75em
      \raise-1.1ex\hbox{$\sim$}}\; }
 \def\simgeq{\; \raise0.3ex\hbox{$>$\kern-0.75em
      \raise-1.1ex\hbox{$\sim$}}\; }

\def\nref#1{(\ref{#1})}

\begin{document}

\begin{titlepage}
 % \rightline{}
  \bigskip

  \bigskip\bigskip

  \bigskip

\begin{center}
 
\centerline
{\Large \bf { Diving into traversable wormholes }}
 \bigskip

 \bigskip
%\centerline
{\Large \bf { }} 
    \bigskip
\bigskip
\end{center}

  \begin{center}

 \bf {Juan Maldacena$^1$, Douglas Stanford$^1$  and Zhenbin Yang$^2$   }
  \bigskip \rm
\bigskip
 
   $^1$Institute for Advanced Study,  Princeton, NJ 08540, USA  \\
\rm
 \bigskip
 $^2$Jadwin Hall, Princeton University,  Princeton, NJ 08540, USA

 % \bf {Write authors  }
  \bigskip \rm
\bigskip
 
 %   Institute for Advanced Study,  Princeton, NJ 08540, USA  \\
\rm

\bigskip
\bigskip

% \vspace{2cm}
  \end{center}

 \bigskip\bigskip
  \begin{abstract}

  We  study various aspects of wormholes that are made traversable by an interaction beween the two asymptotic boundaries. 
  We concentrate on  the case of nearly-$AdS_2$ gravity and discuss a very simple mechanical picture for the 
  gravitational dynamics.  We derive a formula for the two sided correlators that includes  
  the effect of gravitational backreaction, which limits the amount of information we can send through the wormhole. 
   We emphasize that the process can be viewed as a teleportation protocol where the teleportee feels nothing special as he/she 
 goes  through the wormhole. We   discuss some applications to the cloning paradox for old black holes. 
     We point out that the same formula we derived for $AdS_2$ gravity is also valid for   the simple SYK quantum mechanical 
     theory, around the  thermofield double state.
      We present a heuristic picture for this phenomenon in terms of an operator growth model. 
     Finally, we show that a similar effect is present in a completely 
     classical chaotic system with a large number of degrees of freedom.

 \medskip
  \noindent
  \end{abstract}
\bigskip \bigskip \bigskip

  \end{titlepage}

  %  \starttext \baselineskip=17.63pt \setcounter{footnote}{0}
   \tableofcontents

 % \sc

\section{Introduction and motivation} 

It is well known that traversable wormholes are forbidden in general relativity. This says that we cannot send a signal through the wormhole faster than 
we can send it through the outside. This includes the extended Schwarzschild solution which contains a wormhole. 

If we view gravity with $AdS$-like boundary conditions as the gravity dual of a quantum system, then the $AdS$-Schwarzschild wormhole can be viewed as dual to the thermofield double state of the quantum system. This is a state in the Hilbert space of two copies of the original system. In this situation nobody forbids us from coupling the two quantum systems. Interestingly, it was shown by Gao, Jafferis and Wall  \cite{Gao:2016bin} that simple couplings between the two sides can render the wormhole traversable! One can view this effect as giving a protocol for sending information from one system to the other.

What is interesting is not so much that information  {\it can}  be transferred, since after all we are explicitly coupling the two systems. What is interesting and surprising is {\it how}. We care about the ``feelings'' of the information as it gets transferred. And this information feels that it passes through the wormhole, rather than the explicit two-sided couplings. It sails through a smooth classical geometry between the two asymptotic regions. 

This wormhole protocol can be understood as a very simple realization of the Hayden-Preskill scenario \cite{Hayden:2007cs} of information transfer in quantum systems, where a few seemingly uninformative bits (in this case, the two-sided couplings), together with a large prior share of entanglement, are enough to transport information from one system to another. In the geometrical setup, the prior entanglement is important because it provides the wormhole that the information passes through.\footnote{Previous discussion of quantum teleportation and wormholes appeared in \cite{Marolf:2012xe,Susskind:2014yaa,Susskind:2016jjb,Gao:2016bin}.}

Another interesting feature is that this effect  is letting us explore the interior of the wormhole, or the interior of the black hole. 
This is because the perturbation induced by the coupling of the two systems  effectively moves the horizon back a little bit, exposing more of the interior to outside view.

It is with these motivations in mind that we have studied the phenomenon of \cite{Gao:2016bin} more detail. The goal of this paper is to study it at higher orders, describe it in simple toy models in both quantum and classical systems, and use it to address cloning paradoxes that were a cause for concern in the context of information recovery from black holes. We will concentrate on describing the process for nearly-$AdS_2$ geometries where the gravitational dynamics is particularly simple.

This paper is organized as follows. 
In section two we consider traversable wormholes by summing a dominant series of diagrams which capture the important physics. 
We start by recalling the basic setup of Gao, Jafferis and Wall \cite{Gao:2016bin} and introduce a minor variation which makes it the same as teleportation. 
We then sum the dominant diagrams, and study the resulting answer in various limits. First,  we discuss the probe limit where we ignore the backreaction of
the signal. We then discuss some of the effects of backreaction. 
We point out that these effects of backreaction are sufficient to ensure that we cannot send more information through the wormhole than the 
information we have to send in order to set up the interaction between the two sides.
 Finally, we briefly discuss the effects of stringy corrections. 

In section three, we present a very simple picture for the gravitational dynamics of nearly-$AdS_2$ spaces. 
The whole dynamics reduces to the motion of the boundary in a rigid $AdS_2$ space. 
Gravitational effects arise from taking into account that when we send matter into the interior, or matter bounces off the boundary, we have a small change in the 
boundary trajectory. In the thermofield double we have two boundary trajectories and the interaction \nref{Koper} can lead to an attractive force between the two boundaries, 
bringing them into causal contact and making the wormhole traversable.

In section four, we connect this  to the Hayden-Preskill analysis \cite{Hayden:2007cs} of the recovery of information from Hawking radiation. 
The traversable wormhole setup can be viewed as a particular physical implementation that gives a geometric interpretation for how the information is recovered. 
 We also make some remarks on the resolution of the cloning paradox. The point is that the information recovery operation generates a new spacetime that is connected
 to the original black hole through the interior. In this connected picture information is never cloned. We also point out that the Hayden-Preskill experiment can be 
 done by measuring only classical aspects of the new Hawking radiation. 
 
 In section five, we discuss some quantum mechanical models that display a phenomenon like traversability. First we discuss the SYK model, 
 where we can get exactly the same correlation functions that we obtained in the nearly-$AdS_2$ analysis. This is simply 
 because the low energy dynamics of SYK \cite{KitaevTalks} (see \cite{Maldacena:2016hyu} for some details) is dominated by a mode that has the same dynamics as gravity in $NAdS_2$ \cite{Almheiri:2014cka,Jensen:2016pah,Maldacena:2016upp,Engelsoy:2016xyb}. 
  We then discuss an even simpler and more generic operator growth model that also displays similar features. 
  
  In section six, we show that even classical physics displays a phenomenon similar to traversability.

\section{Traversable wormholes from boundary interactions } 

\subsection{Basic setup} 

We consider the Gao, Jafferis, Wall \cite{Gao:2016bin}  setup,  which we now review.
We   focus mainly on the case that the boundary theory is a quantum mechanical theory, 
as opposed to a quantum field theory. 
  We start from the thermofield double state of the quantum system. 
   We then perturb the system with the product of two simple operators, one on each side. 
In other words, we add the following interaction in the path integral 
\be \la{basi}
 e^{ i \tilde g V } = e^{ i \tilde g O_L(0) O_R(0) }.
 \ee
 For simplicity we choose the two times to be equal here.  We can also imagine replacing $O(0)$ by an integral over a short range of times to remove 
 possible contributions from very high energy states. It will be useful for making the effect large, and for simplifying some computations, to take the case where we have $K$ such operators and we insert 
 \be \la{Koper}
  e^{ i g V } = e^{ i g { 1 \over K } \sum_{j=1}^K O_L^j(0) O^j_R(0) } 
  \ee
  where we assumed that we have a large number of light bulk fields. This is approximately a product of operators like \nref{basi} with $\tilde g = g/K$.  We will consider the limit of large $K$ with $g$ fixed, so that $\tilde{g}$ is small. This limit is useful because it suppresses particle creation in the bulk. 

We have in mind that the $O^j$ are ``simple'' operators in terms  of the basic degrees of freedom of the quantum system. Such operators also have a simple gravity description, corresponding to bulk field operators acting near the boundaries. More precisely, the gravity dual of the thermofield double state is a two-sided black hole \cite{Israel:1976ur,Maldacena:2001kr} with two asymptotic boundaries. Denoting also by $O $ the gravity field dual to the operator $O$, then 
  \nref{basi} corresponds to adding a similar term with $O_L O_R$ in the gravitational path integral, where now 
  $O_L$ and $O_R$ represent bulk field operators\footnote{ We assume some familiarity 
   with the basic AdS/CFT dictionary \cite{Gubser:1998bc,Witten:1998qj,Banks:1998dd} that relates suitably rescaled 
  values of the bulk operators near the boundary with ``single particle''  local operators of the boundary theory. }  near the respective boundaries, and at time $t_L = t_R = 0$.

The authors of \cite{Gao:2016bin} argued that, for a suitable sign of the coupling $g$, the operator $e^{igV}$ generates negative null energy in the bulk. If we then act with $\phi_R$ to send in {\it any} probe 
  particle from the right side at early times, this particle will suffer a time advance, rather than a time delay, 
  as it goes through the central region and can therefore emerge on the left side. See figure 1.

Some readers might be worried about our use of a large number of light fields, $K\gg 1 $, since these do not appear in usual examples of 
$AdS/CFT$. We can easily introduce them via a  
 small variant of the usual examples where we introduce $K$ additional Wilson loops and their corresponding 
bulk strings. The fields along the strings can be used for generating the interaction \nref{Koper}. See appendix \ref{LargeNumber}.  
Alternatively, it seems 
 possible to  use the HKLL \cite{Hamilton:2006az} construction to produce a bulk operator localized near the horizon so that we can also use more
massive fields, but we have not analyzed in  detail the  potential pitfalls of this strategy.

\subsubsection{ Standard teleportation setup} 
\la{Teleport}

The operator $e^{igV}$ is an entangling operator that can exchange quantum information between the two sides. However, for the purposes of sending messages from the right boundary to the left, we can replace this by an operation that transfers only classical information from the right to the left. We measure the value of $O_R$ on the right system, and get one of several possible values $o_j$. We then apply the unitary operation $e^{igO_Lo_j}$ to the left system. In other words, we apply  the protocol 
\be \la{Tele}
{\rm Measure} ~ O_R ~,~~{\rm get }~~ O_R \to o_j ~,~~~~{\rm act ~with~ } ~~ e^{ i g O_L o_j }. 
  \ee
This leads to precisely the same {\it left } density matrix as if we had applied the quantum operation (\ref{basi}). This is argued as follows. 
 First note that, since we are tracing over the right system in defining the left density matrix, we are free to imagine that  arbitrary measurements  were made on the right system immediatley after applying $e^{igV}$. Finally, 
 because measurements of $O_R$ commute with $e^{igV}$, we can equivalently measure immediately before implementing the interaction, which leads to (\ref{Tele}).
Note  that the final global state of the two systems is indeed different depending on whether we implement the unitary operator
\nref{Koper} or we do a measurement of the right operator $O_R$. Also, the final right density matrix is different. 
 It is only the {\it left } density matrix that is the same in both cases. 
Using this protocol where we measure the right operator,   the process becomes a more standard teleportation experiment \cite{bennett1993teleporting}.
   
  \begin{figure}[t]
\begin{center}
\includegraphics[width=.95\textwidth]{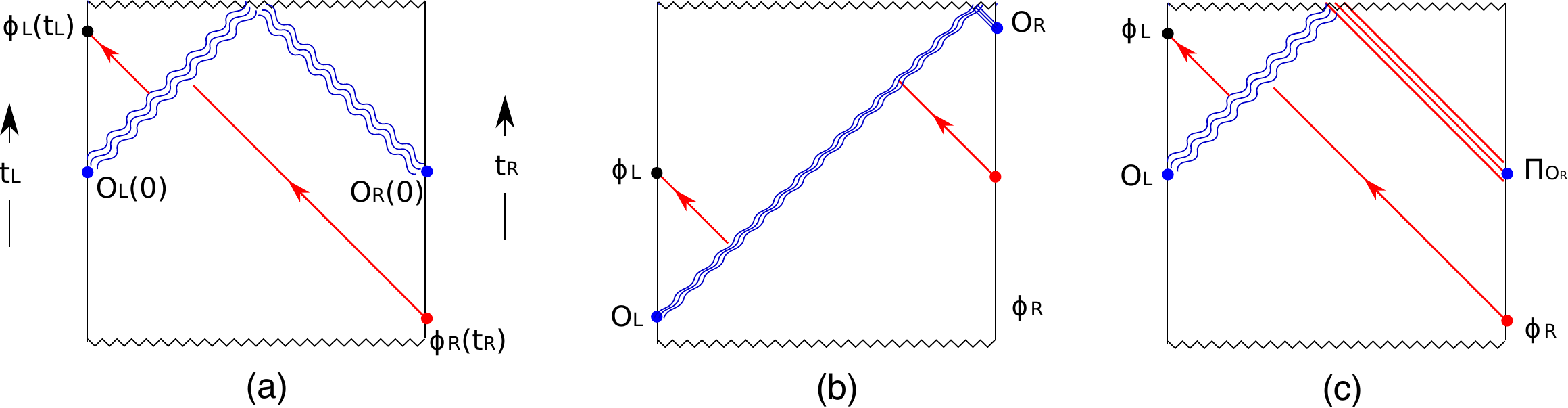}
\caption{ Basic setup. In (a) a message created by $\phi_R$ propagates into the black hole, gets a time advance as it crosses the negative energy associated to $O_LO_R$, and emerges on the left side. In (b) the same configuration has been boosted. In (c) we show the case where $O_R$ is measured, resulting in some positive energy shooting in from the right but the same negative energy on the left. The diagrams are drawn in discontinuous coordinates so that a continuous worldline gets a backwards null shift as it crosses the blue negative energy. }
\label{Setup}
\end{center}
\end{figure}

\subsection{Gravity computation } 
\la{GravComp}

We would like to understand this effect in a more precise way, by studying two-sided correlation functions that can probe traversability. One useful observable is the commutator 
\be
\langle [\phi_R,\phi_L]\rangle_V \equiv 
%\langle e^{-igV}[e^{igV}\phi_Re^{-igV},\phi_L]e^{igV}\rangle 
\langle[ \phi_R , e^{-igV}\phi_Le^{igV}]\rangle = \langle\phi_Re^{-igV}\phi_Le^{igV}\rangle - \langle e^{-igV}\phi_Le^{igV}\phi_R\rangle,
\ee
where we have omitted the time arguments, but in all cases $\phi_L = \phi_L(t_L)$ and $\phi_R = \phi_R(t_R)$. The angle brackets indicate expectation value in the thermofield double state. This object gives the leading response of $\phi_L$ to a unitary perturbation to the R system by $e^{i\epsilon_R \phi_R}$, once we include the operation $e^{igV}$ as part of the time evolution:
\be
\langle e^{-i\epsilon_R \phi_R}e^{-igV}\phi_L e^{igV}e^{i\epsilon_R\phi_R}\rangle = \langle e^{-igV}\phi_Le^{igV}\rangle -i\epsilon_R \langle [\phi_R,\phi_L]\rangle_V+O(\epsilon^2).
\ee
Therefore, a nonzero value for the commutator $\langle [\phi_R,\phi_L]\rangle_V$ indicates that some kind of signal can be sent through the wormhole, since the 
expectation value of $\phi_L$ ends up depending on $\epsilon_R$,  which is a perturbation to  the right system. 

\begin{figure}[t]
\begin{center}
\includegraphics[width=.55\textwidth]{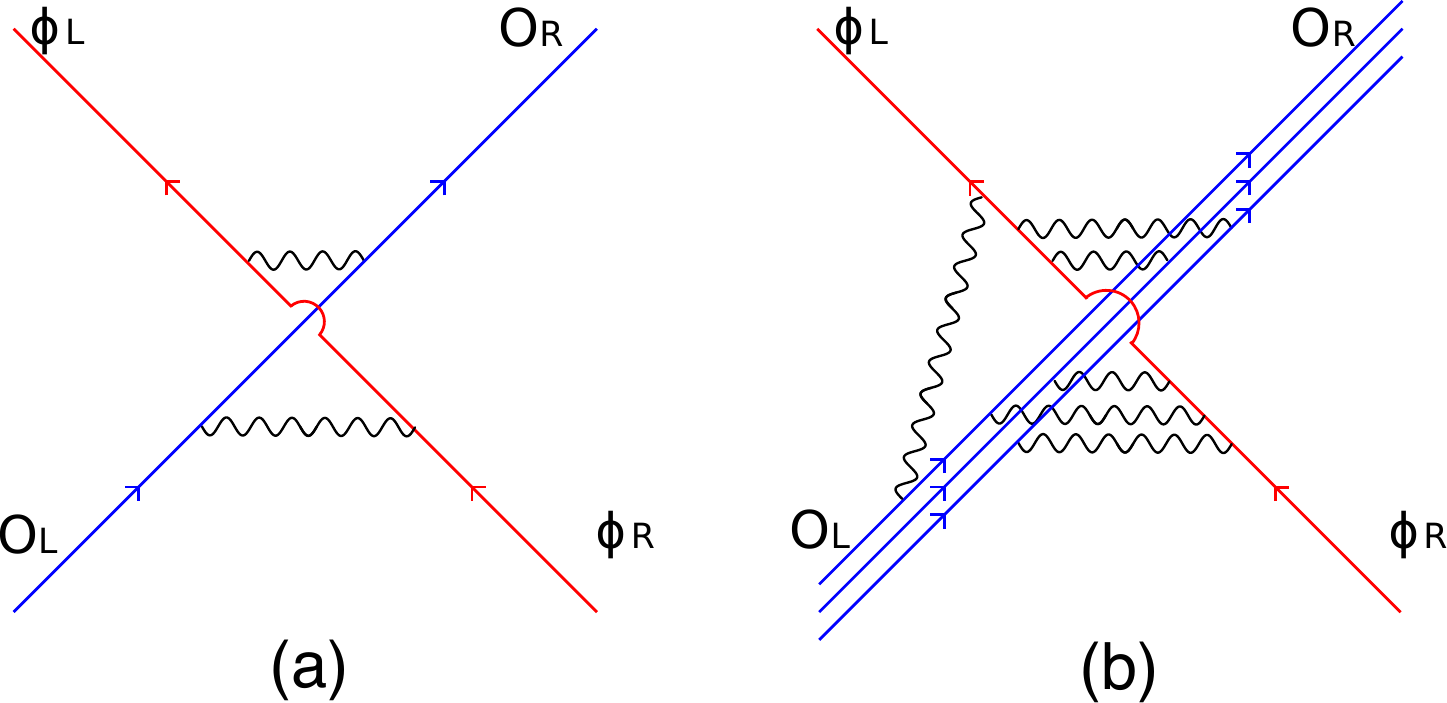}
% \begin{subfigure}{.6\textwidth}
%   \centering
%   \includegraphics[width=.6\linewidth]{scatteringa}
%   \caption{a}
% \end{subfigure}%
% \begin{subfigure}{.6\textwidth}
%   \centering
%   \includegraphics[width=.6\linewidth]{scatteringb}
%   \caption{b}
% \end{subfigure}
\caption{In (a) we show a two-graviton exchange that contributes at leading order in $g$. 
This is proportional to $g\cdot(G_Ne^t)^2$. At higher orders in $g$ we have more $O$ quanta, and we sum diagrams including those in (b).}
\label{Diagrams}
\end{center}
\end{figure}

Setting $-t_R = t_L = t$, we notice that to first order in $g$, the commutator involves a double commutator of $\phi$ and $O$:
\be
\langle [\phi_R,\phi_L]\rangle_V = ig\langle[\phi_L(t),O_L][\phi_R(-t),O_R]\rangle + O(g^2).
\ee
Such commutators have been studied recently in the context of quantum chaos and 
  gravitational scattering near black holes \cite{Shenker:2013pqa,Shenker:2013yza,kitaevfundamental,Shenker:2014cwa}. An important lesson of those studies is that such squared commutators initially grow exponentially in time $t$. In theories with a gravity dual,  this growth is due to the standard growth with energy of gravitational scattering.
   For relatively large values of $t$ ($ 1 \ll { t \over \beta} \ll \log N $)  the wavefunctions of the state created by $\phi$ and the one created by $O$ have a large, but not
   too large,  relative boost.  This implies that  we can approximate the scattering via a shock wave amplitude \cite{tHooft:1987vrq,Amati:1987uf,Verlinde:1991iu} which has the form 
  \be \la{GravPhase} 
  S_{grav} =   e^{i G_N e^t p_+ q_- } 
    \ee
    where $p_+$ is a component of the momentum of the $\phi$ particle, and $q_-$ is a component of the momentum of $O$.\footnote{Also, we find it convenient to use $p_+$ and $q_-$ to label momenta in two different Lorentz frames, in which each of the particles are unboosted. These frames differ by a boost factor $e^t$, so Mandelstam $s$ is proportional to $p_+q_-e^t$.} 
    This formula is equivalent to an eikonal resummation of a series of gravitational exchanges, see figure \ref{Diagrams}(a) and \cite{Kabat:1992tb}.  In this equation, and elsewhere in the paper, we are setting our units for time so that $\beta = 2 \pi$. Factors of the temperature can be restored by changing $t \to t { 2\pi \over \beta } $ in all formulas. Equation (\ref{GravPhase}) is correct in the regime that $G_N \ll 1$ and $t$ is large so that $G_N e^t$ is of order one.  Here $G_N$ stands for a measure of  the strength of the gravitational effects. 
     In this regime we only need to resum the diagrams that increase 
    with time.  For example, diagrams that involve gravitational self interactions of the field $\phi$ can be neglected since they are not enhanced by $e^t$. 

    As was emphasized in \cite{kitaevfundamental,Roberts:2014ifa,Shenker:2014cwa}, these exponentially growing contributions are present only for a particular ordering of the operators. This is 
    the ordering that is  natural for a scattering computation on the wormhole geometry.

In what follows, we will find it convenient to study a slightly simpler quantity, which is a single one of the terms appearing in the commutator:
 \be \la{basico} 
  C \equiv \langle e^{ - i g V}  \phi_L(t_L) e^{ i g V } \phi_R(t_R) \rangle.
  \ee
If we take $V$ and $\phi_{L,R}$ to be hermitian, then we can recover the commutator $ \langle [ \phi_R , \phi_L]\rangle_V $ by taking the 
imaginary part of $C$.

We now turn our attention to \nref{basico} and we expand the exponentials. 
We first consider a correlator of the form 
\bea
 \langle e^{- i g V } B \rangle &= &\sum_{ n=0}^\infty  { ( -i g)^n \over n!} \left\langle  \left( { 1 \over K } \sum_{j=1}^K O_L^j O_R^j  \right)^n  B \right\rangle
 \cr
 & \approx &  \sum_{ n=0}^\infty  { ( - i g)^n \over n!}  \left( { 1 \over K } \sum_{j=1}^K \langle O_L^j O_R^j  \rangle \right)^n \langle B \rangle = 
e^{ i g \langle V \rangle } \langle B \rangle 
\eea
Here we have used a few properties. First we have expanded the exponential. Then we note  the form \nref{Koper} of the operator  and use large $K$ to 
 to simplify the contractions to those involving operators within the same sum.  We are neglecting any gravitational interaction because the $O_L$ and 
$O_R$ are next to the thermofield double.  In this ordering of the operators, there is no scattering between the particles created by $O$ and the ones created by $B$, and therfore no $e^t$ factors to enhance $G_N$-suppressed gravitational corrections. This means that we can rewrite \nref{basico} as 
\be \la{TiCdef}
C = e^{ - i g \langle V \rangle } \tilde C ~,~~~~~ \tilde C \equiv \langle \phi_L e^{ i g V } \phi_R \rangle.
\ee
A more direct argument is the following. First we compute $\langle e^{ - i g V } \rangle \sim  e^{ - i g \langle V \rangle } $, which is true in the large $K$, small $G_N$ limit. 
We then notice that this implies that $ \langle TFD| e^{ - i g V  } \sim  e^{ - i g \langle V \rangle } \langle TFD| $ as a state, since two states whose overlap is a phase have 
to differ precisely by that phase. 

We can now expand the exponential in $\tilde C$. Now the ordering of operators is such that the scattering between the $\phi$ and the $O$ excitations 
is exponentially enhanced. For a term $\langle \phi_L (O_L O_R)^n \phi_R \rangle $ we view these as $n$ separate and independent scattering events and multiply 
together the resulting phases, each of which involves a factor \nref{GravPhase}. As in \cite{Shenker:2014cwa}, the same $\phi$ 
particle can be expressed in terms of a superposition of 
particles of momenta $p_+$. The operator $O$ can similarly be expressed in terms of functions with momenta $q_-$ and then we get that 
\be
\langle \phi_L (O^j_L O^j_R)^n  \phi_R \rangle =\int d p_+ \langle \phi_L |p_+ \rangle \langle p_+|\phi_R \rangle   \left[    \int d q_- e^{ i G_N e^t p_+ q_- } 
 \langle O^j_R | q_-\rangle \langle q_- |O^j_L \rangle\right]^n\label{havesummed}
\ee
where $  \langle p_+|\phi_R \rangle$ are the coefficients of the decomposition of the single-particle wavefunction of $\phi$ in terms of states of definite $p_+$, and similarly for other operators.  
In (\ref{havesummed}), we have effectively summed the diagrams indicated in figure \ref{Diagrams}(b) for a fixed number $n$ of $O$ quanta. To get a final formula for $\tilde{C}$, we sum over $n$, finding 
\bea \la{FirstRe}
\tilde C &=& \int dp_+  \langle \phi_L |p_+ \rangle \langle p_+|\phi_R \rangle \exp\left[ i\tilde g \int d q_- e^{ i G_N e^t p_+ q_- }   \langle O^j_R | q_-\rangle \langle q_- |O^j_L \rangle 
\right]
\\
&=&  \int dp_+  \langle \phi_L |p_+ \rangle \langle p_+|\phi_R \rangle \exp\left[  i \tilde g     \langle O^j_R     e^{ i G_N e^t p_+ \hat P_- } O^j_L \rangle 
\right]. \la{SecondRe}
\eea
Here $\hat{P}_-$ is the momentum operator acting on $O_L$. Note that the TFD is invariant under this action, so we are simply performing an $x^-$ translation of the wavefunction by an amount that involves $ \Delta x^- \sim G_N e^t p_+$. 

The formula \nref{FirstRe}  is quite general. It holds in higher dimensions as well if we perform some replacements, like adding extra labels $r$ $r'$ for the transverse form of the wavefunctions $|p_+\rangle \to |p_+ , r\rangle $ and $|q_-\rangle \to |q_-,r'\rangle$, and we also consider $G_N \to G_N h(r-r')$ with $h$ given by the proper expression for 
the shock wave transverse profile. 

But, to be really concrete, let us imagine the situation of a nearly-$AdS_2$ space. In this case $\hat P_-$ is one of the $SL(2)$  symmetry generators and we can easily compute the expectation value in the exponent in \nref{SecondRe}:
\be \la{dispop}
 \langle O_R(0)     e^{-  i a^- \hat P_- } O_L(0) \rangle  =   { 1 \over ( 2 + { a^- \over 2 }   )^{ 2 \Delta}   }  
 \ee
with $a^- = - G_N e^t p_+$. How this follows from the symmetries is explained in detail in appendix A. 
Note that a physical $\phi$ particle, which has $p_+<0$, leads to  $a^- > 0$ and the $\langle O_RO_L\rangle$ correlator is suppressed. This is the total effect of the backreaction in this approximation. See \cite{Shenker:2013pqa,Shenker:2014cwa}. 
 
When $\phi$ is a local operator, the wavefunctions are also simple to write down. They are determined by conformal symmetry to be powers of $p_+$ times an exponential function. These can be 
simply obtained by Fourier transforming an expression like \nref{dispop} (but with $a^- \hat P_- \to a^+ \hat P_+$). Setting
the $\phi$ operators at $t_L = -t_R = t$, we obtain (see appendix \ref{app:kinematics}, eqn. \nref{WF2})
 \be \la{ResGen}
 \tilde C = { 1 \over \Gamma( 2 \Delta) } 
  \int_{-\infty}^0 { d p_+ \over (- p_+)} ( 2 i p_+)^{ 2 \Delta } e^{ -4 i  p_+}   \exp \left[  { i g \over ( 2-  p_+ G_N e^t /2)^{ 2 \Delta }  } \right].
 \ee
 The factor of $e^{ -4 i p_+} $ arises from the fact that the operators  are on opposite sides of the thermofield double. Note that $p_+$ is negative. In writing this formula, we have assumed that the dimensions of $O$ and $\phi$ are equal, and that $t_L = -t_R = t$. More generally,
\be 
\tilde C={1\over (\cosh{t_L+t_R\over 2})^{2\Delta_{\phi}}}{1\over \Gamma(2\Delta_{\phi})}
\int_{-\infty}^0 { d p_+ \over (- p_+)} ( 2 i p_+)^{ 2 \Delta_{\phi} } e^{ -4 i  p_+}   \exp \left[  { i g \over \left( 2- { p_+ G_N e^{t_L-t_R\over 2}\over 2\cosh{t_L+t_R\over 2}}
\right)^{ 2 \Delta_O }  } \right] 
\ee
We see that the main difference is the replacement $e^t \to { e^{t_L-t_R\over 2}\over  \cosh{t_L+t_R\over 2} }$. Since the final result depends non-trivially only on this 
combination, we can easily restore the dependence on the two times, from the expression for the particular case $t = t_L=-t_R$. Therefore we will consider only this 
particular case from now on. 

Another interesting generalization involves sending many $\phi$ particles. In that case we continue to get an expression like \nref{ResGen} but now we will 
have the wavefunctions of the multiparticle state, schematically
\be \la{ResMult}
\tilde C = \int \prod_l  d p_+^l  \psi_L(p_+^l) \psi_R(p_+^l) \exp\left[ { i g \over ( 2 - p_+^{\rm total} G_N e^t/2)^{ 2 \Delta } } \right]  ~,~~~~p_+^{\rm total} = \sum_l p_+^l.
\ee
The prefactor of the exponential is the inner product of the left and right multiparticle states. We want to emphasize that the scattering amplitude depends only on
the total $p_+$ momentum of this state. 

\subsection{Ignoring the backreaction of the $\phi$ signal } 

Before analyzing the properties of this result in detail, let us take a limit with large $g\gg 1$ and relatively small $G_N e^t \ll 1$, with $g G_N e^t$ finite.  
Then we can expand the denominator
in the exponential of \nref{ResGen} to first order to obtain 
\bea \la{ResFi} 
 C_{\rm probe}  & = & e^{ - i { g  \over 2^{ 2 \Delta }}  } \tilde C =  { 1 \over \Gamma( 2 \Delta) }  
  \int_{-\infty}^0 { d p_+ \over (- p_+)} ( 2i p_+)^{ 2 \Delta } e^{ -4 i  p_+  + i {  g \over 2^{ 2 \Delta+1 } }  p_+ G_N e^t  \Delta } 
  \cr
 &  = & 
   \langle \phi_L  e^{ - i a^+ \hat P_+} \phi_R \rangle =  { 1 \over ( 2 + {a^+ \over 2 } )^{ 2 \Delta } } 
    ~,~~~~~~~~~ a^+ = -   \Delta   { g \over 2^{ 2 \Delta+1 } } G_N e^t.
\eea
For general $t_L$ and $t_R$, we  have
\bea
C_{\rm probe}={1\over (2\cosh{t_L+t_R\over 2}+{a^+\over 2}e^{t_L-t_R\over 2})^{2\Delta}}.
\eea
These formulas represent the effect discussed in \cite{Gao:2016bin},  here for the particular case of nearly-$AdS_2$ \cite{Almheiri:2014cka,Jensen:2016pah,Maldacena:2016upp,Engelsoy:2016xyb}. 
In the approximation \nref{ResFi}, we are ignoring the backreaction of the $\phi$ particles on the $OO$ correlator. The $OO$ insertions, together with  their interaction with gravity, create a background on which the $\phi$ field moves as a probe.

It is clear from (\ref{ResFi}) that the effect of the $OO$ insertion is to implement an $x^+$ translation of the $\phi_R$ wavefunction. When $g$ is positive, so that $a^+$ is negative, this translation is a time advance. It has the effect of increasing the correlator between the $\phi$ operators, effectively shortening the distance between the two boundary points. For a sufficiently large magnitude of $a^+$,  $a^+<-4$, the effective separation between the two boundary points become timelike.  The correlator diverges and then it picks up an imaginary part for later times, or  $a^+ < -4$. This  
 implies that $\langle [\phi_L,\phi_R]\rangle_V$ becomes nonzero and we have a signal being transmitted from the left  to the right sides of the wormhole. Of course, the interpretation of this is that the $SL(2)$ transformation  generated by $\hat P_+$ has moved the right insertion point into the region causally related to the left insertion. This $SL(2)$ transformation is the effect of crossing the blue negative energy in figure \ref{Setup}. 

Note that the correlator becomes infinite when $a^+ =-4$. This infinity comes from the large $p_+$ region, and it
can be removed by smearing the $\phi$ insertion points over a small range of times. We will also see that even if
we do not smear the insertion points, this infinity is removed by gravitational corections or stringy corrections.\footnote{ 
In this respect, this infinity is similar the ``bulk point singularities'' discussed in \cite{Heemskerk:2009pn,Okuda:2010ym,Maldacena:2015iua}.}

In principle, when we put the $V$ interaction at $t=0$ the $\phi$ signal could have come out immediately.  However, in this probe approximation,
 it comes out only when we send it early enough 
that the shift  $a^+$ is enough to pull it out of the left  horizon. 
This delay in the emergence of the particle is related to the fact that the particle is going through the wormhole in the bulk, rather than going through the boundary interactions. 

We should emphasize the simplicity of the final answer \nref{ResFi}. It is saying that the right particle is simply translated by a symmetry which could make it emerge from 
the horizon with a modified velocity. If we send a collection of $\phi_R$ particles, or a multiparticle state, then all these particles are translated by the same amount 
($p_+$ in \nref{ResFi} is replaced by $p_+^{\rm total} $). 
This means that they do not feel any forces, not even a tidal force.\footnote{
This lack of tidal force is special to the nearly-$AdS_2$ case. In higher dimensions we  have tidal forces if the operators $O_L O_R$ are localized in the 
extra dimensions.} Their experience  is as uneventful as it could possibly be: free propagation in empty space.

\subsection{Including the backreaction of the $\phi$ signal } 
\label{Backreaction}

Let us go back to the more general  expression \nref{ResGen}. This general expression shows that $\phi$ particles 
with large values of $p_+$ decrease the correlator between $O_R O_L$, which in turn reduces their ability to open the bridge.

\begin{figure}
\begin{center}
\includegraphics[width=0.31\textwidth]{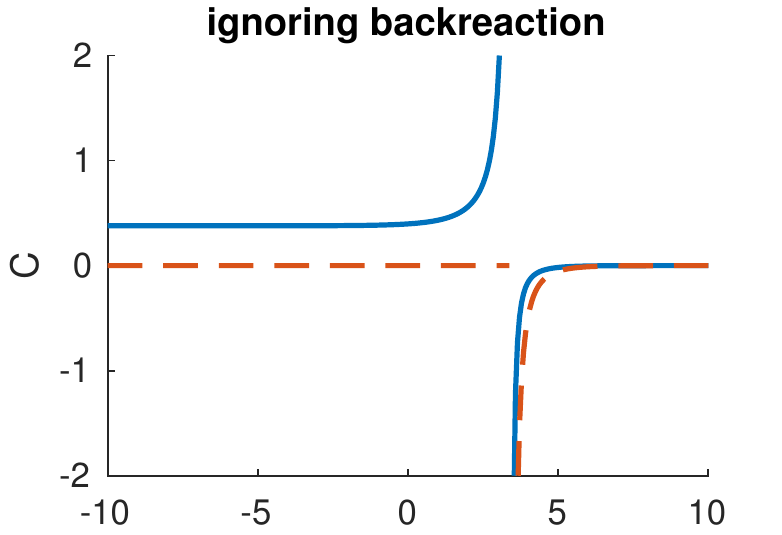}
\includegraphics[width=0.31\textwidth]{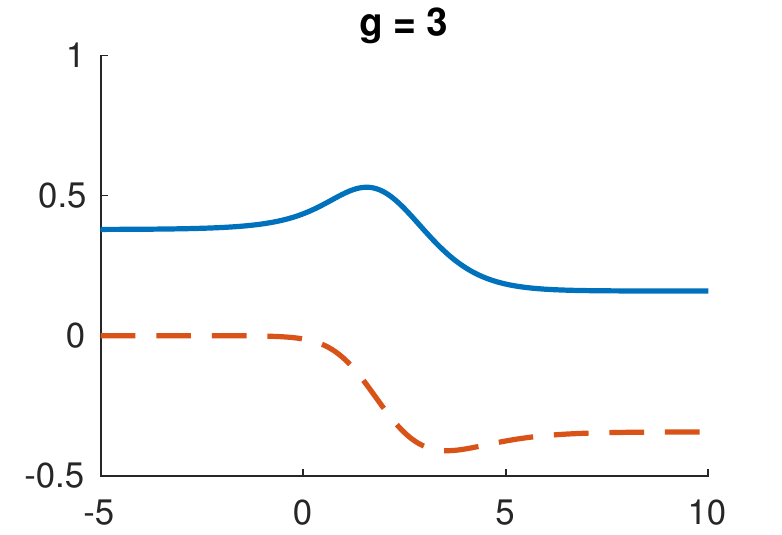}
\includegraphics[width=0.31\textwidth]{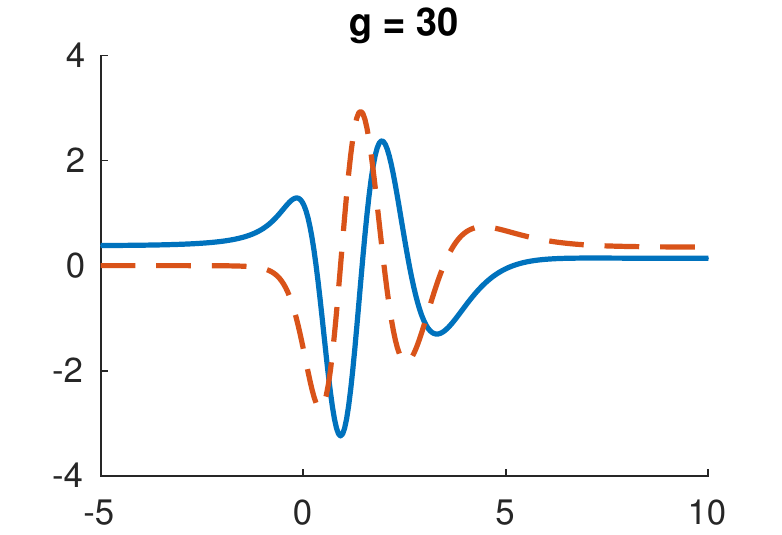}
\\~~~~~~~(a) ~~~~~~~~~~~~~~~~~~~~~~~~~~~~~~~~(b) ~~~~~~~~~~~~~~~~~~~~~~~~~~~~~~~~~(c)
\caption{Real part (blue, solid) and imaginary part (red, dashed) for the correlator $C$, with $\Delta = 0.7$. 
(a) correspond to the probe approximation \nref{ResFi} and (b), (c) to the full result \nref{ResGen} (after getting $C$ via  \nref{TiCdef}).  The horizontal axis is $t + \log(G_N)$.}\label{gfigure}
\end{center}
\end{figure}

We now consider the full function and plot it in some special cases. We plot the real and imaginary parts of $C$ in figure \ref{gfigure}. Note that the imaginary part is nonzero for all times, in contrast to what we found at leading order. An important point is that there is a sweet spot for traversability, in the sense that the magnitude of the imaginary part has a maximum at an intermediate time. The appearance of a non-zero commutator immediately after the introduction of the double trace interaction will be explained in section 
\ref{GravityThreePoint}. 

We would like to understand in what sense the full answer \nref{ResGen} reproduces the probe result, \nref{ResFi}, for large $g$. It is clear that this will be a good approximation if the momenta that dominate the integral are such that $p_+ G_N e^t \ll 1$. When is this true? The analysis of this question is somewhat complicated, and we give details in appendix \ref{PropertiesCorrelator}. The essential points can be summarized as follows. For large $g$, we can think about the integral in a saddle point approximation. There is an endpoint contribution near  $p_+ = 0$ that gives a good approximation to the probe answer. This is all we have for $t<t_d$. However, for $t>t_d$, in addition to the desired endpoint contribution, one finds that there is also a larger oscillating contribution from a ``contaminating'' saddle point. This saddle point corresponds to a momentum of order $-p_+ \sim g e^{-\#(t-t_d)}$, where $\# = \frac{2\Delta}{1  +2\Delta}$. For large $g$, this is a large momentum that reflects high-frequency components of the $\phi$ operator. By considering wavefunctions for $\phi$  containing  a  gaussian envelope,  $e^{ - (p_+)^2 \sigma^2 }$, we will suppress the contribution of this saddle, leaving only the endpoint contribution, which gives a good approximation to the probe answer. 

In fact, there is one further wrinkle: for times much greater than $t_d$, the saddle point momentum will become small enough that the wavefunctions will not suppress the contribution. This causes $\tilde{C}$ to deviate from the probe answer, oscillating and eventually approaching a nonzero constant. We can understand the particular constant value in a simple way from the expression \nref{ResGen}.  As  we  send $t\to \infty $ the argument of the exponential in \nref{ResGen} goes to zero. This removes all $g$ dependence  from  $\tilde C$ and makes it    equal to the original two point function  $\langle \phi_L \phi_R \rangle $.
  What has happened is that the scattering with the highly boosted $\phi$ quanta has effectively destroyed the correlations 
 between $O_L $ and $O_R$ that were important to give rise to the effect. It seems suprising that they only do this,  
 without destroying also the $\langle \phi_L \phi_R \rangle $ correlator. 
   We will later discuss some quantum mechanical  models that display this effect. 
 In any case, this implies that the large time expectation value of $C$ is 
 \be \la{LargeT}
  C \sim e^{ - i g\langle V\rangle } \langle \phi_L \phi_R \rangle ~,~~~~~~~{\rm for } ~~~  G_N e^t \gg ( g)^{ 1\over 2 \Delta }.
  \ee
  It is rather surprising that we get a non-zero phase. This phase implies that the commutator is still  non-zero! It is proportional to $\sin g$ and and it is not 
  exhanced relative to the original value of the two point function.

 This non-zero value of the commuator is produced by a kind of inteference effect between the case that we do  not send the $\phi$ particle and the case where we send the $\phi$ particle. Notice that we can introduce a kind of ``calculus'' that is valid for very large times \nref{LargeT}.   In this regime, we can say that if $V$ is next to the TFD state, then it has the usual expectation value, but if it is between two (or any other number of ) $\phi$s, as in $\langle \phi_L e^{ i g V } \phi_R \rangle $, it can be replaced by $V\to 0$.

  \subsection{Bounds on information transfer}  
 
 In this section we point out that the backreaction effect in the last section is enough to ensure that we cannot 
 send more information through the wormhole than we transferred in order to set up the $OO$ interaction and open the wormhole
 in the first place. We will not provide  sharp bounds,   only  parametric bounds. 
 
 Let us first imagine that we follow the simpler teleportation protocol described in section \ref{Teleport}, where we measure the operator $O^j_R$. 
 For simplicity, let us assume that the operator $O^j_R$ has eigenvalues $\pm 1$, so that each measurement corresponds to one bit. 
 In this case, if we have $K$ such operators, we conclude that the number of bits we need is $N_{bits} = K$. We also had to assume 
 that the coupling $\tilde g$ in \nref{basi} was small. Thus, in particular, we have the bound $ g = \tilde g K < N_{bits} $
 \be \la{nequbits}
   g \lesssim N_{\rm bits}. 
   \ee
  This is a bound on  the amount of information transfer that we need in order to open up the wormhole. We are only after the parametric scaling, and are ignoring constant factors.
 Returning to the case of general operators, we can imagine that we smear them enough so that the same sided correlator is $\langle O_L^2 \rangle \sim 1$ and is
 of the same order as the two sided correlator $\langle O_L O_R \rangle$. In this situation, it seems reasonable to model the operator as a single qubit so that 
 \nref{nequbits} is still valid. 
   
 We now imagine sending   quantum information that consists of some qubits produced by $\phi$. If we send them as qubits with very high 
 momentum $p_+$ we will close the wormhole. So we want to send them with a $p_+$ as small as we can. 
Recall  that the classical shift in the $x^+$ position produced by the shock wave is of order \nref{ResFi}
 \be \la{xmshift}
|a^+ | \sim   g G_N e^t.
 \ee
 The initial wavepacket produced by the operator $\phi_L$ is contained in the region $x^+ > 0$. In order for most of the wavepacket to make it through we need
 that its initial spread, call it $\Delta x^+$ should be of the same order or less than \nref{xmshift}. By the uncertainty principle, $\Delta x^+ \Delta p_+ \geq 1$, 
 this gives a constraint on the momentum of each particle (see appendix \ref{app:inf} for a more rigorous version of this, using \cite{Blanco:2013lea})
 \be \la{peachb}
  -p^{\rm each}_+ \gtrsim { 1 \over \Delta x^+ } = { 1 \over G_N g e^t }. 
  \ee
 If we are sending several particles with this momentum, we will have some total momentum $p^{\rm total}_+$. 
 This total momentum   suppresses  the correlator between the left and right operators $O$ as indicated in the exponent of \nref{ResMult}. In order not to have a 
 significant suppression we want that   
  \be \la{ptotb}
  -p_+^{\rm total}   G_N e^t  \lesssim 1.
  \ee
  This bound is   setting the boundary of the regime of validity of the probe approximation, which is necessary when we want to think of the different pieces of information 
  travelling independently.\footnote{Technically, this allows us to approximately expand to linear order in $p_+^{\text{total}}$ in (\ref{ResMult}), after which the scattering amplitude factorizes into a product of amplitudes for each of the particles, indicating that the information is indeed being sent independently.}
  This implies that the number of particles we can send is bounded by 
  \be \la{FiBound}
  N_{\rm send} \sim { p_+^{\rm total} \over p_+^{\rm each} } \leq g,
  \ee
  where we replaced the numerator and denominator using the inequalities \nref{ptotb} and \nref{peachb}. 
  This is good because it is less than our estimate (\ref{nequbits}) of the number of bits or qubits  necessary to open the wormhole. It would be interesting to improve this analysis to include constant factors. In particular, one would like to reproduce the fact that to teleport one qubit, one needs to send at least two classical bits.
  % establish the interaction  \nref{nequbits}.
Note that, as expected, $G_N e^t$ disappeared from \nref{FiBound}.
  
  So far, we have discussed the transfer of information in the regime where the probe approximation is roughly valid. One could also wonder whether we can send much information in the late-time case, where the correlator is described by \nref{LargeT}. In fact, we expect that the late-time effect can only be used to send an order one amount of information between the two sides. One argument is that the effective quantum channel depends on $g$ only through the phase $e^{-ig\langle V\rangle}$, so we can take $0\le g\langle V\rangle< 2\pi$. The interaction with such values of $g$ can be set up with just a few bits of exchange, so we can't send more than a few bits.

  We emphasize that when the traversable wormhole protocol is working well, at or slightly before the scrambling time where the probe approximation is valid, it is exploiting the fact that the time evolution is not a random unitary.
%\revJM{  Note that the state we get at very  late times (as in \nref{LargeT}) 
%  depends on $g$ only through a phase. For this reason we 
%  do not expect to be able to send much information at late times. We only expect to be able to send order
%  one qubits at very late times. } 
  
 \subsubsection{Modeling the quantum coupling } 
 
It is also interesting to  quantify the information transfer that we need in order to set up a quantum coupling of the form $ e^{ i \tilde g O_L O_R } $. 
To do this, we implement the operator using an intermediary  degree of freedom whose number of qubits we can easily quantify. 

Let us first assume  we had access to a single quantum coordinate parametrized by standard cannonical operators $q$ and $p$, with $[q,p]=i$.
Then it is easy to generate this interaction by performing the following sequence of unitary transformations
\be
 e^{ - i \alpha  p O_L } e^{ - i \alpha q O_R} e^{ i \alpha p O_L } e^{  i \alpha  q O_ R} = e^{ - \alpha^2 [p,q] O_L O_R} = e^{ i\tilde g O_L O_R} ~,~~~~{\rm for}~~
  \alpha^2  = \tilde g 
\ee
where we used $ e^{A} e^{B} = e^{ A +B } e^{\half [A,B]} $ and $e^{ -A} e^{ -B} =  e^{\half [A,B]}  e^{ -A -B }  $, which are valid when $[A,B]$ commutes with $A$ and $B$, which is
the case here. The operations in the left hand side correspond to simple unitaries that we perform on each of the sides and the single 
coordinate described by $p,q$. All we need is to transfer this coordinate between the left and the right systems back and forth a couple of times. 
However a single coordinate $p,q$ contains an infinite number of qubits. 

If we can only transfer a finite number of qubits, then we   perform instead the following operation
\be \la{qubits}
 e^{ - i \alpha  \sigma^2 O_L } e^{ - i \alpha \sigma^1 O_R} e^{ i \alpha \sigma^2 O_L } e^{  i \alpha  \sigma^1 O_ R} \sim  e^{ - \alpha^2 [\sigma^2,\sigma^1] O_L O_R} = e^{ i\tilde g \sigma^3  O_L O_R} ~,~~~~{\rm for}~~
  2 \alpha^2  = \tilde g
\ee
where here we have assumed that $\alpha$ is small and ignored corrections that go like $\alpha^3$.
More precisely, we assumed that $\alpha $ times the typical eigenvalue of $O_L$ or $O_R$ is small. If we assume that $\langle O_L^2 \rangle \sim 1$, then we can 
assume that the typical eigenvalue is of order one. 
 %This is justified in the limit $\tilde g \to 0$ and $K\to \infty$ keeping
%$g$ fixed. 
 We can start with a state with $+1$ eigenvalue under $\sigma^3$ and then, by performing the  series of steps in  \nref{qubits},
we get the desired operator. This involves taking a single qubit between the two systems. The qubit goes back and forth two times. 
Again we have $K$ operators, so we need $K$ qubits and this leads to $g \leq N_{\rm qubits} $ as in \ref{nequbits}.

\subsection{Stringy corrections} 
\la{Stringy}

In this subsection we  sketch  the expected modifications due to stringy effects in the bulk.
%Our main motivation   is that
%we expect that the whole stringy answer should interpolate between the weak and strong coupling behavior of the boundary quantum mechanical dual. 
 The proper way to analyze these would be to find a string theory 
realization of nearly $AdS_2$ and then compute stringy correlators in that theory. 
Instead, in this section, we will examine these effects in a heuristic form. 
Namely, we will simply assume that these stringy effects have the same effect on the
scattering amplitude as they have for flat space scattering amplitudes. 
In flat space,  
\nref{GravPhase} is replaced by 
\be \la{StringPhase}
e^{ i G_N p_+ q_- e^t} \longrightarrow   e^{  - G_N ( -i p_+ q_- e^t)^{ 1-a} }   ~,~~~~ 0 \leq a \leq 1.
\ee
where $a$ depends on the transverse momentum transfer. Here we   assume that $a$ is a constant, which is a good model for scattering on a general 
black hole background \cite{Shenker:2014cwa}. 
The gravity case is $a=0$.\footnote{ As argued in \cite{Maldacena:2016hyu}, 
stringy corrections to the Nearly-$AdS_2$ gravity results  are suppressed not by $l_s^2/R_{AdS}^2$, but by the smaller quantity
$ % \be \la{alde}
\alpha  = { l_s^2 \over R_{AdS}^2 }  { (S-S_0)\over S_0 }  
%\ee 
$ 
 where  $S_0$ is the extremal entropy and $S-S_0$ the extra entropy we have away from extremality.  
  So we have in mind that $a$ in \nref{StringPhase} is of order  $\alpha$.}
%
 %And we should think of $a$ as  roughly $\alpha$ in \nref{alde}.
\begin{figure}[t]
\begin{center}
\includegraphics[width=0.31\textwidth]{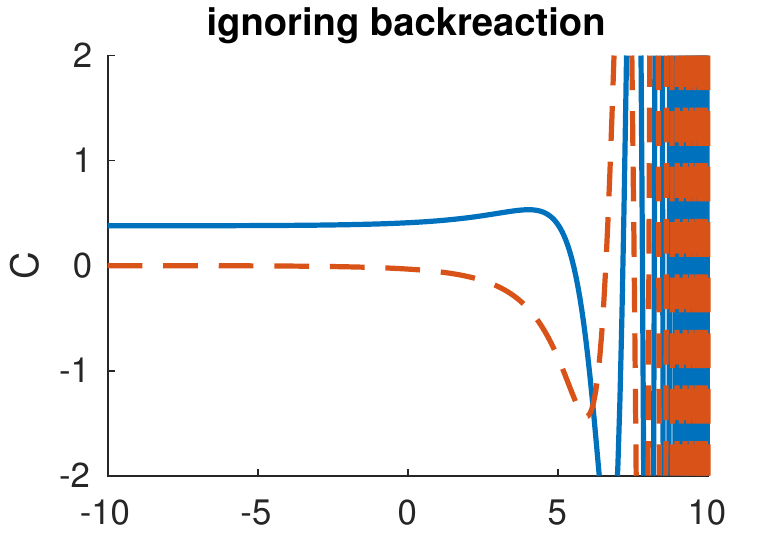}
\includegraphics[width=0.31\textwidth]{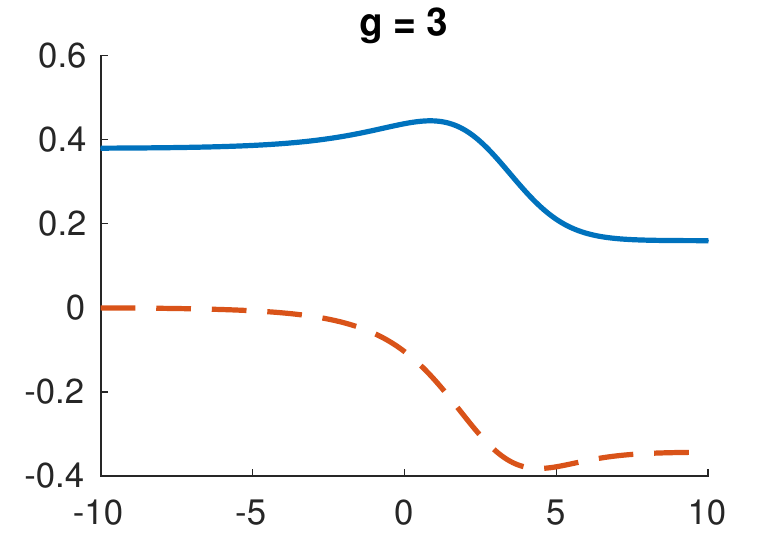}
\includegraphics[width=0.31\textwidth]{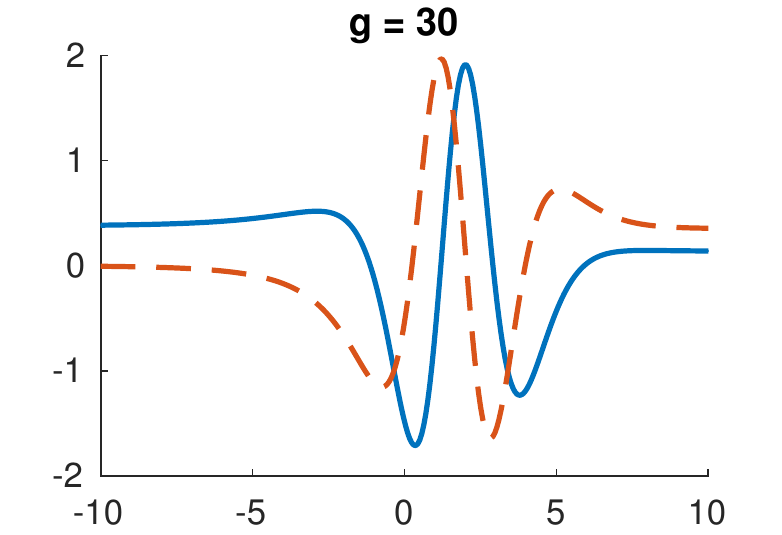}
\\
 ~~~~~(a) ~~~~~~~~~~~~~~~~~~~~~~~~~~~~~~~~~~(b) ~~~~~~~~~~~~~~~~~~~~~~~~~~~~~~~~~(c)
\caption{Real part (blue, solid) and imaginary part (red, dashed) for the correlator $C$, with $\Delta = 0.7$ and the stringy parameter $a = \frac{1}{2}$. The horizontal axis is $t + \frac{1}{1-a}\log(G_N)$. (a) Is the probe approximation \nref{StrCase2}. (b) and (c) describe the full result at finite $g$.  }\label{stringyplot}
\end{center}
\end{figure}

We can easily make this replacement in the general formula \nref{FirstRe}.
However, it is instructive to consider first the probe approximation
 where we ignore the backreaction of $\phi$ on the $O$ correlators. This is obtained by expanding 
the exponential involving $G_N$ in \nref{FirstRe} to first order in $G_N $, after  
%  to obtain \nref{ResFi}.  
making the replacement \nref{StringPhase}. This gives  
\be \la{StrCase1} 
 C^{\rm Stringy}_{\rm probe}= \int dp_+  \langle \phi_L |p_+ \rangle \langle p_+|\phi_R \rangle \exp\left[ - i\tilde g  G_N \int d q_-    ( - i p_+q_- e^t)^{ 1-a}
      \langle O^i_R | q_-\rangle \langle q_- |O^i_L \rangle 
\right]
\ee
We can further simplify this by using the wavefunctions for the $O$ and $\phi$  operators  in \nref{WF} to get 
\bea \la{StrCase2} 
 C^{\rm Stringy}_{\rm probe}
  &=&  \int_{-\infty}^0 dp_+   { 1  \over \Gamma( 2 \Delta) } { (2  i p_+)^{ 2 \Delta } \over ( -p_+) } e^{    -i 4 p_+ }    \exp\left[ - i  g  G_N  { 1  \over \Gamma( 2 \Delta) } 
 \int_{-\infty}^0  d q_-    ( - i p_+ q_- e^t)^{ 1-a} 
      { (2  i q_-)^{ 2 \Delta } \over ( -q_-) } e^{   -i 4 q_- }  
\right]
\cr
&=&  \int_{-\infty}^0 dp_+   { 1  \over \Gamma( 2 \Delta) } { (2  i p_+)^{ 2 \Delta } \over ( -p_+) } e^{    -i 4 p_+ }    \exp\left[ - i  g  G_N  {\Gamma( 2 \Delta + 1 -a) 
 \over \Gamma( 2 \Delta)    2^{ 2\Delta + 1-a} } ( - p_+e^t/2)^{ 1- a} \right]
\eea
This final formula should be compared to the gravity case \nref{ResFi}, which we get as $a\to 0$. 
There are a few interesting lessons in this formula.\footnote{As a curiosity: the integral in \nref{StrCase2} can be done analytically for $\Delta =\half$ and $a=\half$ in terms of error functions.}
\begin{itemize} 

 \item  The imaginary part is non-vanishing for generic times. We can see the difference from gravity as follows: in (\ref{ResFi}), after rotating the contour so that $ip_+$ is real and positive, the whole integrand became real. In (\ref{StrCase2}) it does not, since we have $-p+$ to a fractional power in the exponent.
 \item 
 The imaginary part is non-vanishing for any sign of $g$. 
  \item 
 There is no divergence for any particular time, since after the contour rotation the integral is convergent for any time $t$. 
 In contrast, in the gravity we found that a divergence at $t=t_d$, where $a^+(t_d) = -4$, see \nref{ResFi}. 
 %  Contrast with figure \ref{gfigure}(a). 
 \item
 We can view  \nref{StrCase1} as the propagation of the $\phi$ particles through a stringy background generated by the double trace operator. 
 The fact that the phase has a non-trivial dependence on $p_+$ indicates that the $\phi$ particle wavepackets suffer a distortion that is beyond what
 is expected by the action of a symmetry. This is likely to be painful for an observer made of $\phi$ particles!
  \end{itemize}
  
 The first two points can be understood   by realizing that in string theory 
 we have effective three point interactions between the two fields involved in
 the correlator and general stringy modes. We will discuss this further in section \ref{GravityThreePoint}.

One would have naively expected that for $a\sim 0$  \nref{StrCase1} would reproduce the gravity result, \nref{ResFi}, for all $t$, except for $t$ very close to $t_d$, where the gravity answer was diverging.  This is indeed the case for $t< t_d$. But for $t> t_d$ the stringy answer \nref{StrCase2} is different from \nref{ResFi}, compare figure \ref{gfigure}(a) to \ref{stringyplot}(a). 
 As we discussed for the full gravity answer in section \ref{Backreaction},  this is because we are considering local operators, with high frequency components. In order to get the probe stringy answer to approach the probe gravity answer, we have to include a wavepacket for the $\phi$ operator that suppresses large momenta.

 So far, we have discussed the stringy-corrected correlator $C$ in the probe approximation.  We can also study the full integral (\ref{FirstRe}) after the replacement \nref{StringPhase}. 
It is qualitatively similar to the ones we get for the gravity case ($a=0$), compare     figures \ref{stringyplot}(b,c) and \ref{gfigure}(b,c). 
 
This stringy picture is expected to connect with the results we get at weak coupling in the boundary theory in section \ref{QMModels}.

\section{Nearly-$AdS_2$ gravitational  dynamics  } 

In the above section we have performed the computations by using shock wave methods, which are still valid
in two dimensions. But the gravitational dynamics is particularly simple because these shockwaves, lacking any transverse
directions, lead to no tidal forces. In fact, in nearly-$AdS_2$ gravity, 
  we can view the shock waves as localized near the horizon or also near the boundary, as
  boundary degrees of freedom. In this subsection, we will take the latter point of view and consider the 
dynamics as due to degrees of freedom living at the UV boundary. In this picture, $AdS_2$ is totally rigid, with a 
fixed metric.  Gravitational effects are encoded in a degree of freedom that tells us how the UV boundary 
moves in this fixed $AdS_2$ space. Or, alternatively, how the boundary time is related to the $AdS_2$ time
coordinates \cite{Almheiri:2014cka,Jensen:2016pah,Maldacena:2016upp,Engelsoy:2016xyb}. This gives rise to a rather simple and intuitive picture for the gravitational dynamics. 

%In this subsection we discuss a bit more the gravity picture in the nearly-$AdS_2$ context, where the gravitational dynamics is captured by the 
%Schwarzian action \cite{}. The goal of this section is to present an alternative derivation of the above formulas, as well as a nice lead into the computations in the 
%SYK model at low temperatures. 

Let us see this more concretely. 
%In a nearly-$AdS_2$ theory we have a metric which is exactly $AdS_2$, and bulk fields  propagate on this $AdS_2$ metric.
%It is important to remember that this $AdS_2$ is going to be cut-off by the presence of a UV boundary. The whole gravitational dynamics will be encoded by a 
%degree of freedom that we can view as living on the boundary. We can call it a boundary graviton. 
The gravitational physics of nearly $AdS_2$ spaces is  described by the action 
\be
S = \int_{\mathcal{M}} \Phi ( R + 2) +  2 \Phi_b \int_{\partial\mathcal{M}} K 
\ee
where it is important to note that there is a boundary. At this boundary, we fix the value of the   dilaton field $\Phi$ to  $\Phi_b$.  
The functional integral over $\Phi$ sets  $R+2=0$, fixing the metric to $AdS_2$ and removing the first term from the 
action.
The action then reduces to the extrinsic curvature term.  In principle, we want to integrate over all boundary trajectories of a fixed proper length. 
It is convenient to integrate over all trajectories by introducing a lagrange multiplier  that we could use to fix the length later. Then 
the action looks  like
\be \la{ActRel}
S = 2 \Phi_b \int \sqrt{h} K - M \int \sqrt{h}.
\ee
% It is useful to think of $M$ as a Lagrange multiplier that sets the length of the boundary to the desired value.
 $M$ is related to the energy of the system, 
since the length (in Euclidean space) is the inverse temperature, and energy is the conjugate variable. More precisely, up to a constant shift, $M$ is minus the energy. It is convenient to shift by $2\Phi_b$, so that
\be
M = 2 \Phi_b -   E_s  % ~,~~~~~~~ ds = { d u \over \epsilon } 
\ee
where $0\le E_s\le 4\Phi_b$ is the energy conjugate to proper time on the boundary. 
%It turns that \nref{ActRel} can be viewed as the action of a relativistic massive particle in an electric field (see appendix A).
 Instead of 
discussing this in great detail, 
  for our present purposes we only  need to note the following points. (See appendix \ref{app:kinematics} for more details.)

\begin{itemize} 

\item 
The action \nref{ActRel} is local along the boundary. 

\item 
The action \nref{ActRel} is invariant under the $AdS_2$ isometries. It is $SL(2)$  invariant. 

\item
The time along the boundary theory is proper time along this trajectory. 

\end{itemize} 

\begin{figure}[t]
\begin{center}
\begin{subfigure}{.31\textwidth}
  \centering
  \includegraphics[width=.6\linewidth]{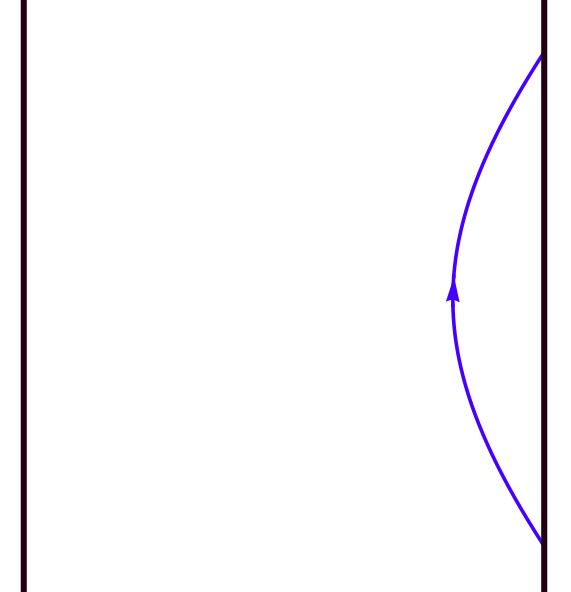}
  \caption{}
\end{subfigure}%
\begin{subfigure}{.31\textwidth}
  \centering
  \includegraphics[width=.6\linewidth]{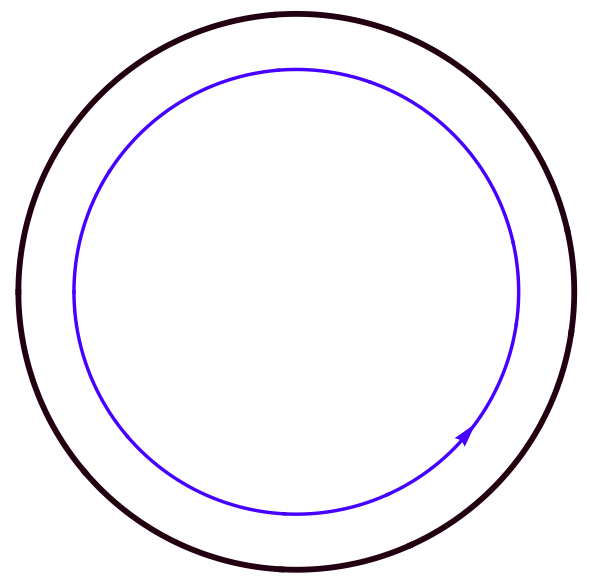}
  \caption {}
\end{subfigure}
\begin{subfigure}{.31\textwidth}
  \centering
  \includegraphics[width=.6\linewidth]{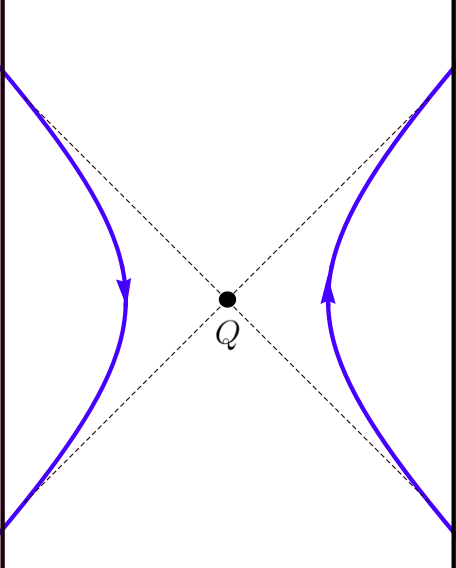}
  \caption{}
\end{subfigure}
\caption{A classical solution for a single boundary trajectory in Lorentzian AdS${}_2$ (a) and Euclidean AdS${}_2$ (b). In (c) we show the pair of boundaries associated to the thermofield double. The projection of the vector $Q$ onto AdS${}_2$ is the bifurcation point.}
\label{Hyperbolas}
\end{center}
\end{figure}

We then conclude that gravity reduces to the dynamics of a boundary degree of freedom. In Euclidean space a classical solution for the action \nref{ActRel} is simply a circle, 
representing $AdS_2$ with a boundary at a finite location. The radius of the circle is specified by the ratio of the values of $\Phi_b$ and $M$. 
Going to Lorentzian time, this circle trajectory gives rise to hyperbolic like trajectories as shown in figure \ref{Hyperbolas}. 
Mathematically, these can be written simply  in terms of embedding coordinates 
\bea \la{traj}
 &~&Y. Q= -2\Phi_b  ~,
 \\
 & ~&  Y^a = (Y^{-1},Y^0 , Y^1) ~,~~~~  Y.Y =- (Y^{-1})^2 - (Y^0)^2 + (Y^1)^2 =-1\notag
\eea
 where $Q^a$ is   a vector in $R^{2,1}$ and the contraction is with the standard $R^{2,1}$ metric, which is diag$(-1,-1,1)$.
 In our regime, with $M< 2\Phi_b$,
  $Q^a$ is a timelike\footnote{  In constrast, in the case of ordinary massive bulk  geodesics we have $Y. \tilde Q =0$,  with $\tilde Q^a$ spacelike.  }  vector in $R^{2,1}$.
See appendix \ref{app:kinematics}.  The fact that solutions can be described by \nref{traj} is easy to see. First we notice that this is true for the circular trajectory in Euclidean space for a
particular choice of $Q^a$. Then we go to Lorentzian signature and by an $SL(2)$ transformation we can choose a generic $Q^a$. 
The solutions to the equation \nref{traj} consist of a pair of hyperbolic-like trajectories, as indicated in figure 
\ref{Hyperbolas}(c). The center of this pair is at a point $Y^a \propto Q^a $. So, when we look at \nref{traj} we can think of the direction of $Q^a$ as specifiying
the point where the asympotic lines of the two hyperbolas meet. This point $Q^a$ is light-like separated from the points where the trajectories hit the boundary. This 
can be seen by noticing that the product $Y.Q$ in \nref{traj} remains finite as the components of $Y$ are taken to infinity. 
 The null lines emanting from this point also show that the two sections of the hyperbolas are causally disconnected.

A useful point is that the vector $Q$ can also be thought of as the associated $SL(2)$ charge of the right boundary trajectory (and minus that of the left one).
In other words, the left and right $SL(2)$ charges are $Q_R^a = -Q^a_L =  Q^a$. This makes it possible to work out the backreaction of matter on the boundary trajectories using $SL(2)$ charge conservation.

\begin{figure}[h]
   
 \begin{center}
\begin{subfigure}{.31\textwidth}
  \centering
  \includegraphics[width=.83\linewidth]{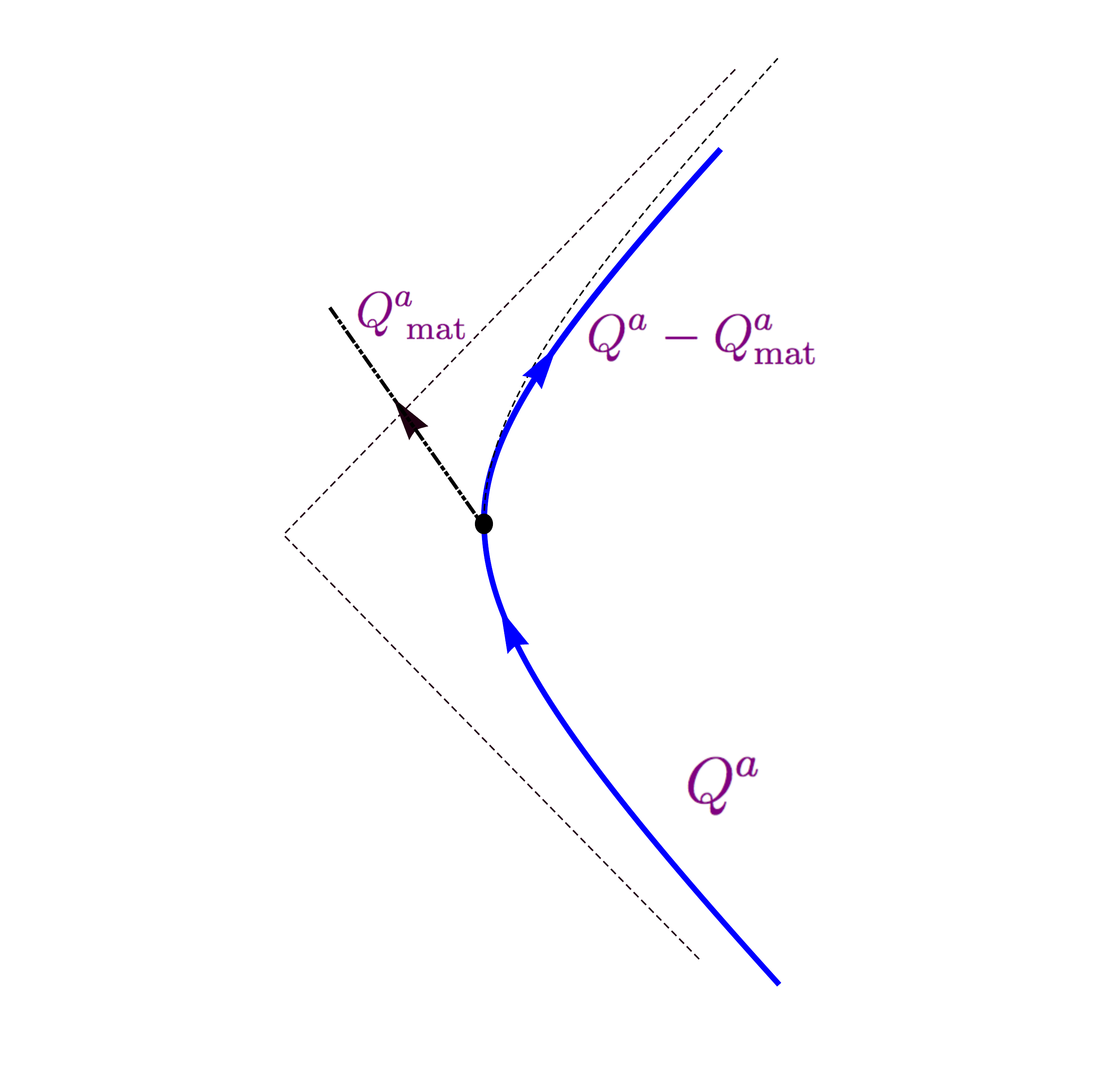}
\end{subfigure}%
\begin{subfigure}{.31\textwidth}
  \centering
  \includegraphics[width=.83\linewidth]{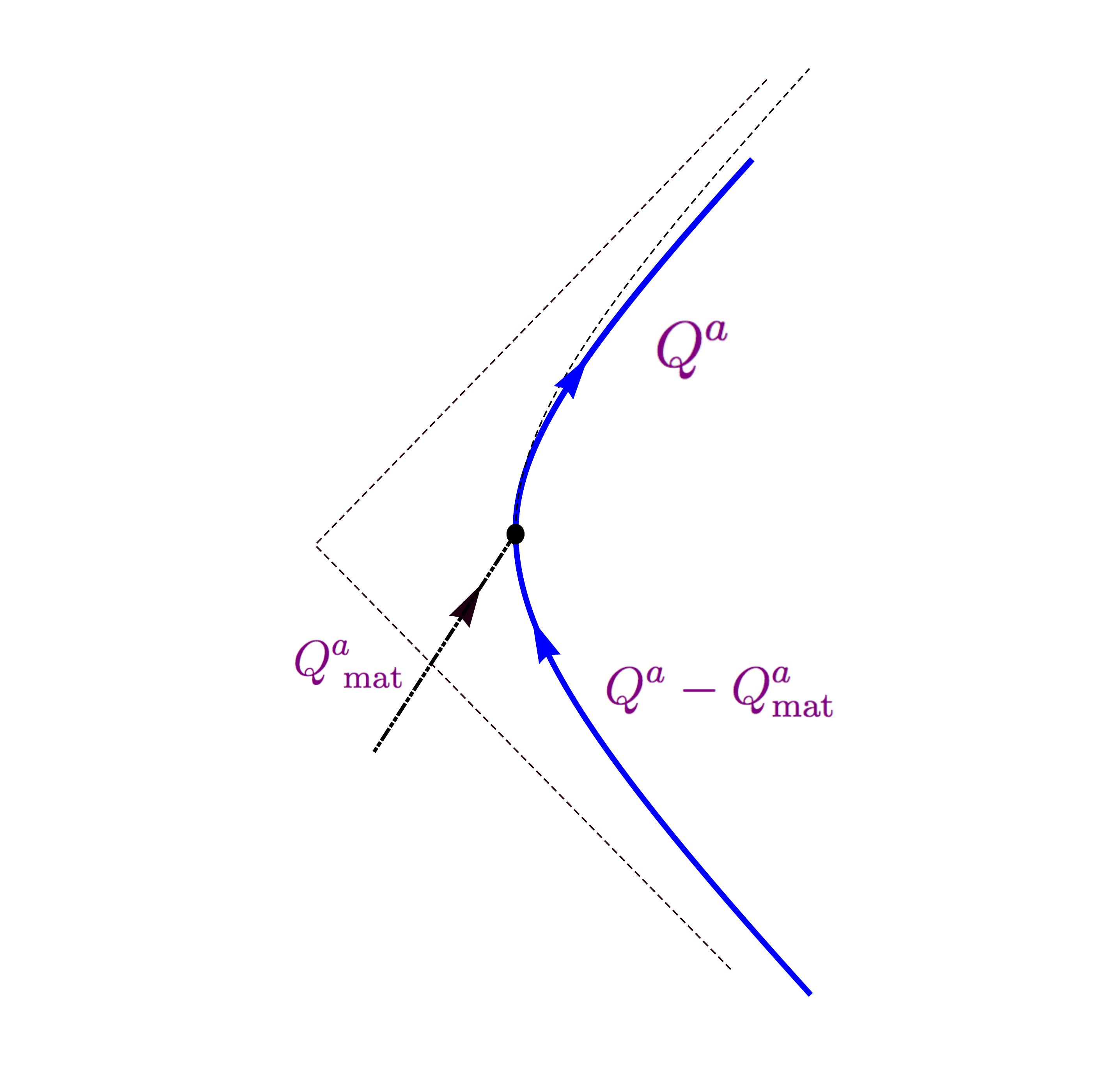}
\end{subfigure}
\begin{subfigure}{.31\textwidth}
  \centering
  \includegraphics[width=.83\linewidth]{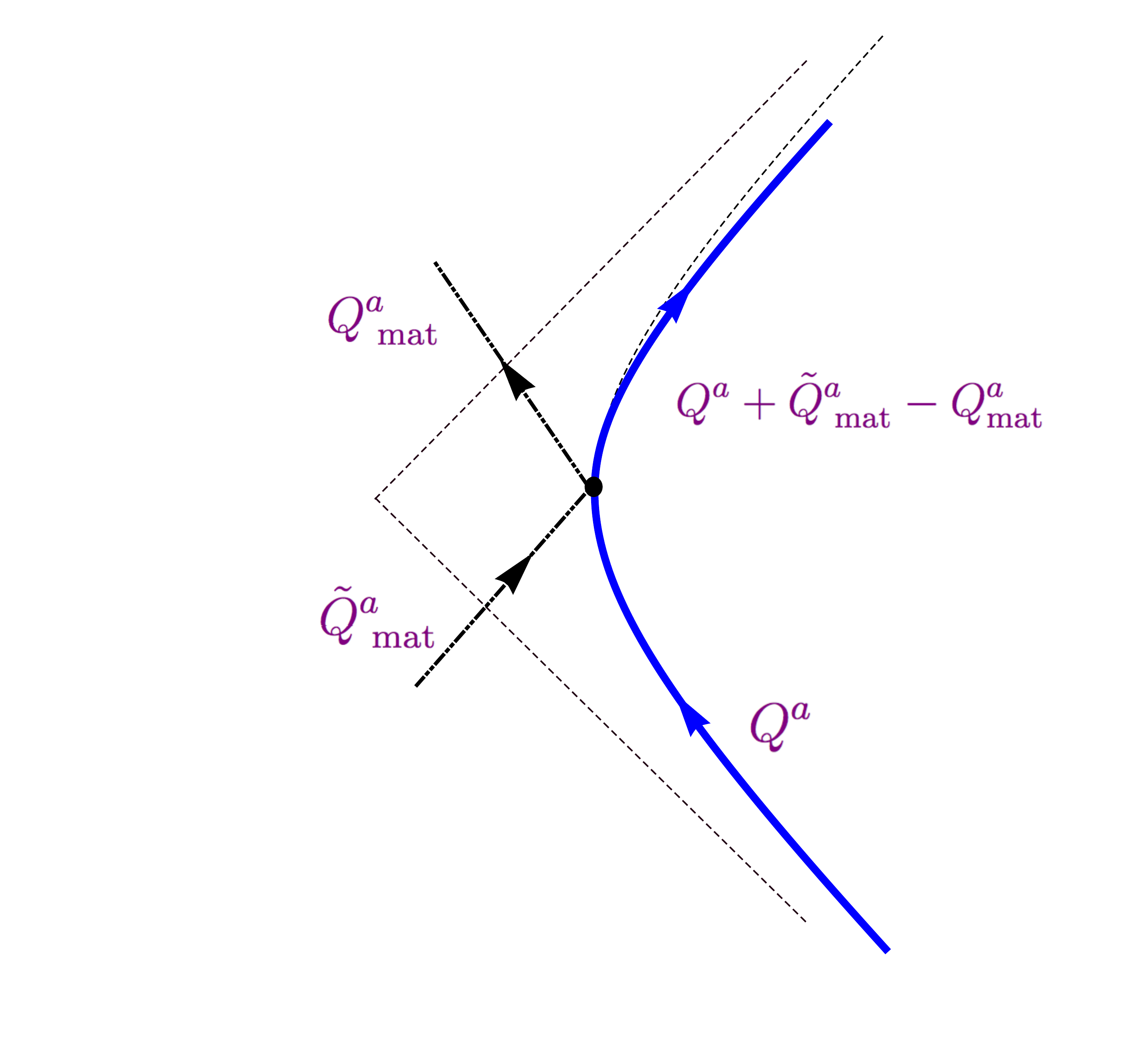}
\end{subfigure}
\includegraphics[scale=0.18]{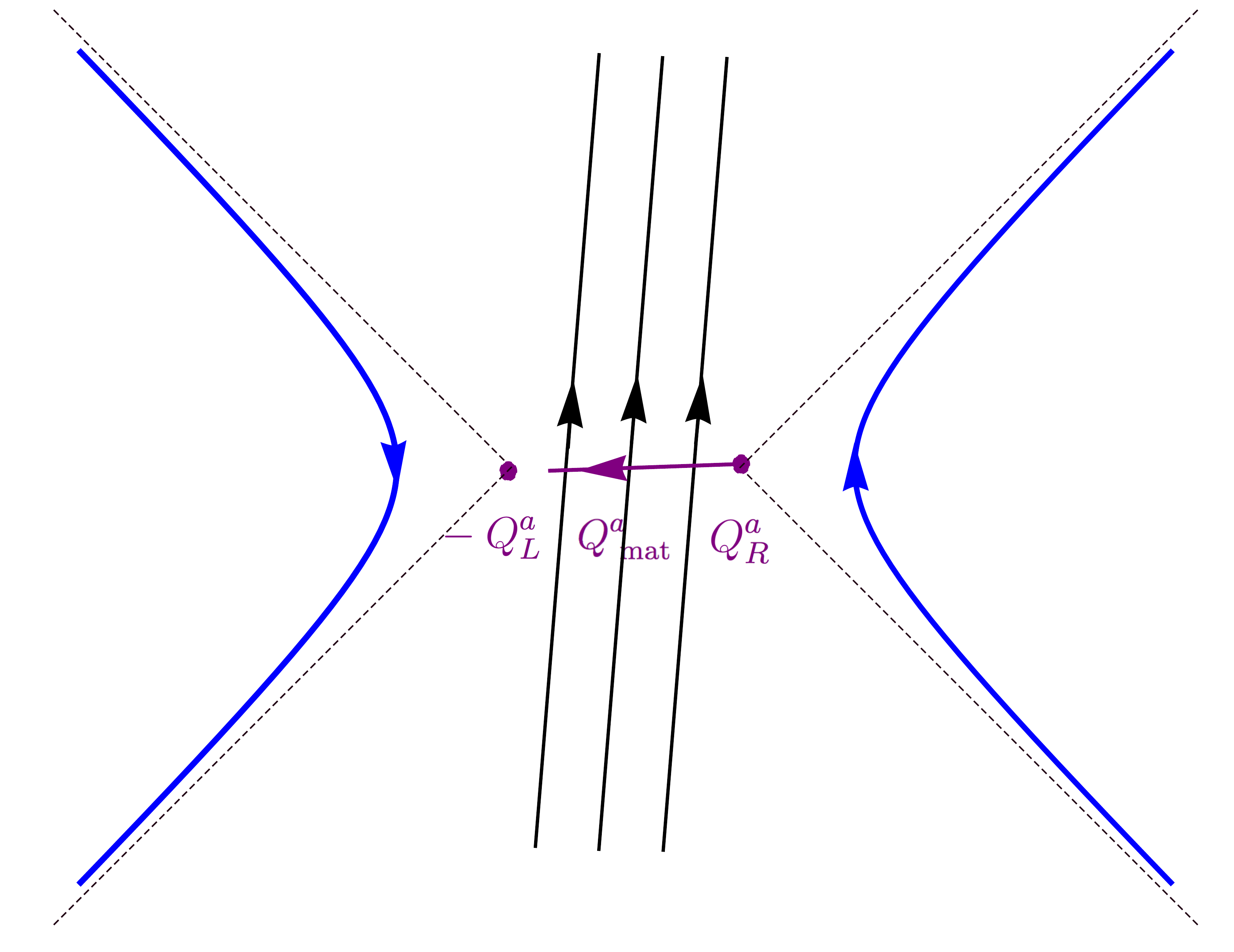}
\caption{{\bf Top:} if we insert an operator that creates,  absorbs or reflects a particle, the boundary trajectory is kicked farther outwards. {\bf Bottom:} In a general two-sided configuration, the horizons will not meet, since $- Q_L^a -Q_R^a  $ is a leftward-pointing spacelike vector that represents the $SL(2)$ charges of the matter.}
\label{MatterInside}
\end{center}
\end{figure}

 In this picture the insertion of a bulk field operator of mass $  m$ at some time along the boundary can be viewed as  a vertex process where  
 a particle of mass $M$  spits out a bulk massive particle of mass $  m$
 and    the boundary becomes a particle of mass $M'$.
  Note that the ADM energy is not conserved, the difference 
 $ M-M'$ is proportional to the energy we injected.
  Importantly, the whole process should conserve the $SL(2)$ charges.  This implies, in particular, that 
   two dimensional energy-momentum is conserved at
 the vertex. Therefore, the emission of the particle leads to a kick in the boundary trajectory which is in the direction that we expect from energy momentum conservation at
 the vertex, see figure \ref{MatterInside}. 
  The new trajectory is determined by equating the $SL(2)$ charges before and after $Q_{\rm after}^a+Q_m^a = Q^a_{\rm before} $, where $Q_m^a$ are the three 
  $SL(2)$ charges of the particle emitted and $Q^a_{\rm before/after}$ are the charges of the boundary trajectory. 
  We can similarly evaluate the kicks that result from the absorption of the particle and the reflection of a particle from the boundary. In all these cases the kick is 
  towards the outside, in the direction that pushes the boundary further away from the interior.  
   
 It is also interesting to consider a two-sided situation where we have a state which contains some extra particles on top of the thermofield double. In such a state we 
 have charges $Q^a_{\rm mat}$ associated to the matter fields in the interior. Since the total state should be $SL(2)$ invariant we conclude that  
 $Q_L^a + Q_{\rm mat}^a +Q^a_R =0$. Now, it is important to note that the vectors $-Q_L$ and $Q_R$ are the points where the past and future horizons of the boundary 
 trajectories meet. Matter in the interior has total positive $Q_{\rm mat}^2$ and this means that 
  $-Q_L$ and $Q_R$ are never timelike relative to each other. In addition, positivity of the null momenta $-P_-$ and $-P_+$ ensure that $Q_{\rm mat}^a$ actually points to the left, so in general $Q_R$ is to the right of $-Q_L$, see appendix \ref{app:kinematics}. If the matter inside bounces off or is absorbed by the boundaries, the resulting kicks will send these  the boundaries even further appart.  This means that no signal can be sent between the boundaries.  See figure \ref{MatterInside}.

\begin{figure}[t]
\begin{center}
\begin{subfigure}{.25\textwidth}
  \centering
  \includegraphics[width=.9\linewidth]{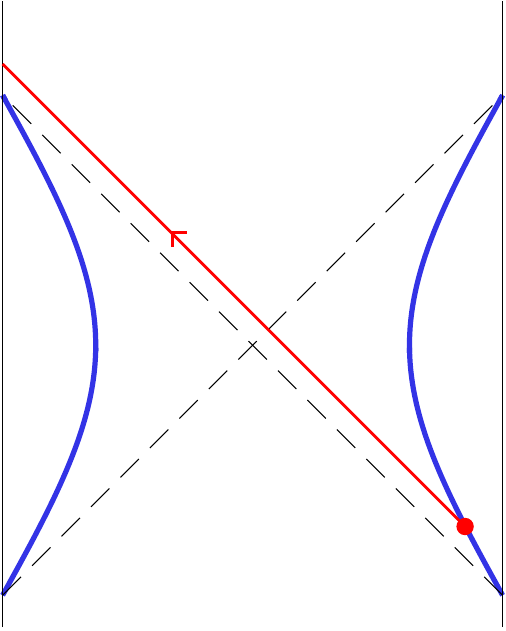}
  \caption{}
\end{subfigure}\hspace{10pt}
\begin{subfigure}{.25\textwidth}
  \centering
  \includegraphics[width=.9\linewidth]{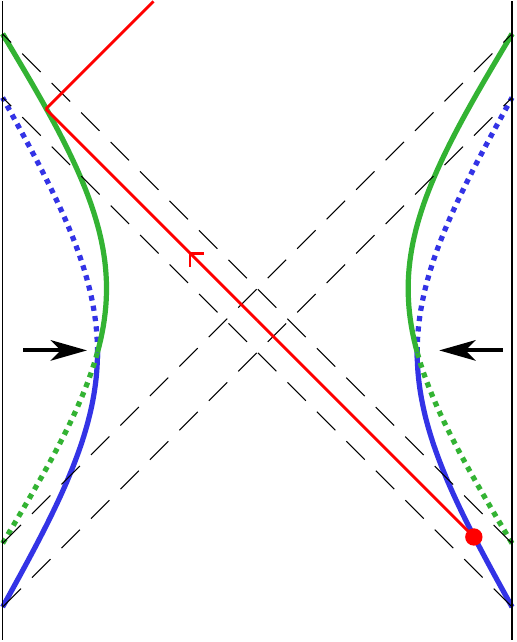}
  \caption {}
\end{subfigure}\hspace{10pt}
\begin{subfigure}{.27\textwidth}
  \centering
  \includegraphics[width=.9\linewidth]{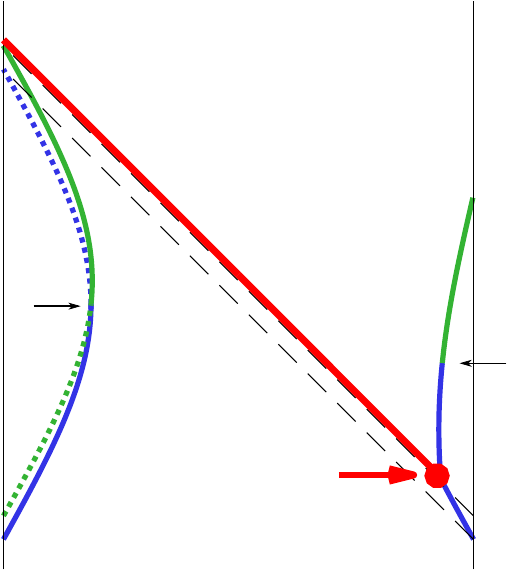}
  \caption{}
\end{subfigure}
\caption{In (a) we show the thermofield double configuration. A message sent from the right boundary does not reach the left. In (b) we act at time zero with $e^{igV}$, which exerts a force indicated by the arrows, pushing the boundaries onto the green trajectories. The message now makes it to the left boundary. In (c) we try to send a big message. Creating the message exerts a force (red arrow). This doesn't affect the left trajectory directly, but it makes the two boundaries farther apart at time zero, so  $e^{igV}$ generates only a tiny kick (black arrows) that is too weak to make the wormhole traversable.}\label{UVparticlesForce}
\end{center}
\end{figure}

% SHOULD WE SAY THAT IS IT LIKE A QFT WITH A NUMBER OF PARTICLES GIVEN BY THE DENSITY OF STATES?. 

 If we want to send a signal between the boundaries we need to bring them closer.
 This can be done with an attractive force between the boundaries. In fact, this is precisely what 
 the Gao-Jafferis-Wall interaction does, as we explain below. 
 We consider an interaction between two bulk field operators at the two boundaries,
 $ V = \tilde g O_L O_R$. For small $\tilde g$ the leading effect comes from the expectation value of this 
 operator,   $\langle O_L O_R \rangle$. This expectation value depends on the distance between the 
 UV boundary particles. So the effective potential is  $V_{pot} = - \langle V \rangle \propto -\tilde{g}e^{-m\rho}$, where $\rho$ is the distance in $AdS$. By including this term in the action at a particular time, we are briefly turning on a potential energy that gives rise to a force between the
 two particles. We can view it as a force due to the exchange of scalars and it is an attractive force if $\tilde{g}> 0$. Since we are turning on 
 the interaction  only briefly, this is an impulsive force. If we start near the thermofield double state, this small kick will be enough to pull together the two accelerating UV particles so that they can now send signals
 to each other, see figure \ref{UVparticlesForce}. From this perspective, it is clear why we get traversability for one sign of $g$ and not for the other; for $g<0$ the potential would be repulsive and we would kick the boundaries farther apart! As described above, we can increase the size of this effect by considering $K$ bulk $O$ fields, with large $K$.

Note that if we consider the state after the force is applied and evolve it backwards and forwards with the two decoupled left and right Hamiltonians, then we get a full history where the boundaries remain causally disconnected (the solid plus dotted green lines in figure \ref{UVparticlesForce}(b)). However, the history that is relevant for our setup is one where we begin with the original thermofield double state, and we perturb by $OO$ at $t = 0$, switching from the blue to the green boundary trajectories in figure \ref{UVparticlesForce}(b). Here, the configuration is traversable. 
 
This picture of the attractive force (for $g> 0$) also helps us understand what happens to the energy, or ADM mass of the particles. A force that is switched on 
 at times $t_L = t_R=0$ is orthogonal to the velocity and it   gives no change in the energy. 
 If the interaction is turned on at $t_L=t_R > 0$, when the UV particles are moving away from each other, the force  opposes their motion and will cause the energy to 
 decrease. This also decreases the entanglement between the two sides \cite{Gao:2016bin}. On the other had, if 
  the force acts at $t_L = t_R < 0$, when the particles are approaching each other, then the energy will be increased, and consequently also the entanglement increases. 
 We expect that the $OO$ perturbation gives a negative null energy in the region after it acts, $- P_-= 
 \int dx^- T_{--} \leq 0 $. However, the actual ADM energy, or killing energy, 
 is given by $E = \int dx^- x^- T_{--}$. The fact that this can be positive,  negative or zero
  can be understood as follows. The idea is that the operator insertions create a pulse of negative energy followed by a smaller 
  pulse of positive energy. The fact that these two are separated along $x^-$ implies that 
  these two pulses can then lead to various values for the energy $E$, even though $(-P_-)$ is always negative.\footnote{A comment related to the previous paragraph is that the full integrated null energy $\int_{-\infty}^\infty dx^+ T_{++}$ is non-negative in the alternate history obtained by taking the state after $OO$ is applied and evolving {\it backwards} and forwards with the decoupled Hamiltonians.}

 We can also pictorially understand the backreaction effects that limit the amount of information we can send. 
  As we increase the number of  $\phi_R$ particles that we send from the right, we will increase 
their momentum along the null direction.  
This  will ``kick'' the trajectory of the right boundary particle away from the left particle. As this kick becomes stronger the 
distance between the insertion points for $O_L$ and $O_R$ becomes larger. This makes the potential term smaller, and the associated force weakens. Eventually, the force is too weak to open the wormhole, see figure \ref{UVparticlesForce}(c).

The whole description we have given here is equivalent to what we would obtain from the Schwarzian action. More precisely, the Schwarzian action is  obtained 
 when $M$, $Q$ both go to infinity in a particular scaling limit, 
\be \la{ScalMQ}
\Phi_b = { \Phi_r \over \epsilon } ~,~~~ M = 2 \Phi_b -  \epsilon E_u ~,~~~~~~~~ ds = { du \over \epsilon } ~,~~~~~ \epsilon \to 0 
\ee
with $\Phi_r$, $u$ and $E_u$ remaining constant. See appendix \ref{app:kinematics} for some details.  In this limit,  the pictures we have drawn correspond 
precisely to the result \nref{ResGen}  in section \nref{GravComp}, with $G_N = { \beta \over 2 \pi \Phi_r} \ll 1$.

 The conclusion of this section is that gravitational dynamics in nearly-$AdS_2$ has a very simple mechanical 
 description in terms of the motion of the UV particles. 
 In particular, 
  traversability can be understood as the result of a force between the UV particles. Let us emphasize that this is simply a rewriting of the gravitational dynamics, 
we have not used any duality to a boundary theory in this discussion. 

As a side remark, note that the complexity = volume conjecture \cite{Susskind:2014rva,Stanford:2014jda} here would say that complexity = (distance between the two boundaries).

 \section{Relation  to ``black holes as mirrors''  and cloning} 
 \subsection{Black holes as mirrors } 
 
 In \cite{Hayden:2007cs} Hayden and Preskill noticed an interesting property of general chaotic quantum systems, including the ones suposedly describing 
 black holes. 
  They imagined an observer, called Bob, who has  access to a quantum state that is maximally entangled with a black hole. Bob knows the full quantum state of the black hole
  and his system but he can do experiments only on his system. 
   (Bob can get to this state by  watching matter collapse into a black hole and then collecting the Hawking radiation that comes out for a time
  long enough so that half of the original system has evaporated. He knows the state of the full system from his knowledge of the exact equations of quantum gravity.)
 In this situation if a second observer, called Alice, drops a qubit into the black hole, then Bob can quickly recover it as follows. 
 He waits for a scrambling time, and then he collects a few new Hawking radiation modes. After adding these to his quantum computer, Bob can perform an operation to distill Alice's qubit. %Hayden and Preskill prove that such an operation exists, but they do not give a recipe. In general, the operation has been argued to be very difficult \cite{Harlow:2013tf}.

How many additional radiation modes does Bob need? If he captures the new Hawking modes coherently, as qubits, then he needs a few more than the number of qubits in Alice's message. On the other hand, if he captures only classical information about the new Hawking modes (but of course maintains quantum control over the early radiation in his computer), then he needs a few more than twice the number of qubits of Alice's message. See appendix \ref{app:HP}.
 
    We can interpret the traversability phenomenon as a specific and conceptually straightforward way to visualize how Bob can achieve this recovery. It is giving one explicit algorithm that can be followed.  
    The algorithm is the following. We assume  that Bob knows the initial state of the matter that collapsed into the black hole as well as the laws of physics. After enough evaporation has taken place, the density matrix of the black hole will be close to a thermal state, so Bob can act on his large supply of Hawking radiation to create a second black hole which is approximately in the thermofield double state with respect to the first black hole. It might be necessary to 
   consider black holes in $AdS$ and to turn on and off some boundary interaction to achieve this. For simplicity, we will consider here only $AdS$ black holes. 
   This process was argued in \cite{Harlow:2013tf} to be computationally hard (exponentially hard in the entropy of the system), 
   but it would be possible for the boundary observer, 
   an observer who views the $AdS$ geometry as a dual description to a system he has in his well equipped lab.
   
   Once Bob has produced the thermofield double, we expect that the geometry would be that of a wormhole 
   \cite{Israel:1976ur,Maldacena:2001kr,Maldacena:2013xja}. This is the only assumption we make. 
    We view the right  black hole as Alice's black hole and the
   left  black hole as the one that Bob has access to. 
  The process is now the same as  in  the traversable wormhole setup.  
     Alice dropping the qubit is the same as 
   the insertion of the perturbation at early times produced by $\phi_R$. Bob collecting a few qubits and feeding them to his quantum computer is  similar to 
   the insertion of the operator $V$ that couples the left and right black holes. This coupling results in some transfer of information between both black holes. 
   The signal reaching the left side is the success of Bob's decoding operation. More precisely, instead of adding the double trace interaction we can imagine
   that we measure the right operator on Alice's side and that we transfer this classical information to the left side (Bob's side) 
    where we then act with the right operator in a way that depends on the classical information, as in (\ref{Tele}).   This is
   exactly the version of the Hayden Preskill experiment where Bob collects classical information. In this version, we have a teleportation process. 
   A qualitative   discussion of teleportation and  wormholes can be found in  \cite{Susskind:2014yaa,Susskind:2016jjb}, see also \cite{Numasawa:2016emc}.
    
    Of course, the fact that information can go from one side to the other is not surprising, since we are indeed sending information. What is interesting is
    the manner in which the information goes.  The point is that the signal feels that it travels through empty space. If we were to follow this protocol, with a suitably 
    large black hole and suitably large classical information exchanged between the two sides so that instead of a small signal we can send an observer, then 
    the observer would not feel anything dramatic as he/she is being teleported.  In standard teleportation, one has the feeling that the qubit getting teleported is 
   seriously affected by the classical measurement. This would make us think twice before agreeing  to be teleported. However, in this wormhole configuration, it 
    is quite pleasant, as pleasant as floating in empty space. 
    
    \subsection{Comments on the cloning paradox } 
   
\begin{figure}[ht]
\begin{center}
\includegraphics[width=.4\textwidth]{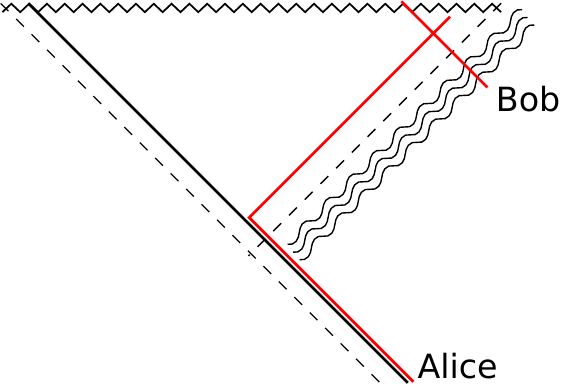}
\caption{A cloning puzzle \cite{Hayden:2007cs}. Alice throws a message (red) into a black hole in a machine that shoots it out to the right after crossing the horizon. Bob decodes the message (also red) from the Hawking radiation and then jumps in. Does he find a second copy inside?}
\label{CloningUsual}
\end{center}
\end{figure}

    One of the points of the Hayden and Preskill paper was to sharpen a potential cloning paradox  credited to Preskill in \cite{Susskind:1993mu}. 
    This paradox is the following. Suppose that Alice sends a message into the black hole. If we assume that black holes preserve information, this information 
    can be recovered by looking at the Hawking radiation. If Bob recovers this information and then jumps into the black hole, will he be able to see Alice's message too?
     See figure \ref{CloningUsual}.
     The problem is that quantum states cannot be copied, so   Bob should not be able to see both. What prevents Bob from seeing both? It seems allowed by the geometric
    picture one would associate to this process, as in figure \ref{CloningUsual}.
    One possible problem is that Alice would have to send her message with very high energy in order to reach Bob \cite{Susskind:1993mu}. Hayden and Preskill drastically shortened the time required for Bob to do the decoding, but even so, Alice's message might need to be sent with planckian energy \cite{Hayden:2007cs}. However, we will not find it necessary to use this fact.

    Traversable wormholes give us an interesting new perspective on this decoding process and show us  explicitly how the paradox is avoided. 
    The main point is that the paradox is avoided by thinking about ordinary general relativity in a wormhole geometry. We do not need to appeal to other new physics. 
    In what follows, we discuss a few simple examples that illustrate this. 
    
\begin{figure}[ht]
\begin{center}
\begin{subfigure}{.25\textwidth}
  \centering
  \includegraphics[width=.9\linewidth]{figures/1a}
  \caption{}
\end{subfigure}\hspace{10pt}
\begin{subfigure}{.25\textwidth}
  \centering
  \includegraphics[width=.9\linewidth]{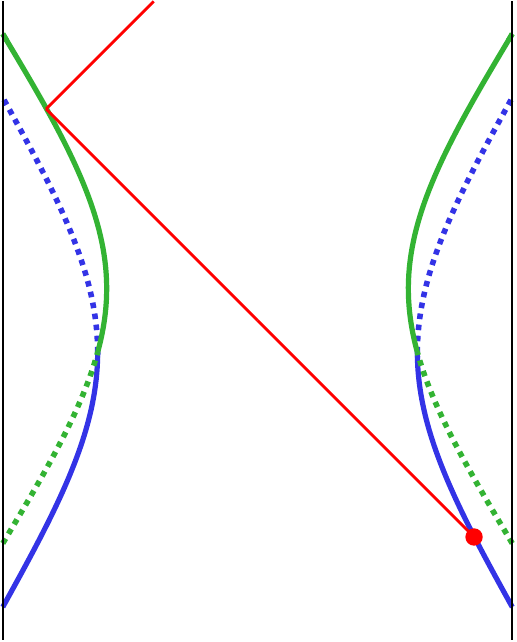}
  \caption{}
\end{subfigure}\hspace{10pt}
\begin{subfigure}{.25\textwidth}
  \centering
  \includegraphics[width=.9\linewidth]{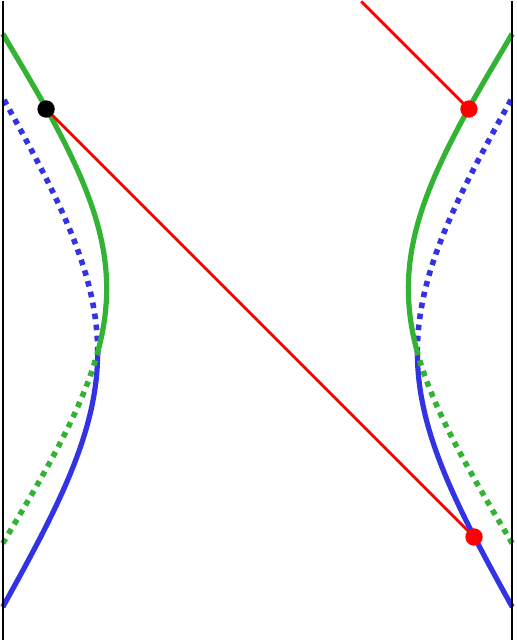}
  \caption{}
\end{subfigure}
\caption{Simple experiment: (a) Alice adds her message. (b) After applying $e^{igV}$, the message propagates from the R boundary to L. If we evolve backwards in time without undoing the $e^{igV}$ operation, then we follow the dotted backward extrapolation of the green boundaries, and the message does not intersect the R system. Bob can either allow the message to bounce off the L boundary (b), or capture it and (if he likes) carry it over to the R system and then jump in with it (c).}
\label{SimpleMessage}
\end{center}
\end{figure}

    First let us consider the simple message transmission problem we have discussed in previous sections of the paper. Alice drops the message on the right black hole, then we act with the double 
    trace interaction that couples the left and right black holes. In the state that results after this interaction, Bob can read the message in the left system simply by waiting, by evolving towards the future. 
    In that same state,  Alice now has lost the message, in the sense that even if she attempts to evolve her system backwards in time she would not be able to retrieve it. 
    See figure \ref{SimpleMessage}.  
    Now Bob can switch over to the right side and jump in. If Bob jumps in after having previously collected
     the message, the message will not be present in the bulk any more.   Alternatively, Bob could have decided not to read 
     the message and instead let it bounce off the left boundary and back into the black hole. Then, when he jumps in from the right side he could see it in the bulk. Either way, there is only one copy of the message.

 %   If we picture the message as a ball that can be in either side of  a basketball court, the it seems that what happened is that we have moved the lines of the court, the
  %  ball did not feel anything special.  
    
         \begin{figure}[ht]
\begin{center}
\begin{subfigure}{.25\textwidth}
  \centering
  \includegraphics[width=.9\linewidth]{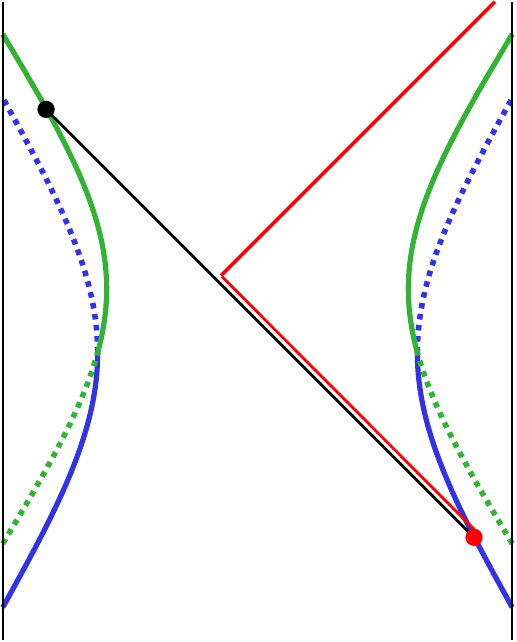}
  \caption{}
\end{subfigure}\hspace{10pt}
\begin{subfigure}{.25\textwidth}
  \centering
  \includegraphics[width=.9\linewidth]{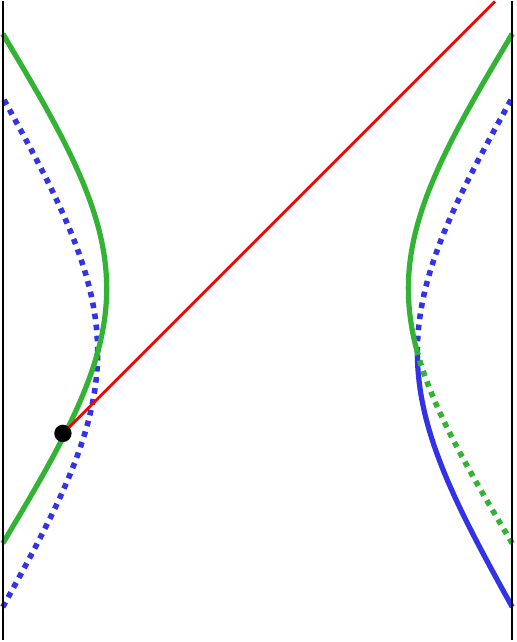}
  \caption{}
\end{subfigure}\hspace{10pt}
\begin{subfigure}{.25\textwidth}
  \centering
  \includegraphics[width=.9\linewidth]{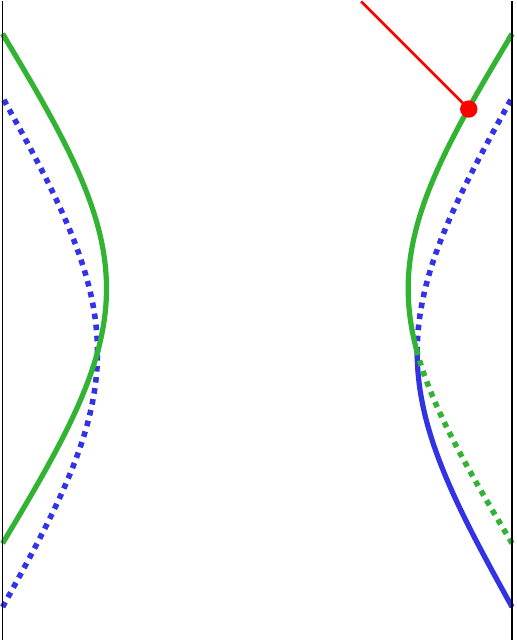}
  \caption{}
\end{subfigure}
\caption{More complicated experiment: Alice sends the message (red) inside a machine (black) that spits the message out to the right after falling into the black hole. In (a) Bob has applied $e^{igV}$ and recovered the empty message-sending machine at the left boundary. He then runs the L system back in time, and recovers the message at an early time (b). Finally, he jumps into the R black hole along with the message (c). Note that there is no second copy of the message behind the horizon, because Bob removed it in step (b).   }
\label{MessageEvolveBack}
\end{center}
\end{figure}

    Now, let us consider a slightly more elaborate configuration similar to the setup in figure \ref{CloningUsual}. This time, Alice sends the message inside a machine that will spit it out in a direction parallel to 
    Alice's black hole horizon, but just behind it. In this case, when Bob tries to retrive the message on the left side, he will just find an empty machine, without the message, see figure \ref{MessageEvolveBack}(a). If he then decides to go to Alice's side and
    jump in, he can find Alice's message behind the horizon. This is fine, because Alice's machinery foiled his plan and he doesn't have a second copy himself.
    
    Let us now consider the same situation. But now, when Bob finds the empty machine, he extracts it, and runs his side of 
    the system backwards in time. If we evolve Alice's message back in time in a bulk with the machine removed, it will sail right past the point where it formerly interacted with the machine, and emerge from the horizon of the left black hole where Bob can retrieve after a suitable amount of backwards evolution. See figure \ref{MessageEvolveBack}(b).
    If he extracts the message, then the message will not be there in the geometry any more. He can then go to Alice's side and jump in. This time, he will have a copy of the message with him, but 
    he will not find a second copy behind the horizon, becuase his extraction process has removed it! See figure \ref{MessageEvolveBack}(c). Of course, instead of Alice's message he might see something else, such as
    some perturbations resulting from his extraction process, but these will not have Alice's information. 
    
    The lesson of these examples is simply that Bob's extraction process removes Alice's message from the interior. This realizes the idea that we cannot talk about the information in the Hawking radiation 
    and the interior as independent from each other, as spacelike separated, commuting observables. The wormhole connects them,  so that extracting information from
    Hawking radiation removes it from the interior. It can be said that the two are complementary  (or non-commuting) observations \cite{Susskind:1993if,Susskind:1993mu}. For this to work   it was crucial to take into account the wormhole geometry that joins the two systems. 
    Previous discussions of the cloning paradox seemed to imply that this complementarity was going to be enforced by some unknown transplanckian physics. 
    As an analogy we could say that it looked like  a  transaction was taking place  where the police was not looking, slightly beyond the reach of the law. 
    Instead we have a perfectly legal transaction. The police is looking and has no complaints.
     In other words, it follows the standard laws of general relativity and quantum fields in the wormhole geometry, once we include the geometry that is associated to Bob's extraction machinery. The discussion here is similar in spirit to the 
    proposed solution to the AMPS$_1$ \cite{Almheiri:2012rt}  paradox  in \cite{Maldacena:2013xja}. 
    It also shares features with the resolution of the Mawell demon paradox. There we had an apparent contradiction with the second law
     of thermodynamics and 
    it was solved by including  the entropy generated by the deamon. Here we have an apparent contradiction with the laws of general relativity and quantum fields and 
   we propose to resolve it by  including  the spacetime geometry generated by the quantum computer doing the decoding.

    \subsection{Going beyond the thermofield double}
    We have seen that the  wormhole geometry has ``legalized'' the information extraction process. However, 
    the use of the thermofield double state might raise some questions, since this state is hard to produce \cite{Harlow:2013tf} and it is inherently unstable. 
    In particular, one would think that Bob does not need to go through the trouble of building the whole thermofield double state in order to extract the message. 
    To explore this question we can  deform the TFD state to see whether the procedure still works. 
    For example, we can introduce shock waves by performing perturbations in the far past. These shock waves are related to the instability of the TFD: a small perturbation
    either in the past or the future can have important consequences. Indeed, if we have these shock waves, Bob's decoding procedure is frustrated, because they both widen the wormhole and decrease the correlation between the two sides, making the double-trace perturbation less effective. 

\begin{figure}[ht]
\begin{center}
\begin{subfigure}{.25\textwidth}
  \centering
  \includegraphics[width=.9\linewidth]{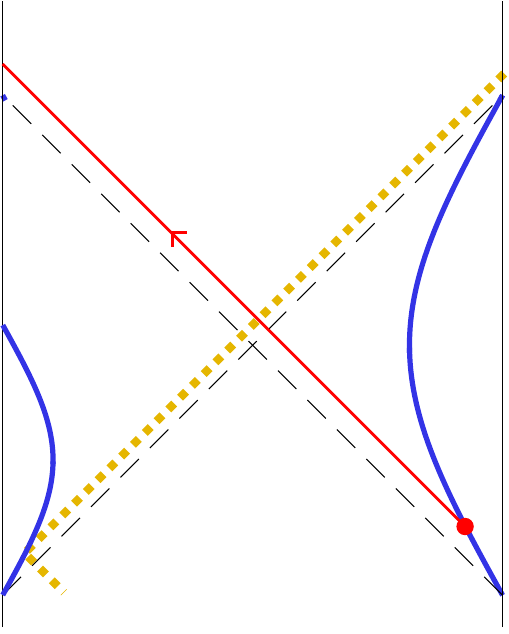}
  \caption{}
\end{subfigure}\hspace{10pt}
\begin{subfigure}{.25\textwidth}
  \centering
  \includegraphics[width=.9\linewidth]{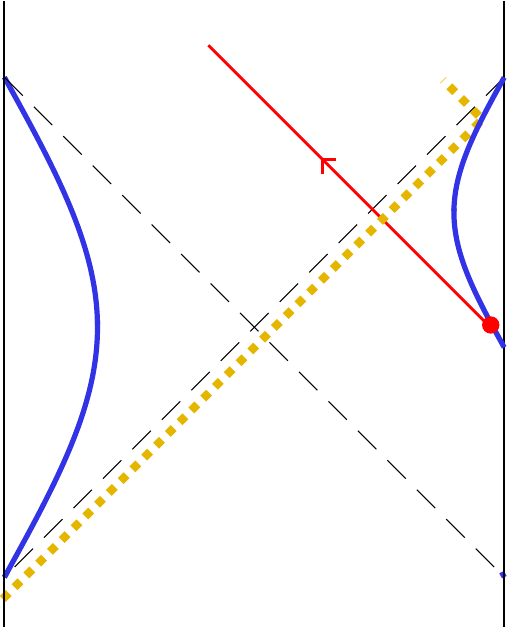}
  \caption{}
\end{subfigure}
\caption{Small perturbations at late or early times introduce shock waves denoted by orange dotted lines. If the state is perturbed by a shock wave from the left (a), Bob can act on the left system to remove it, returning to the thermofield double state and following the standard protocol. If the state is perturbed by a shock wave on the right side (b), and Bob reads some qubits at intermediate times,  Bob cannot easily remove it, and in general the traversable wormhole protocol will not work. }
\label{ExtractionWithShockWaves}
\end{center}
\end{figure}

    We can consider shock waves sent either from the left boundary or the right. If they are sent from the left, see figure \ref{ExtractionWithShockWaves}(a), Bob can act on the left system to remove the perturbation, returning to the starting point for the thermofield double protocol and then proceding from there. One would like to conclude from this example that Bob really needs to produce something close to the TFD to extract the information. 
    But another possible conclusion is that we have not been imaginative enough to think about  {\it other } methods that Bob could use to extract the information\footnote{
    A possible method discused in \cite{kitaevyoshida} also involves producing something close to a thermofield double.}.

If the shock waves come from the right instead, the situation is more complicated. If the shock wave doesn't involve too many particles, and if it is released at roughly the same time as Alice's message, then Bob can view it as part of a somewhat larger message, and (at the cost of more double trace interactions) perform the same protocol. However, if the shock wave is sent at a very different time, see figure \ref{ExtractionWithShockWaves}(b), then the standard protocol will no longer work: the two sides are never correlated enough to make the double trace interaction effective. 
%More explicitly, in figure \ref{ExtractionWithShockWaves}(b) we have added a perturbation at very late time on the right, then evolved the system backwards in time
%and then added the message. If Bob extracts some qubits from the right at intermediate times, it is not clear what procedure he should follow. 
The logic of Hayden-Preskill shows that there should be {\it something} we can do to recover the message, but from the geometry it is not clear, and a more complicated decoding procedure may be necessary.

     Another issue we can discuss is the famous Page curve (see figure 1  in \cite{Page:2013dx}). 
     The Page curve shows that the entanglement entropy of a black hole with the Hawking radiation can 
     increase, but after half the system has evaporated, we expect it to start decreasing. How can we visualize the evaporation 
     process when it decreases? As shown in 
     \cite{Gao:2016bin} the double trace interaction can decrease the entanglement entropy of the black hole. From one of the sides this is just a coupling that 
     lets the black hole evaporate, extracting energy from the black hole. However, due to the correlations present with earlier radiation, which are manifested by 
     the correlations with the second side of the wormhole, we get that the entanglement entropy can decrease. What looks from one side like an ordinary evaporation
     process is actually decreasing the entanglement entropy. 
     We can prolong this process by acting with the double trace operator, and the evolving the system backwards on the left side and then adding another double trace 
     deformation again. Repeating this process it looks like we can decrease the entropy by a large amount. From the right point of view we simply have an ordinary evaporation 
     process. But it is a rather elaborate process on the left side. It is elaborate because the algorithm effectively checks that the entanglement decreases at each stage.

 \section{Quantum mechanical models } 
 \label{QMModels}

  In this section we will discuss various quantum mechanical models that display phenomena related to wormhole traversability.

   \subsection{Traversability in the SYK model } 
   
   The SYK model \cite{Sachdev:1992fk,KitaevTalks} consists of $N$ Majorana fermions with a random $q$-local Hamiltonian. See e.g. \cite{Maldacena:2016hyu} for a definition. 
   The main point of this section is to show that we have this traversability effect in this model at strong coupling or low temperatures, 
     $ 1 \ll \beta J \ll N $. In this regime, the interesting part of the dynamics is dominated by a quantum mechanical degree of freedom that is 
     captured by a Schwarzian action \cite{KitaevTalks,Maldacena:2016hyu}. 
     This is the same as the action for the  boundary graviton in neary-$AdS_2$. This equality of the actions implies
      that the SYK computation reduces to the computation we have discussed before.

     More precisely, we can consider an interaction between the two boundaries of the form 
     \be \la{inve}
      g V = i \tilde g  \sum_{j=1}^K \psi^j_L(0) \psi_R^j(0) ~,~~~~~~g = K \tilde g.
      \ee
      This interaction involves $K$ of the $N$ fermions (with $K< N$).  Here we are assuming that the $\psi$ operators should be interpreted as true fermions, so that they anticommute in the two systems. Note that the interaction (\ref{inve}) cannot be replaced by simple classical measurement as in (\ref{Tele}), since it would involve the measurement of a fermionic operator. In order to have a coupling entirely consistent with the previous discussion, we can use an interaction of bosonic operators, like $(\psi^i_L\psi^j_L) (\psi^i_R\psi^j_R)$. However, we will stick to the interaction (\ref{inve}) for simplicity. 

      Let us first discuss traversability at weak coupling $\beta J \ll 1$. In order to quantify the transmission of information we compute the anti commutator 
      \be \la{AntiCo}
      Re(C)=  \frac{1}{2}\langle \{ e^{ - i g V } \psi^l_L(t_L)  e^{ ig V}, \psi^l_R( t_R) \} \rangle \sim  \frac{\tilde g}{2} \sum_{j=1}^K \langle \{ \psi^l_L(t),\psi^j_L(0) \} \{ \psi^l_R(t_R),\psi^j_R(0) \} \rangle 
      \ee
      where we assumed $t_L>0$ and $t_R<0$ so that the double trace interaction only acts on the $\psi_L(t_L)$ operator. 
      We see that we get anti commutators of operators purely on the left and anti commutators of operators purely on the right. This means that the non-zero answer will 
      come from the correlator between the left and the right anticommutators. 
      For simplicity, we can assume that the index $l$ is not among the $K$ indices summed over in \nref{inve}. In this case we will get a non-zero answer only 
      due to the interactions. For example, to first order in the interaction Hamiltonian, each of the anticommutators above has a structure 
      \be
       \{ \psi^l(t),\psi^j(0) \} \sim   i t  \left\{ [ H , \psi^l(t)],\psi^j(0) \right\}   = - i t  J_{ljkm} \psi^k \psi^m 
       \ee
       This is both true for the left and the right, so $C$ in \nref{AntiCo} 
       is equal to 
       \be \la{FastCom}
       Re  ( C )=  \frac{\tilde g}{2}  t_L t_R  \sum_{j=1}^K \langle    J_{ljkm} \psi_L^k \psi_L^m J_{ljk'm'} \psi_R^{k'} \psi_R^{m'} \rangle =   { g  \over 2N } J^2  t_L t_R 
       \ee
       We can extract a few lessons from this. First note that the answer is non-zero for any $t_L>0$, $t_R<0$. As opposed to the gravity computation, we do not have
       to wait to be able to send the signal, it can be done immediately. The effect is small and suppresed by $1/N$. Note that it is an effect that grows with time, but in this 
       perturbative computation we can only trust the expression for $ J t \ll 1 $.  We expect higher orders to lead to an exponential growth proportional to $\frac{1}{N}e^{(const.)J t}$. The time dependence of the exponential is much slower than the $\frac{2\pi t}{\beta}$ behavior that we have at strong coupling. 
       Note that in this method of computation, the result comes manifestly from correlations between
       the operators on the left and the right. 
       
        We can now consider the strong coupling limit $ 1 \ll J \beta , ~J t_{L,R} $. In this limit we get that the fermions have a conformal invariant correlation function of
        the form 
        \be
         \langle \psi^l (t) \psi^l(t') \rangle \propto { 1 \over \left[i\sinh\frac{\pi( t - t')}{\beta}\right]^{ 2 \Delta } },
         \ee
         which is the same as the correlators of boundary values of bulk  fields in $AdS_2$. 
         In addition,  the dynamics of the model is dominated by a degree of freedom that can be described by a Schwarzian action. This degree of freedom 
         arises from a spontaneously and explicitly broken time reparametrization symmetry. Both the action, and the coupling to the fermions, is identical to what we have 
         for the boundary gravition in $AdS_2$. 
         This implies that the final answer for the correlators is identical to what have discussed in general in section \ref{GravComp}. 
         In particular we get a result identical to \nref{ResGen} but with $G_N  \to { 1 \over \pi\alpha_S} { \beta \mathcal{J} \over N } $, where $N \alpha_S/\mathcal{J}$ is the constant multiplying the 
         Schwarzian action
         \be
I = -N \frac{\alpha_S}{\mathcal{J}}\int_0^\beta d\tau\, \text{Sch}(f,\tau) ~,~~~~~~~Sch(f,\tau) = { f'''\over f' } - { 3 \over 2 } \left( { f'' \over f'} \right)^2.
         \ee
       The reason we do not have to do   extra work   is that the full answer is determined both in gravity and in SYK by the structure of the symmetries. 
       In particular, by the spontaneous and explicitly broken time reparametrization symmetry. 
         
      For intermediate couplings it seems harder to do the computation, but it could be done in principle. We expect an answer which is conceptually similar to the 
      ``stringy'' answer discussed in section \ref{Stringy}. Note, in particular, that the ``stringy'' corrections were also giving a non-zero value for the (anti) commutator inmediately 
      after the double trace operator has acted, as in \nref{FastCom}.

\subsection{Another perspective on the bulk computation} 
\label{GravityThreePoint}

In the gravity computation, 
we have seen that to leading order in $G_N$, or $G_N e^t$, we do not get a commutator immediately, we have to wait for a while until the signal tranverses the wormhole. 
This is a feature of \nref{ResFi}. 
On the other hand, the full gravity result, \nref{ResGen}, leads to a non-trivial commutator immediately if we take into account more than the first order expansion of the 
denominator in the exponent in \nref{ResGen}. 
We would like to have a more direct understanding of this feature. 

 Let us start with the computation of the commutator in the gravity theory. We work to leading order in $g$
 \be \la{ComGra}
 Im(C) = - i \langle [ e^{ - i g V }  \phi_L(t_L) e^{ i g V } , \phi_R(t_R) ] \rangle =g  \langle [\phi_L(t_L),O_L(0) ]   [\phi_R(t_R),O_R(0) ] \rangle 
 \ee
 We are assuming that $\phi_L$ and $O_R$ are different fields. For that reason each commutator would be zero unless we consider some interaction. 
 In gravity, if we are considering four point interactions, these will each be of order $G_N$ and that is why a non-zero commutator will arise only at order 
 $G_N^2$. But once we go to this order, we can have a non-zero commutator immediately. An example of a four point interaction would be 
 $G_N  \int  ( \partial \phi)^2 (\partial O)^2 $ with various possible contractions for the derivatives. 
 Another possibility would be to consider a three point interaction of the form $ \sqrt{G_N}  \int \phi   O \sigma  $ where $\sigma $   is a third field. 
 Then each commutator in \nref{ComGra} gives a $\sigma $ field and the correlator of the $\sigma $ fields gives rise to the commutator.

 \begin{figure}[ht]
\begin{center}
\includegraphics[width=0.45\textwidth]{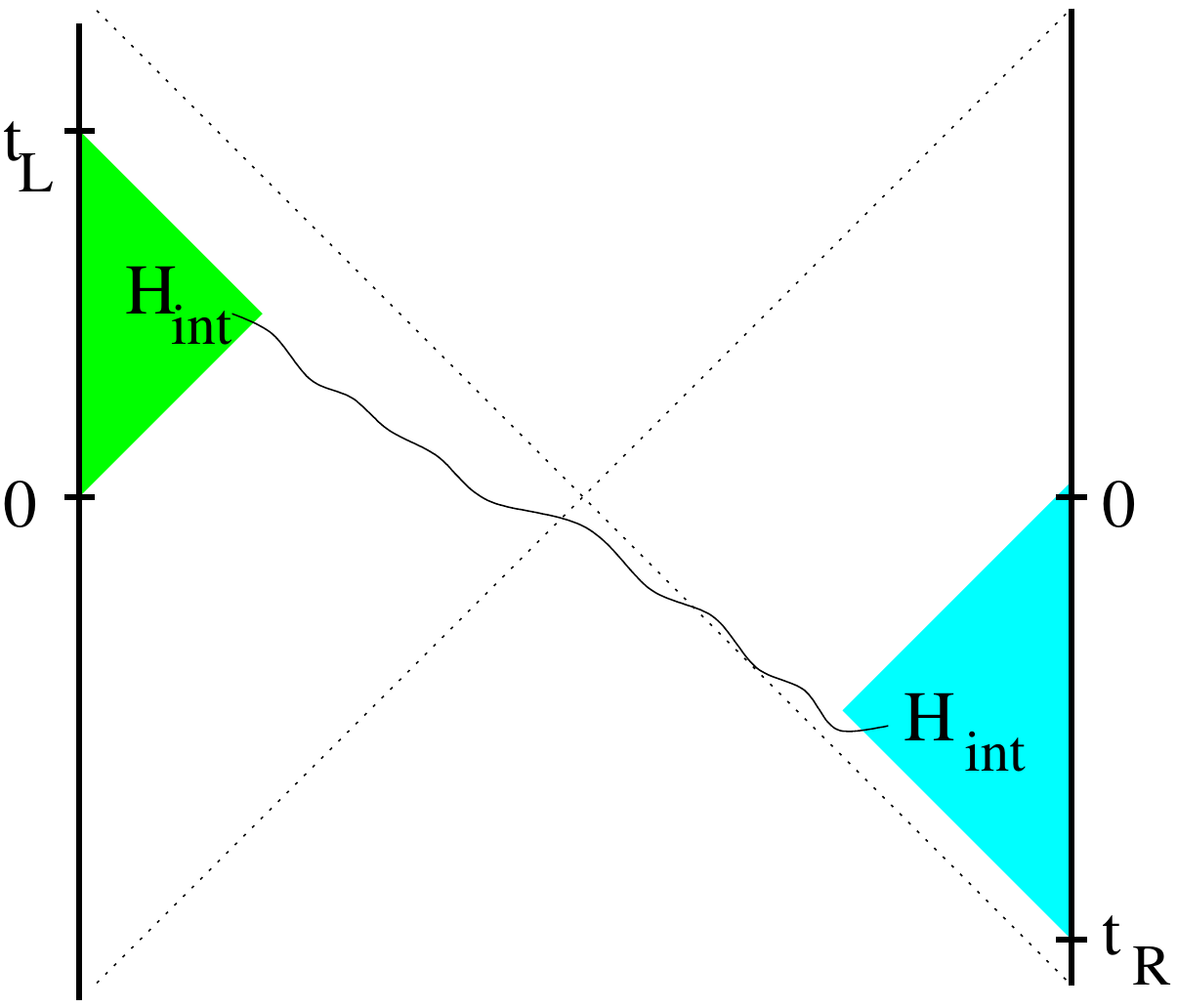}
\caption{ We consider the process in the bulk effective theory that includes gravity, with rigid boundaries. 
 The interaction Hamiltonian in the bulk effective field theory gives something non-trivial only when it is integrated over the shaded wegdes both for the left and the right 
pictures. This makes it manifest that the effect arises from correlations, also in the bulk picture. The double trace operators are inserted at the times labelled by zero.  }\label{Wedges}
\end{center}
\end{figure}

 We should emphasize that, in the computation of the commutators in \nref{ComGra}, the interaction Hamiltonian is only integrated over a wedge that is bounded by 
 the times $t=0$ and $t=t_{R,L}$ see figure \ref{Wedges}. This is true to all orders in the expansion of the bulk interactions. This highlights that we are getting a non-zero answer
 purely from the correlations that we have among the various fields present in the commutators. This computation is conceptually similar to the one we would be doing in the 
 SYK model by performing nested commutators of the interaction Hamiltonian. In both of the computations, the time evolution is producing a complicated operator with many pieces. But, by the magic of the thermofield double,  these pieces are perfectly correlated with the pieces we get from the other side by doing a similar time evolution.  
      
  It is interesting to connect this to the stringy scattering picture.  The stringy corrections were computed by considering a scattering configuration where we were 
  exchanging a graviton (or pomeron) along the $t$ channel. We were getting a non-zero result for the commutator immediately in time  and at order $G_N$. 
  This is similar to what we discussed above, when we exchanged the $\sigma $ field. In fact, we can connect the two. In string theory, the exchanges of highly massive 
  string states along the $s$ channel  (i.e. many $\sigma$ fields)  can be resummed and be viewed as the exchange in the $t$ channel of some effective excitations, sometimes called a ``reggeon''
  (or pomeron, in this case). See \cite{Brower:2006ea} for a nice discussion. The particular exponent given by $a$ in \nref{StringPhase}, would result from this resummation.

\subsection{An operator growth model } 

\newcommand{\sizevar}{s}

In this section, we will try to understand the behavior of the correlator $\langle \phi(t) e^{igV}\phi(t)\rangle$ in terms of the ``size'' of the evolving operator $\phi(t)$. Some rough features of the bulk results can be reproduced in a simple model where we work at infinite temperature and we assume that $\phi(t)$ is a random operator with given size characteristics.

At high temperature, the thermofield double becomes a maximally entangled Bell state, which we write as $|\Phi\rangle$. We will work in terms of Majorana fermions, and for the $V$ operator, we choose
\be
V = \frac{2i}{K}\sum_{i = 1}^K\psi^i_L\psi^i_R, \hspace{20pt} \langle \Phi|i\psi_L^i\psi_R^j|\Phi\rangle = \frac{1}{2}\delta^{ij}.
\ee
This satisfies $e^{igV}|\Phi\rangle = e^{ig}|\Phi\rangle$. In order to compute $\langle \phi(t) e^{iV}\phi(t)\rangle$, we would like to have an expression for $\phi(t)$ operator, which we take to be a boson. For the moment we write it in the general form
\be
\phi(t) = \sum_{\sizevar = 0}^N \sum_{i_1<...<i_\sizevar}\gamma_{i_1...i_\sizevar}(t)\psi^{i_1}...\psi^{i_\sizevar}.
\ee
Here the sum over $\sizevar$   is a sum over operators of different sizes, where size is defined as the number of basic fermions appearing in a given product. Note that $\sizevar$ is even since we assume $\phi$ is a bosonic operator. Using anticommutation relations, one can also show that $e^{igV}\psi^{i_1}...\psi^{i_s}|\Phi\rangle = e^{ig(1 - \frac{2c}{K})}\psi^{i_1}...\psi^{i_s}|\Phi\rangle$, where $c$ is the number of indices in common between the sets $\{1,...,K\}$ and $\{i_1,...,i_\sizevar\}$. This implies
\be
\langle \phi(t) e^{iV}\phi(t)\rangle = \sum_\sizevar \sum_{i_1<...<i_\sizevar}|\gamma_{i_1...i_\sizevar}(t)|^2 e^{ig(1 - \frac{2c}{K})}.
\ee

In order to evaluate this sum, we will assume that the $\gamma_{i_1...i_\sizevar}$ coefficients are random, subject to the condition that the total weight in operators of length $\sizevar$ is given by some function $w(\sizevar,t)$ that encodes the size distribution of the operator $\phi(t)$:
\be
w(\sizevar,t) = \sum_{i_1<...<i_\sizevar}|\gamma_{i_1...i_\sizevar}(t)|^2.
\ee
This makes it possible to determine the statistics for the $c$ variable defined above by computing the probability $P(c,K,\sizevar,N)$ of getting $c$ colllisons when we separately draw $\sizevar$ indices and $K$ indices from $N$ total. This probability is
\be
P(c,K,\sizevar,N) = \frac{{K\choose c}{N-K\choose \sizevar-c}}{{N\choose \sizevar}} \approx {K\choose c}\left(\frac{\sizevar}{N}\right)^c\left(1 - \frac{\sizevar}{N}\right)^{K-c}.
\ee
where the last expression is valid for $K,c\ll s,N$. We now evaluate the correlator as
\begin{align}
\langle \phi(t)e^{iV}\phi(t)\rangle &= \sum_\sizevar w(\sizevar,t)\sum_{c = 0}^{\text{min}(\sizevar,K)}P(c,K,\sizevar,N)e^{ig(1 - \frac{2c}{K})}\\
&\approx  e^{ig}\sum_{\sizevar}w(\sizevar) \left(1  - \frac{\sizevar}{N} + \frac{\sizevar}{N}e^{-2ig/K}\right)^K \approx \sum_{\sizevar}w(\sizevar,t) e^{i g(1 - \frac{2\sizevar}{N})}.
\end{align}
In the final step we assumed that $K$ is large compared to one.

Let's now try to understand this formula. Suppose that $\phi$ was originally a simple operator. Then at early times, $w(\sizevar,t)$ will be supported for small values of $\sizevar$, and the above expression will be close to $e^{ig}$; this means the left-right commutator will be small. At late times, we expect the operator $\phi(t)$ to become rather random, so that it has significant support on operators of length near $\sizevar = \frac{N}{2}$. In this case the correlator approaches the value of one, as in the gravity discussion.

In fact, the whole expression for the correlator is vaguely reminiscent of the bulk expression, if we think of $\sizevar$ as being related to the momentum and $w(\sizevar,t)$ as being the square of the wave function of the $\phi$ field. An important difference is that in this infinite temperature setting, it does not  matter whether we correlate $\phi_L$ or $\phi_R$, since they act in the same way on $|\Phi\rangle$.

\section{A classical model } 

Traversability looks rather unexpected, and is a sharp signature of the connectivity of the bulk. One natural question is whether something like this exists in the physics of classical chaotic systems. A classical version of the traversability experiment would look something like the following: instead of the thermofield double, we can imagine two identical classical systems prepared in a state where the positions agree at $t = 0$, but the momenta are opposite. We start our experiment at an early time (long before $t = 0$) by making a perturbation to one of the particles in the $R$ system, let's call it $\phi_R$. We then wait until $t= 0$. We then select some other particle $O_R$, and measure how much it has been affected by $\phi$. Using the result of this measurement, we move over to the $L$ system and make a corresponding perturbation to the state of $O_L$, without touching any of the other particles. We then evolve for time $t$ and hope to find that $\phi_L$ has been affected in a way that is simply correlated to the way we initially perturbed $\phi_R$.

Let us now analyze this in more detail. We consider a classical system with phase space coordinates $x^i,p^i$, where $i=1,...,N$, and in a state such that
\be\label{refstate}
x^i_R(0) = x^i_L(0), \hspace{20pt} p^i_R(0) = -p^i_L(0).
\ee
Eventually, we will consider an average over states of this type, but for the moment we focus on a single configuration. For the $\phi$ perturbation, let's imagine making a perturbation at time $t_R<0$ by moving particle one by a small amount, $\delta x^1_R(t_R)$. Because the system is chaotic, when we evolve the system up to $t = 0$, the positions and momenta of all of the other particles will be affected to some extent that depends on the size of the perturbation and the time $t_R$. Let's focus on some other particular coordinate $x^2_R$ (the analog of $O$). If the perturbation by $\delta x^1_R$ was small enough, and $t_R$ wasn't too long, then the perturbation to $x^2$ will be small and given approximately by
\be
\delta x^2_R(0) \approx \frac{\partial x^2(0)}{\partial x^1(t_R)}\delta x^1_R(t_R).
\ee
In the second step of the experiment, we briefly couple the $R$ and $L$ systems together, by evolving with a new Hamiltonian
\be
V = (x^2_R - x^2_L)^2
\ee
for a ``time'' interval $g$. This leads to a perturbation to $p^2_L$ given by
\be
\delta p^2_L(0) \approx g \left[x^2_R(0) - x^2_L(0)\right] = g \delta x^2_L(0).
\ee
In the final step, we evolve forwards to time $t_L$ and attempt to read off the signal by looking at the momentum of particle one. If the perturbations are small enough and the time is short enough, we find
\be\label{prodofder}
\delta p^1_L(t_L) \approx \frac{\partial p^1(t_L)}{\partial p^2(0)}\delta p^2_L(0) \approx  \frac{\partial p^1(t_L)}{\partial p^2(0)}g\frac{\partial x^2(0)}{\partial x^1(t_R)}\delta x^1_R(t_R).
\ee
The three factors that appear here represent the three steps of the protocol:
\be
\delta x_{R}^{1}(t_R) \xrightarrow{R\text{ time evolution}} \delta x_{R}^2(0)\xrightarrow{\text{evolution with }V}\delta p_L^2(0)\xrightarrow{R\text{ time evolution}}\delta p_L^1(t_L).
\ee

It is clear that if the system is chaotic and the times $t_R,t_L$ are large enough, the final effect on $p^1_L$ will be large. However, one might expect the sign of the perturbation (which depends on the product of the signs of the derivatives in (\ref{prodofder})) to depend on fine-grained details of the state. In this situation, it would not be possible to read off the sign of $\delta x^1_R$ from $\delta p^1_L$ alone; one would also need to know other details of the state. This would be very different from the simple traversability discussed previously. However, an important point is that if we choose $-t_R = t_L = t$, then the sign of the product of derivatives is always positive, because
\be
-\{p^1(t),x^2(0)\}  = \frac{\partial x^2(0)}{\partial x^1(t)} = \frac{\partial p^1(t)}{\partial p^2(0)}.
\ee
This implies that the two derivative factors in \nref{prodofder} are actually equal, so their product is positive. This means that our traversability experiment succeeds: the perturbation we make to $x_R^1(t)$ gets recorded in a simple way in $p^1_L(t)$, even if we don't know the original reference state, or if we average over it. Notice that although the sign of the effect on $p^1_L(t)$ is definite, the sign of the intermediate step $\delta x^2_R(0)$ will depend in a complicated way on the state. So the information we transfer between the two systems is ``encoded.'' In the final protocol, the second derivative factor ``decodes'' the first one.

We will now consider some generalizations of this basic effect. First, it is interesting to see what happens if $-t_R \neq t_L$. For the moment, we continue to work in the approximation where all perturbations are small, so the effect is given by a Poisson bracket
\be
\{p^1_R(t_R),p^1_L(t_L)\} = \frac{\partial p^1_L(t_L)}{\partial x^1_R(t_R)}=\frac{\partial p^1(t_L)}{\partial p^2(0)}g\frac{\partial x^2(0)}{\partial x^1(t_R)}.
\ee
When $t_R = t_L$, we have seen that this has a definite sign for all states, given by the sign of $g$. Now we ask what happens if we hold $t_R$ fixed and vary $t_L$. For a particular configuration, we expect this Poisson bracket to become large as $t_L$ grows. However, its sign will no longer be fixed by the argument above, and so if we take an average over phase space, we expect cancelations to take place. A simple guess is that the average of this function is actually peaked near $-t_R = t_L$.

As a second generalization, we can keep $-t_R = t_L =t$, but make the size of the perturbation $\delta x^1_R(t)$ and/or the time $t$ large enough that the derivative formulas discussed above are no longer applicable. Now we ask what happens to $\delta p^1_L(t)$. Again, in an individual configuration, the magnitude of $\delta p^1_L(t)$ will grow with $t$, but once we are outside the Lyapunov regime where the perturbations are given by the simple derivative expressions above, we can no longer argue that the sign is fixed. So, again, we expect that an average over phase space will lead to cancelations that make the expression small for sufficiently large $t$. This is a classical version of the effect described by gravity analysis in section \ref{Backreaction}.

\begin{figure}[t]
\begin{center}
\includegraphics[width=0.45\textwidth]{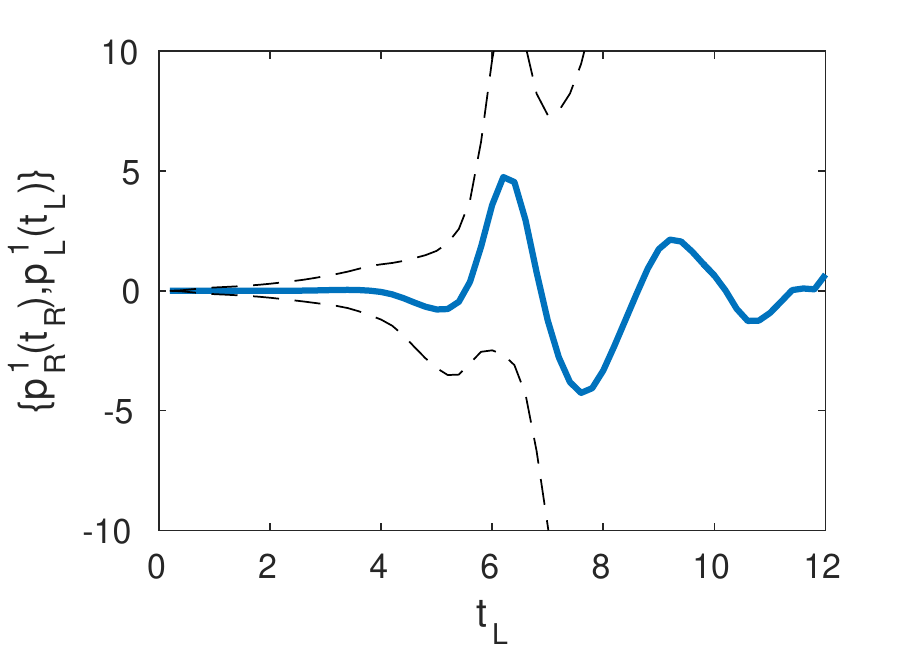}
\includegraphics[width=.45\textwidth]{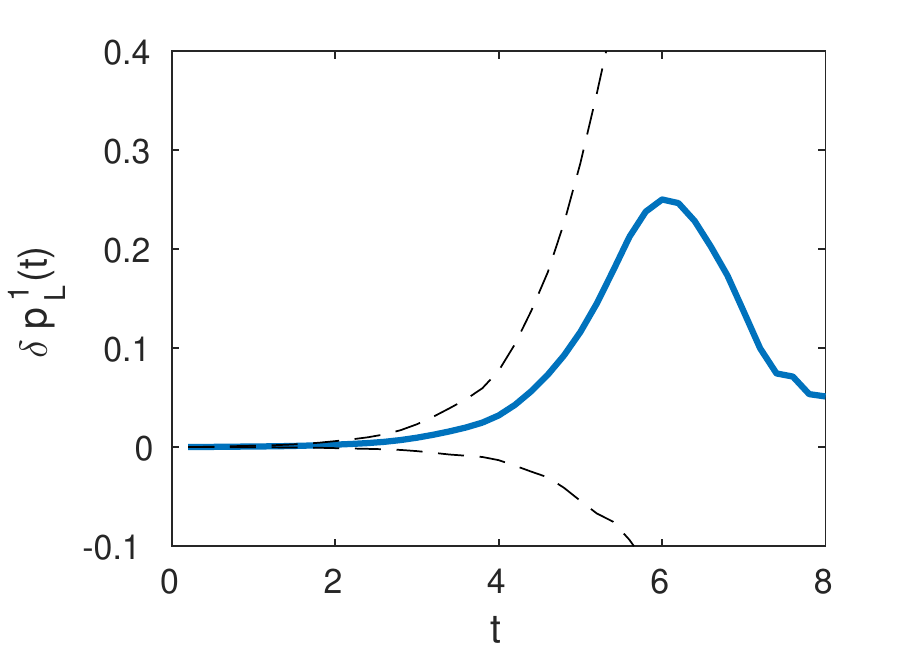}
\caption{{\bf Left:} the Poisson bracket $\{p_R^1(t_R),p_L^1(t_L)\}$ for fixed $t_R = -6$ and varying $t_L$. The solid curve shows the average over phase space with a thermal distribution at $\beta = 1$, and the dashed curves show the average plus or minus the standard deviation. {\bf Right:} similar averages for $\delta p_{L}^1(t)$ with finite initial $\delta x_{R}^1(t)= 0.1$. Exponential growth is visible at early times, but at late times cancelations lead to a decay in the mean. For the simulation we used $N = 200$ and $g = 1$.}\label{figclassicalplots}
\end{center}
\end{figure}
We can check these expectations in a simple system by numerically studying a collection of one dimensional particles with Hamiltonian
\be
H = \sum_i\frac{1}{2}(p^i)^2 + \sum_{a = 1}^N \left(\sum_{ij=1}^NJ_{aij}x^ix^j\right)^2.
\ee
Here the $J$ coefficients are random, with $\langle J^2\rangle = \frac{1}{N}$. In figure \ref{figclassicalplots} we give plots of the Poisson bracket for fixed $t_R$ and varying $t_L$, and the average finite displacement $\delta p^1_L(t)$ in the case of a finite initial perturbation of fixed sign. In both cases, we average over the thermal ensemble with fixed couplings, and we find behavior consistent with the expectations above.

	Of course, the fact that we can send classical information via signals that apparently contain no information is no surprise. This is 
	standard shared key cryptography. Namely, Alice and Bob share an identical random bit string. Alice does a binary sum of the message plus
	her bit string and sends it to Bob. Moreover, it is possible to encode the signal so that only a few more bits than the message suffice \cite{Hayden:2007cs}. 
	What is interesting about the classical model discussed here is not this aspect, but the fact that 
	 we have the same chaos-fueled amplification of the signal that we had in the quantum 
	case. Also, once we have the thermofield double, the decoding is easy: the signal emerges on the other side at a particular time.

\section{Conclusions} 

In this paper we have investigated the traversable wormhole protocol of \cite{Gao:2016bin}. This is particularly simple to describe in nearly $AdS_2$ gravity, which is the case that we focused on. By looking at the dynamics at higher orders in $G_N$, we found that the backreaction of signals we try to send limits the number that can pass through the wormhole. This implements a rough bound that the number of quanta one can send through the wormhole is less than some constant times the number of bits that one exchanges in setting up the interaction between the two sides.

We emphasized that the traversable wormhole is a simple explicit example of the Hayden-Preskill \cite{Hayden:2007cs} operation for extracting information from a black hole. In this application, one thinks of one side of the thermofield double as the black hole, and the other as the quantum computer of whoever is trying to do the decoding. After a message falls into the black hole, the interaction that leads to traversability causes it to emerge on the other side, e.g. in the quantum computer. 

Notice that the decoding is trivial here: we just wait for roughly a scrambling time as the particle propagates through the wormhole. This is surprising because, in general, the Hayden-Preskill decoding operation is very difficult \cite{Harlow:2013tf}. Much of this difficulty is avoided by starting with the thermofield double state, but even still the decoding operation appears to be hard for large messages \cite{kitaevyoshida}. It seems that the traversable wormhole protocol is taking advantage of the fact that the time evolution of the black hole near the scrambling time is not a random unitary. Indeed, we saw that if we let the black hole evolve for longer than the scrambling time, so that the time evolution is a more generic matrix, our protocol fails.

Having an explicit realization of the Hayden-Preskill decoding allowed us to address cloning paradoxes. These paradoxes involve a thought experiment where Bob uses Hawking radiation to decode some information that fell into the black hole, and then jumps in and also finds the original copy behind the horizon, violating the no-cloning property of quantum mechanics. In fact, we found that Bob's decoding operation (if successful) removes the original copy from behind the horizon, in such a way that there is never more than one copy. Of course, the fact that makes this possible is that Bob's quantum computer is connected to the interior of the black hole via the wormhole.

The traversable wormhole protocol gives a picture for how information can escape from a black hole, when it is in the special thermofield double state. To address the full black hole information problem, we would need to understand how this works for more general states of the Hawking radiation plus black hole system. The ideas of ER = EPR \cite{Maldacena:2013xja} suggest that we can act with the quantum computer on the Hawking radiation to form a state resembling the thermofield double, and then use the traversable wormhole protocol for that simple case. Of course, we would like to have a bulk understanding of how acting on the Hawking radiation can produce the simple thermofield double wormhole. In general, this operation has been argued to be very complex, just from the perspective of quantum computation \cite{Harlow:2013tf}, so perhaps it is naive to hope for a simple bulk picture. But in any case, it seems important for the black hole information problem to have some understanding of this point.
In particular, {\it after} we have made the wormhole it is easy to understand the resolution of the cloning paradox. However, we did not 
explain how the wormhole gets produced when Bob produces the thermofield double state.  The fact that the thermofield double state is very special is also highlighted by the presence of a similar effect in an ordinary classical system, 
when it is started in the classical analog of the thermofield double state.

We have emphasized that the traversable wormhole gives a smooth ride for the information passing from one side to the other. In other words, for this form of quantum teleportation, the experience is perfectly pleasant for the teleportee. In coming to this conclusion, we are assuming that the dual geometry of the thermofield double state is actually the two-sided black hole. It has been suggested that there might be more than one bulk interpretation of this state \cite{Marolf:2012xe}, and we caution the adventurous reader that if this is the case, the teleportation experience might not always be pleasant.

{\bf Acknowledgements } 

We thank A. Almheiri, P. Gao, D. Harlow, D. Jafferis, A. Kitaev, J. Preskill, L. Susskind, A. Wall, B. Yoshida  for discussions. 
J.M. is supported in part by U.S. Department of Energy grant
de-sc0009988.
 D.S. is supported by the Simons Foundation grant 385600.

\appendix

\section{Some kinematics of $AdS_2$ }\label{app:kinematics}

It is convenient to represent $AdS_2$ in embedding coordinates $Y^M$ obeying 
 \bea\label{metricEmb}
 &&    -Y^+ Y^- - Y_{-1}^2 =-1~,~~~~~~~ ds^2 = - dY^+ d Y^- - (dY_{-1})^2 ~,~~~~
 \cr
  && Y^\pm = Y^0 \pm Y^{1} ~,~~~~ Y.W = - { 1 \over 2} ( Y^+ W^- +Y^- W^+)  - Y_{-1} W_{-1}  
 \eea
 The usual bulk rindler like coordinates can be obtained by taking 
\bea
 && Y^+ = \sinh \rho e^t ~,~~~~Y^- = - \sinh \rho e^{ -t } ~,~~~~~~Y_{-1} = \cosh \rho ~,~~~
 \cr
 && ds^2 = d\rho^2 - \sinh^2 \rho dt^2  \la{BHcoord}
 \eea
 Taking $\rho$ positive or negative we get both sides of Rindler space. 
 The boundary coordinates can be found by taking  $\rho \to \infty $ and rescaling to define the projective coordinates 
 $X^M$, obeying $X^2 =0$. A representative for a right boundary point is 
\be
 X^+ = e^{t_R} , ~~X^{-} = - e^{ -t_R} ~,~~~~X_{-1} = 1
 \ee
  We can get a point on the left by taking $t_R  \to - t_L + i \pi $ in these formulas. 
  
 The $P_-$ translation, which near the origin translates $Y^-$ acts as 
\be \la{PplusAct}
Y^- \to Y^- + a^- Y_{-1} -  ( { a^- \over 2 } )^2 Y^+ ~,~~~~~~Y^+ \to Y^+ ~,~~~~~~~Y_{-1} \to Y_{-1} - { a^- \over 2 } Y^+ 
 \ee
 Notice that the origin has $Y^\pm =0$ and $Y_{-1} =1$. 
The correlators are defined by $\langle O(X) O(X') \rangle = ( - 2 X . X')^{-\Delta } $. 
We can now easily derive the formula 
\be
\langle O(X_R) e^{ -i a^- \hat P_- } O(X_L') \rangle = \left(  2 \cosh{ t_L  + t_R  \over 2 } +  { a^- \over 2 } e^{ { t_R - t_L  \over 2 } } \right)^{ - 2 \Delta } 
\ee
by acting with $P_-$ as in \nref{PplusAct} on $X'$. For $t_L = t_R =0$ we obtain \nref{dispop}. 

We can also use this formula to obtain the momentum space wavefunctions. In an unboosted frame with equal external times, we have
\bea \la{WF}
\langle O_R |q_- \rangle \langle q_- | O_L \rangle & =&  \int_{-\infty}^{\infty} { da^- \over 2 \pi } e^{i a^- q_-}
\langle O_R e^{ - i a^- \hat P_- } O_L \rangle     = \int_{-\infty}^{\infty} { da^- \over 2 \pi } 
e^{  i a^- q_- } { 1 \over ( 2 +   a^-/2 +i\epsilon)^{ 2 \Delta } } 
\cr
& =& { 1  \over \Gamma( 2 \Delta) } { (2  i q_-)^{ 2 \Delta } \over ( -q_-) } e^{    -i 4 q_- }  \Theta(-q_-).
\eea
Similarly, in a different frame that is unboosted with respect to $t_L = -t_R = t$, we have
\bea \la{WF2}
\langle \phi_L |p_+ \rangle \langle p_+ | \phi_R \rangle & =& { 1  \over \Gamma( 2 \Delta) } { (2  i p_+)^{ 2 \Delta } \over ( -p_+) } e^{    -i 4 p_+ }  \Theta(-p_+).
\eea

\subsection{Dynamics of  particles in $AdS_2$ } 

\subsubsection{Massive particles}

The trajectories of massive   geodesics in $AdS_2$  are given by 
the condition $Y . A =0$, where $A$ is a vector. It also turns 
out that $A$ is proportional to the $SL(2)$ charges of the particle. Namely, we can define a vector 
\be \la{SLCh}
Q_a =m \epsilon_{abc} \Sigma^{bc} = m \epsilon_{abc} Y^b d_s Y^c  ~,~~~~~~~d_sY^a \equiv  { d Y^a \over d s } 
\ee
where $s$ is proper time,  $ d_s Y . d_s Y=-1$. Note that $Q^2=m^2$.
The geodesic equation can  be written as $d_s^2 Y + Y =0$ and it implies that $Q_a$ is constant. We can then see that the trajectory is determined by the
equation  $Q.Y=0$. 

Notice that for a particle at $Y^\pm= 0$, $Y^{-1} =1$, we 
get that $Q_+= m\dot Y^-/2 $, $Q_- = -m\dot Y^+/2$, $Q_{-1} =0$ (here we used $\epsilon_{+-(-1)} =-1/2$). In particular this implies that 
$Q_+ =-  p_+$ and $Q_-= p_-$ where $p_\pm$ are the standard null components of the momentum relative to the coordinates $Y^\pm$ at the origin. 
In particular,   for any state of the quantum fields in $AdS_2$ that $-P_+$ and $-P_-$ are both positive. This implies that 
$Q^+<0$ and $Q^->0$ (note we raised the indices using the metric in (\ref{metricEmb})). This implies that $Q^a$ is a spacelike vector pointing from the right to the 
left
boundary. This is the vector of $SL(2)$ charges associated to particles in the interior  or, more generically, to matter in the interior. 

\subsubsection{Massive charged particles}

It turns out that the dynamics of the UV particle described by the Lagrangian in \nref{ActRel} is equivalent to 
that of a particle in an electric field \cite{kitaevIASChaosWorkshop}. This can be seen by using the expression for the topological Euler number to write
\be
\int dx \sqrt{h} K =- \int d^2x \sqrt{g} {R \over 2}  + 2 \pi =  \int d^2x \sqrt{g}  + 2 \pi ~,~~~~~~~{\rm when} ~~R = -2
\ee
For the dynamics, the  $2\pi$ can be ignored as a constant and the first term is the area enclosed by the boundary trajectory\footnote{
This $2\pi$ should be included when we compute the on shell action, which is finite in the limit \nref{LimPar}, thanks to the inclusion of this $2\pi$.}.
 This formula is equivalent to the
one we would get for a particle in an electric field, since the integral over the electric flux enclosed by the trajectory is also given by the area. 
This problem is SL(2) invariant and the conserved SL(2) charges are given now as 
 \be \la{Qcha}
 Q_a = m \epsilon_{abc} Y^b d_s Y^c - q Y_a ~,~~~~~~~~~~~ Q^2 = m^2 - q^2 
 \ee
 The geodesic equation is given by $ m ( d_s^2 Y^a + Y^a) + q \epsilon^a_{~b c} d_sY^b Y^c =0$. It implies that $Q$ is constant. 
 The trajectory is then fixed by the equation $Q.Y = -q $. For example, for the hyperbolic trajectory at $\rho =\rho_0$ in coordinates \nref{BHcoord}, we
 have $Q_{-1} = q/\cosh \rho_0$ and $\tanh \rho_0 = m/q$, see figure \ref{Hyperbolas}(a). 
 
 In our gravity problem we are interested in taking the limit where the trajectory is very close to the boundary.  In that situation 
 we can define a rescaled  proper time, $u$, rescaled parameters, and rescaled coordinates via 
 \be \la{LimPar}
  ds = { du \over \epsilon } ~,~~~~~Y = { 1 \over \epsilon } X ~,~~~~~~q = 2 \Phi_b = { 2 \Phi_r \over \epsilon } ~,~~~~~m =q - \epsilon E 
  \ee
  In this limit the equation for the trajectory becomes $Q.X = - 2 \Phi_r$. We also have that $Q^2= - 4 \Phi_r E$. This implies that $Q$ remains finite in the limit 
  (there are   cancellations between  the two terms in \nref{Qcha}).  This already fixes the shape of the trajectory, together with the equations $X^2=0$, $ d_u X^2 =-1$. 
  One can deduce from these equations that 
  \be \la{Cos}
  - { Q^a \over 2 \Phi_r} = d_u^2 X^a - { E \over \Phi_r} X^a
  \ee
  The equation \nref{Cos} can be derived by expressing $d_u^2 X$ in terms of the linearly independent vectors $X, d_u X, Q$. Note that the inner products among these
  vectors and $d_u^2 X$  can be deduced from the ones given after eqn \nref{LimPar} together with their derivatives. These inner products also allow us to determine that 
  \nref{Cos} holds. 
  
   We can view\nref{Cos} as an expression for the charges. Alternatively, we can assume $Q_a $ is constant and obtain an equation for $X$. 
% It turns out that  in this limit,  the charges $Q^a$ remain finite and can be expressed as 
 % \be \la{eqnsch}
 % Q_a =  2 \Phi_r \epsilon_{abc}d_u^2 X^b d_u X^c ~,~~~~~~~ d_u^3 X^a + \lambda d_u X^a =0 ~,~~~~~  (d_u X)^2 =-1 ~,~~~~X^2 =0
 %  \ee
 %  where we expressed the scaling limit of both the conserved charge as well as the equation of motion. CHECK . DERIVE ? 
%We have that $Q^2 = 4 \Phi_r^2 \ddot X^2 $. 
%We can solve the last two equations in \nref{eqnsch} by setting 
To make contact with the Schwarzian description, we can solve $X^2=0$, $d_uX^2 =-1$ by writing 
\be
 X = ( X^+,X^- , X^{-1} ) = \left( { 1 \over f' } , - { f^2 \over f' } , { f \over f' }  \right)
\ee
% Then the expressions for the charges \nref{eqnsch} agree with those in \cite{Maldacena:2016upp}. 
Taking a further derivative of \nref{Cos} we find $d_u^3 X - { E \over \Phi_r} d_u X=0$. This equation implies  the equations of motion 
of the Schwarzian theory plus the condition that $ E/\Phi_r = - 2 \{ f,u\} $, which is also a constant of integration for the equations of the Schwarzian theory. 
Using this expression for the energy one can also check that the charges  \nref{Cos} agree with the ones given in \cite{Maldacena:2016upp}.

 \section{Exploring properties of the correlator } 
 \label{PropertiesCorrelator}
 
 \def\Ghat{ T } 
 
\begin{figure}[t]
\begin{center}
\includegraphics[width=0.45\textwidth]{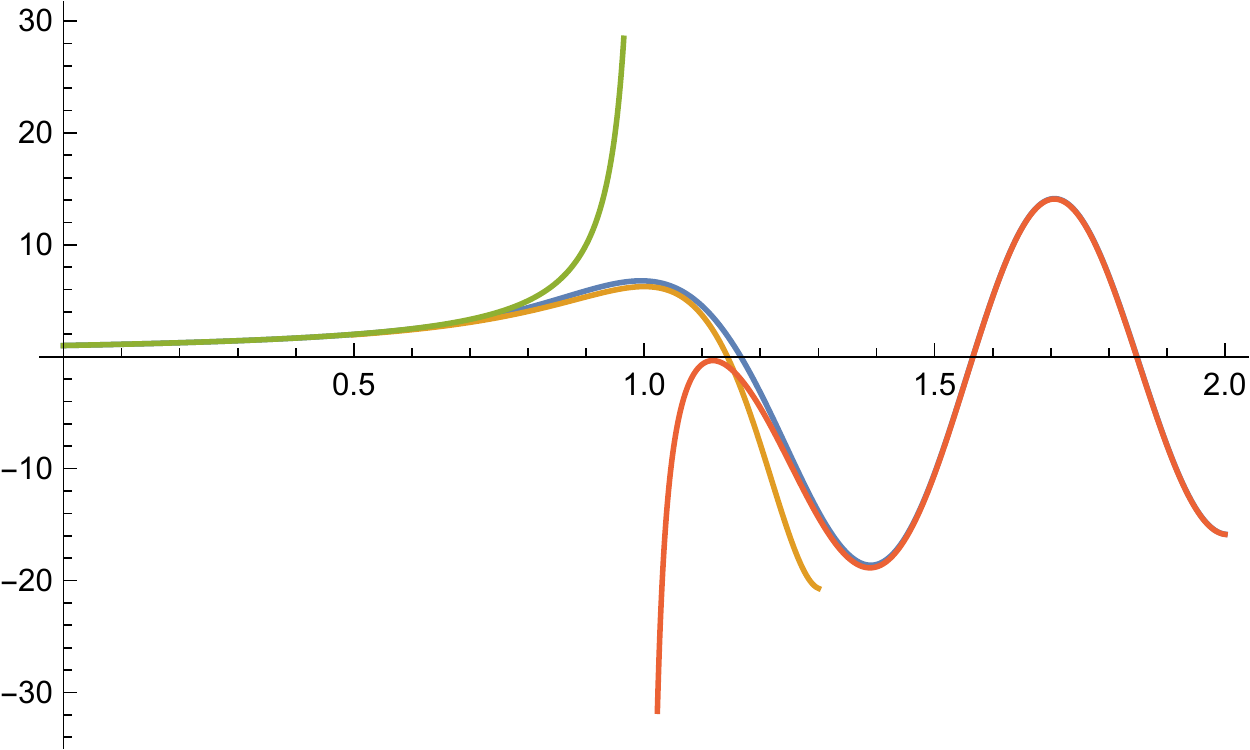}
\includegraphics[width=0.45\textwidth]{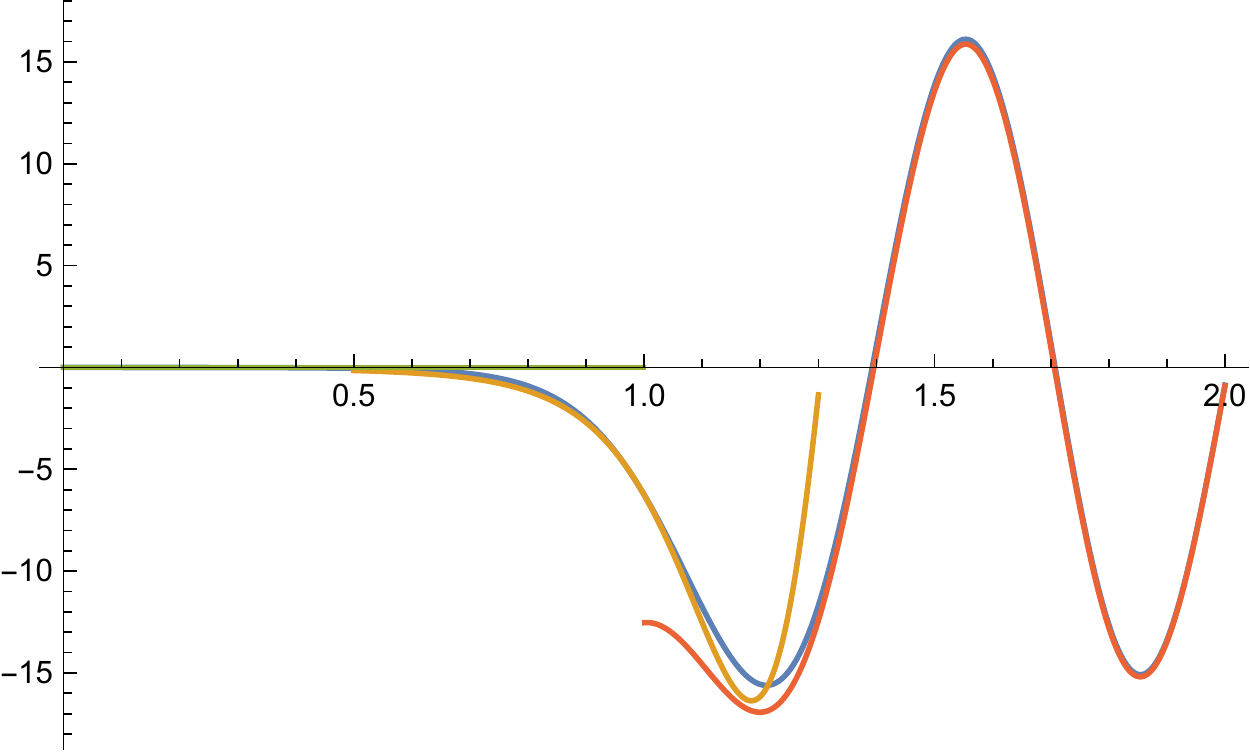}
\caption{ The real (left) and imaginary (right) parts of the function in various approximations for $\hat g=100$.
The horizontal axis is $\Ghat$. The blue line is the actual integral\nref{Intu}. In green we see \nref{SmallT}. 
In red we see \nref{LargerT}. Finally, in organge we see \nref{Errorf}.  (Color online). }\label{Approx}
\end{center}
\end{figure}

 In this appendix we give an analysis of the properties of the function \nref{ResGen}. Our goal 
 is to give an analytic understanding for various properties of this function. 
 For simplicity we consider this correlator in the special case $\Delta =1/2$, and comment on the more general case later. 
 In that case the correlator can be written as (after redefining the integration variable,  $ u = - 4 p_+/\hat g $) 
 \bea \la{Intu}
 \hat C &=&    - i     { \hat g  }  \int_0^\infty  du \exp \left[    i \hat g \left(-1+ u + { 1 \over ( 1 +u \Ghat ) }   \right) \right]  
 \\
  &~&  \hat C \equiv 2 C = 2 e^{ - i \hat g } \tilde C     ~ ~~~~\hat g \equiv { g \over 2} ~,~~~~\Ghat \equiv  { g G_N e^t
 \over 32 }  = e^{ t -t_d} 
 \eea
 where we redifined the variables to get rid of some simple factors. 
 We would like to explore the properties of this integral for large values of $\hat g$.
 Just to get oriented, let us first evaluate it by expanding the denominator in \nref{Intu} to first order in $T$ and doing the $u$ integral after rotating the contour
 to $u \to i y$. We get 
 \be \la{SmallT}
 \hat C \sim { \hat g   } \int_0^\infty dy e^{ - \hat g y ( 1 - \Ghat ) } =   { 1 \over ( 1 -   \Ghat  ) }  ~,~~~~~~~~ \Ghat < 1 
  \ee
  We will later see  that  this is an approximate value of the integral for $\Ghat < 1$. It is clear that for $\Ghat > 1$ we have a bit of trouble with\nref{SmallT}
   since the integral is not supressed for large $y$. Note that 
    the integrand has the form $e^{ \hat g f(u) } $ so that for large $g$, it can 
 be done using the sadlle point method. The saddle points, and the values of the function and the derivatives at the saddle, are   
 \be \la{Saddles}
 u_{\pm } ={ ( -1 \pm \sqrt{\Ghat} ) \over \Ghat  } ~,~~~~f(u_\pm) =  -i { (-1 \pm  \sqrt{ \Ghat} )^2 \over \Ghat } ~,~~~~f''(u_\pm) = \pm { 2 i \sqrt{ \Ghat } } ~,~~~~
 f'''(u_\pm) = - 6 i\Ghat 
 \ee
% Only the saddle point $u_+$ is along the contour.
 It turns out that for $\Ghat < 1$ the steepest descent contour starting at $u = 0$ and going to infinity  does not pass through any saddle point. 
 It can also be smoothly deformed to the defining contour. In this situation we do not get any saddle point contributions, and the integral is well approximated by the endpoint contribution at $u=0$ that gives \nref{SmallT}. For $ \Ghat> 1$ the steepest descent contour 
 starting at $u = 0$ ends at the singularity at $ u \Ghat =-1$. To get a contour that can be deformed to the defining contour, we also have to add a second piece that goes from the singularity to infinity, passing through the $u_+$ 
saddle point. Adding the contributions from these two components of the contour, we get 
\be \la{LargerT} 
\hat C  \sim  {\sqrt{ \pi \hat g} e^{- i \pi/4} \over (\Ghat )^{1/4} } \exp \left[ - i \hat g \left( 1 - { 1 \over \sqrt{ \Ghat } } \right)^2 \right]  +   { 1  \over ( 1 -   \Ghat  ) }     ~,~~~~~~ 1 < T \ll   \hat g^2.
 \ee
Both \nref{SmallT} and \nref{LargerT} contain subleading  terms in the  $1/\sqrt{\hat g}  $ expansion multiplying each term. 
The  $  \Ghat \ll \hat g^2$  bound comes form the requirement that 
 we can neglect the cubic term around the saddle point. 
 For $ T > \hat g$ we can compute \nref{Intu} by replacing $1/(1 + u T) \to 1/(u T) $ which gives us a Bessel function of the form 
 \be \la{Hankel}
 \hat C \sim  i{ \pi \hat g  e^{ - i \hat g }  \over \sqrt{ \Ghat } } H_1(  { 2 \hat g \over \sqrt{ \Ghat } } )  ~,~~~~~~~~   \hat g \ll \Ghat
 \ee
 where $H_1$ is a Hankel function. 
 Notice that this expression has an overlapping regime of validity with \nref{LargerT} and it also matches on to the $  \hat g^2 \ll \Ghat  $ behavior which is simply a constant 
 phase 
 \be
 \hat C \sim e^{ - i \hat g} ~,~~~~~~~~~~ \hat g^2 \ll  \Ghat 
 \ee
 This gives the behaviour of the function in almost the whole range. One small detail that we   now should clear up is the behavior of the function 
 when $ | \Ghat -1 | \ll 1$. In this regime the above expressions diverge, while the original function was convergent. We can get the proper behavior in this limit
 by expanding \nref{Intu} to quadratic order around $u=0$ to find 
 \bea 
 \hat C &=& - i { \hat g  } \int_0^\infty du \exp\left[ i \hat g \{ u ( 1 - \Ghat ) +  u^2 \}  \right]  
 \cr
 &=&  \sqrt{ \hat g } e^{ - i\pi/4}  e^{ - i \alpha^2 } { \sqrt{\pi } \over 2}  \left( 1 - {\rm erf}
 ( e^{ - i {  \pi/4} } \alpha )  \right) ~,~~~~~\alpha = { ( 1 - \Ghat )\sqrt{ \hat g } \over 2  } \la{Errorf}
 \eea
 where we have set  $T=1$ in the $u^2$  term. When $\alpha \gg 1$ it approaches $1/( 1 -\Ghat )$. 
 Even the relative exponential factor in the limit $T\sim 1$ between \nref{SmallT} and \nref{LargerT} matches in this regime\footnote{
 We use that $ \sqrt{ \pi } ( 1- {\rm erf}(z))  \sim  { 1\over z } e^{ - z^2} $ for $z\gg 0$ and
  $ \sqrt{ \pi } ( 1- {\rm erf}(z))  \sim  2 \sqrt{\pi} + { 1\over z } e^{ - z^2} $ for $z\ll 0 $.}
   Thus \nref{Errorf} gives the right approximation 
 for $ |\Ghat -1| \lesssim 1/\sqrt{ \hat g } $. 
 Putting all these approximations together we get a good picture of the function everywhere, see 
 figure \ref{Approx}.

   \subsection{Smeared wavefunctions} 
   
\begin{figure}[t]
\begin{center}
\includegraphics[width=0.45\textwidth]{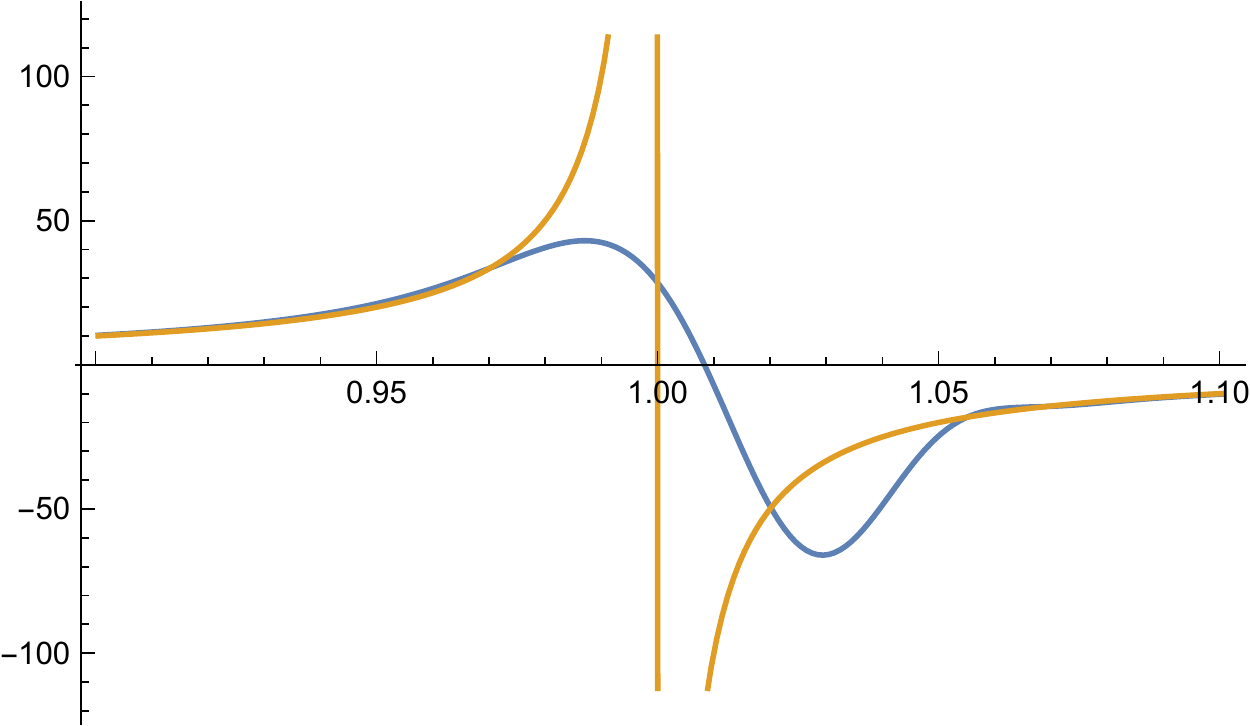}
\includegraphics[width=0.45\textwidth]{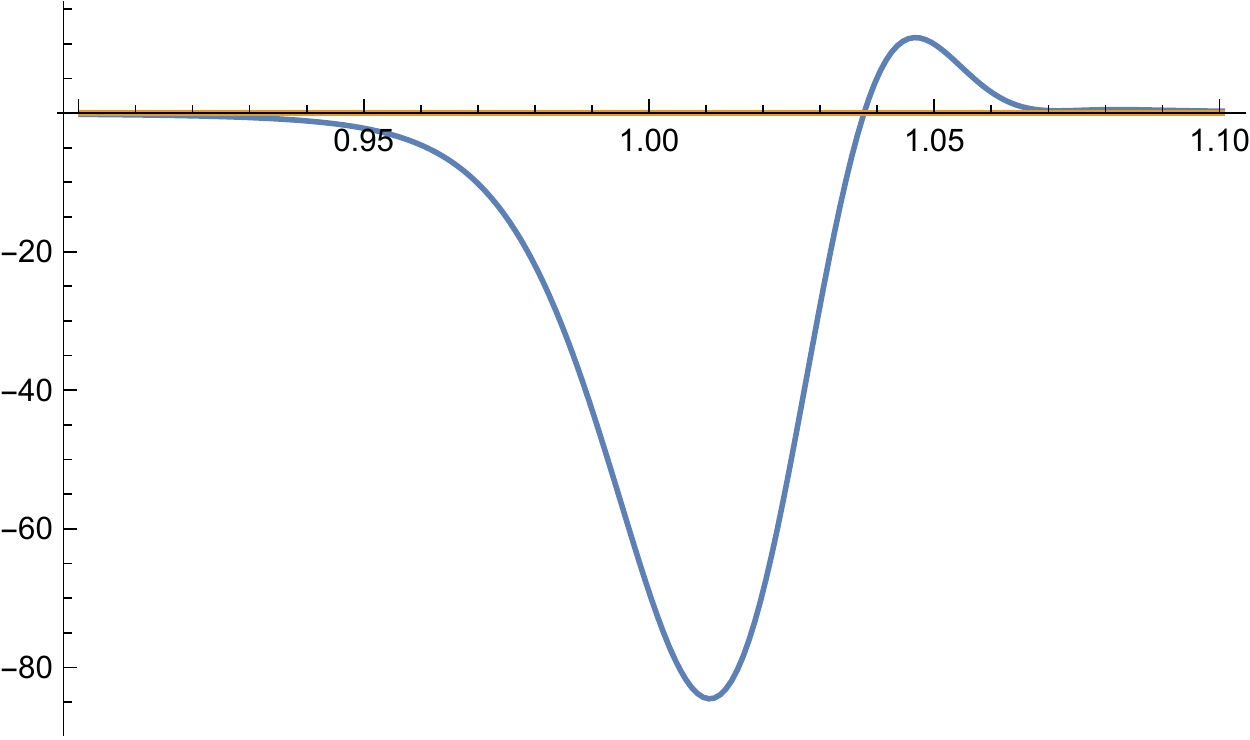}
\caption{ The real (left) and imaginary (right) parts of the smeared function \nref{Smeared}  for  $\hat g=10^4$ 
with $\epsilon=1$. The horizonal axis is $\Ghat$. Blue the function \nref{Smeared}  and orange is the approximation \nref{Smearedap}. We have blown up 
the region $|\Ghat - 1| \sim { 1 \over \sqrt{g} } $  where they are different. For the rest of the values of 
$\Ghat $ that are of order one, the function is approximated by the orange line. 
   }\label{SmearedPlots}
\end{center}
\end{figure}

\begin{figure}[t]
\begin{center}
\includegraphics[width=0.45\textwidth]{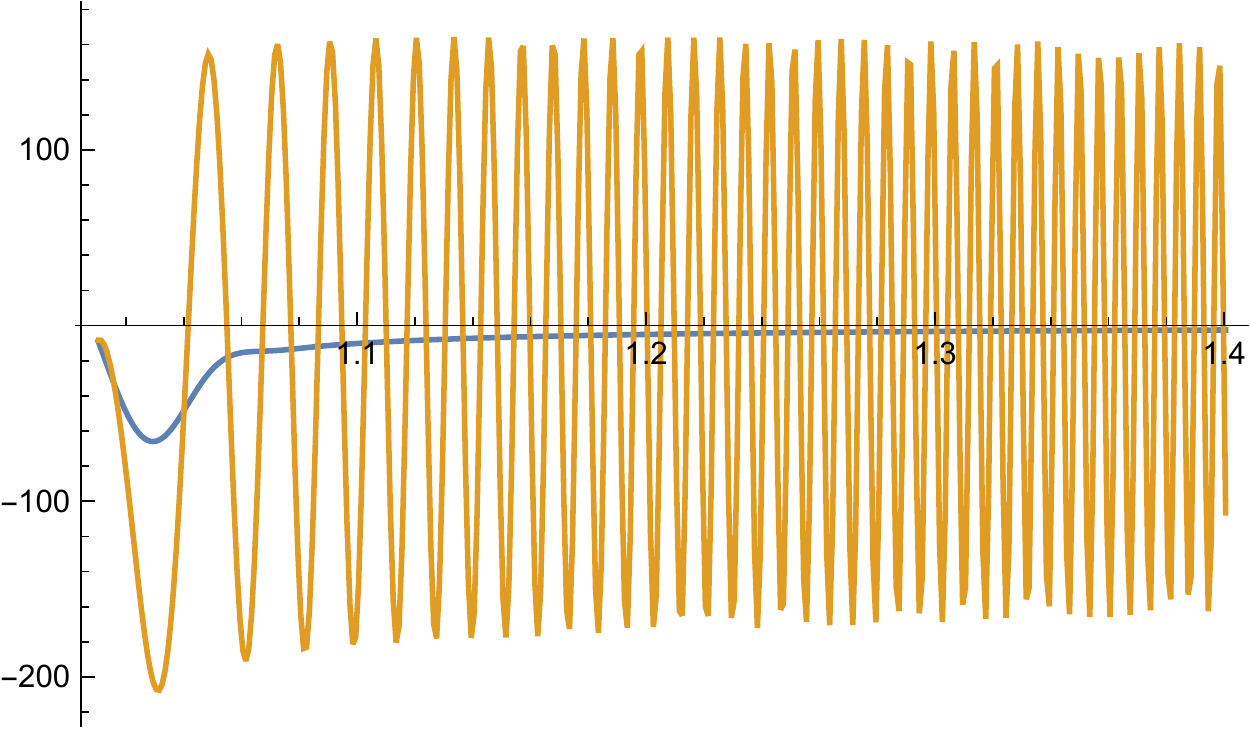}
\caption{ The real part of the original \nref{Intu} (orange) and the smeared \nref{Smeared} results (blue) for 
$\hat g = 10^4$ and $\epsilon =1$.  The important point is that the oscillations are gone and we recover the result 
\nref{Smearedap}. 
   }\label{SmearedvsNot}
\end{center}
\end{figure}

   Note that the oscillating behavior we see at $\Ghat > 1$ is due to the high momentum components of the 
   wavefunction. Namely, we see that, for $\Ghat$ of order one (say $\Ghat =2$ for example), the 
   momentum at the saddle point $u_+$ in \nref{Saddles} is $p_+ \propto g $, which is very large. Note
   that momenta of order one correspond to thermal scale features. We want to smear the wavefunctions of
   the operators so that we consider operators that are localized to timescales much less than the temperature, but
   still much larger than $1/g $. One way to do this is to add a term of the form
    $e^{ - \sigma^2 p_+^2 } $ to the integrand in 
 \nref{ResGen}, with $ { 1 \over g } \ll \sigma \ll 1 $\footnote{ In principle, we should smear with a function localized in 
 time  within the Rindler patch. This gaussian wavefunction is spreading outside the Rindler path. But since
 $\sigma \ll1 $ the part that is outside is vanishingly small and so it is not a problem.}. 
 For ease of the computation we will assume that $\sigma^2 \propto  { \epsilon \over g }$ , so that the extra factor 
 in the integrand of \nref{Intu} is now 
 \be
 \hat C_{\rm smeared} =    - i     { \hat g  }  \int_0^\infty  du \exp \left[     \hat g\left\{ i  \left(-1+ u + { 1 \over ( 1 +u \Ghat ) }    \right) - \epsilon u^2 \right\} \right].  \la{Smeared}
  \ee
  We imagine that $\epsilon$ is small\footnote{The  small $\epsilon$ is convenient  to argue that it does not modify the 
  location of the saddle points too much. But we see in figures \ref{SmearedPlots} and \ref{SmearedvsNot} that
  the same properties are true for $\epsilon =1$. }
   but independent of $\hat g$. In this case, we can view the extra term as a 
  small modification to the previously found saddle points. This extra term then greatly suppresses the contribution 
  of the saddle point in \nref{LargerT}. 
  This implies that we will get just the endpoint contribution, which is simply
  \be \la{Smearedap}
  \hat C_{\rm smeared} = { 1 \over (1- \Ghat  ) }  - i \pi \delta(\Ghat -1) ~,~~~~~~~~  \Ghat   \ll  \hat g 
  \ee
  which is the naively expected answer. Note that for larger values of $\Ghat$ we do not have the saddle point 
  suppresion and we return to the oscillatory value given by \nref{Hankel}. We have added a delta function contribution which is expected
  from the $i\epsilon$ prescription. It is also an approximation to the imaginary part we see in the exact answer in figure \ref{SmearedPlots}.
  
\subsection{General $\Delta$}
For general $\Delta$ is is convenient to define
\begin{align}
\hat C &= \frac{(-i\hat{g})^{2\Delta}}{\Gamma(2\Delta)}\int_0^\infty du\, u^{2\Delta-1}\exp\left[i\hat{g}\left(-1+u + \frac{1}{(1+\frac{1}{2\Delta}u T)^{2\Delta}}\right)\right] \\
&\hspace{20pt}\hat{C} = 2^{2\Delta}C, \hspace{20pt} \hat{g} = \frac{g}{2^{2\Delta}}, \hspace{20pt} T = \frac{\Delta\hat{g}G_Ne^{t}}{8} = e^{t-t_d}.\notag
\end{align}
The analysis of this integral is very similar to the $\Delta = \frac{1}{2}$ case. The endpoint contribution to the integral now gives the probe approximation appropriate for general $\Delta$, (\ref{ResFi}). The saddle point for large $\hat{g}$ is at 
%\be
%u_+ = \frac{2\Delta}{T^{\frac{2\Delta}{1+2\Delta}}}\left(1 - T^{-\frac{1}{1+2\Delta}}\right).
%\ee
%\revJM{ 
\be
u_+ = \frac{2\Delta}{T }\left(  T^{ \frac{1}{1+2\Delta}} -1\right).
\ee 
As before, this saddle does not contribute to the contour if $T <1$, but it dominates for $T>1$. If we add a Gaussian smearing term as in (\ref{Smeared}), the situation will be improved and the contribution of this saddle will be small provided $T\ll \hat{g}^{\frac{1}{2}+\frac{1}{4\Delta}}$, giving a parametrically large window where the probe approximation is good. For $\hat{g}^{\frac{1}{2}+\frac{1}{4\Delta}}\lesssim T \lesssim \hat{g}^{1+\frac{1}{2\Delta}}$ this saddle point will dominate, giving an oscillating contribution. For very late times, $T\gtrsim \hat{g}^{1 + \frac{1}{2\Delta}}$, the saddle point approximation breaks down, and the function approaches a constant, $\hat{C}\sim e^{-i\hat{g}}$.

\section{Chaos and traversability on stretched strings}

In this appendix we make some comments on both chaos and traversability when we have strings that stretch from the boundary to the 
horizon.

  \subsection{Signals and boundary interactions on a string} 
  
    Let us first consider a general situation where we have a black hole and we have a string ending on the horizon. 
In the thermofield double situation we can consider a string that is extended in the $t$ and $r$ directions and sitting at some location of the boundary 
sphere. 
We can then consider the out of time order correlators on one side (or the two sided ``reasonable order'' correlators), for operators that live on the string. An example is an 
operator that produces a fluctuation of the string along a transverse direction. Another operator can be a fluctuation in a different transverse direction. 
Such fluctuations move along the string and they scatter. The scattering is simply a time delay of the form $e^{ i \alpha' p_+ q_- e^t}$. 
This has a form similar to (\ref{GravPhase}), but with $\alpha'$ instead of $G_N$. 
In the black hole background then an out of time order correlator will have terms that go like 
\be
 1 \pm i { \alpha' \over R^2}   e^t 
\ee
which are a signature of chaos. Here $R$ is some distance scale of order of the radius of curvature of the black hole (in string units). 
 There are a couple of points we want to emphasize. First, notice that the exponential growth is the same; it is the maximal exponent. 
 Second,  the chaos that is occuring involves a subset of all operators. This is because 
 we get ${ \alpha' \over R^2 }$ instead of $G_N$. 
 
 This same fact implies that traversability is easier to obtain for the degrees of freedom on the string. In other words, we can consider the $O$s and $\phi$s of the 
 discussion in the main text to be operators on the same string. 
 
 This effect is related to a similar effect that happens in higher dimensional black holes. There the shock wave profile has some dependence on the transverse dimensions. 
 If $\phi$ and $O$ create wave-packets that are close to each other in the transverse dimensions, then both the out-of time order correlators and the traversability 
 are enhanced. 
 This suggests that each piece of a higher dimensional horizon corresponds to a special subset of degrees of freedom.
 
 \subsection{Using strings to generate a large number of fields } 
 \la{LargeNumber}
  Finally, we can mention that if we start from a standard $AdS/CFT$ example, such as $AdS_5 \times S^5$ dual to a CFT on $S^3 \times R$, we can add 
 many  fundamental strings that stretch across the thermofield double. This gives rise to a large number of fields, proportional to the number of strings. 
 There is a constraint saying that the total net number of oriented  strings entering the black hole on each side should be zero. So we should have both strings and anti-strings. 
 In the above example, we can put $K$  strings at the north pole of the $S^3$ and $K$  oppositely oriented strings at the south pole of $S^3$ (and the same point on 
 the $S^5$). Around the vacuum,  this is a 
 BPS configuration, related to 1/2 BPS Wilson loops, so it is stable.
  When we consider the same asymptotic boundary conditions for the Schwarschild-AdS background the strings go through the wormhole to the 
  other side. We 
 also expect the configuration to be stable (but we did not check it in detail).  This provides a way to produce and example with a large number, $K$, 
 of fields. We can put double trace interactions involving these fields and use them to send a message via the bulk gravity 
 theory, using the bulk gravitational interactions, as in section \ref{GravComp}.

  \section{Hayden/Preskill with bits vs. qubits of radiation} \label{app:HP}
 
 In this appendix we discuss the Hayden and Preskill argument \cite{Hayden:2007cs}, which shows when it is possible to recover a message that has fallen into a scrambling black hole using (i) the first half of the Hawking radiation and (ii) a few additional quanta that were released after the message fell in. First we review their argument for the case where Bob has full quantum information about the additional quanta. Then we discuss a minor variation were Bob uses only classical information about the additional quanta.

\begin{figure}[h]
\begin{center}
\includegraphics[scale=0.6]{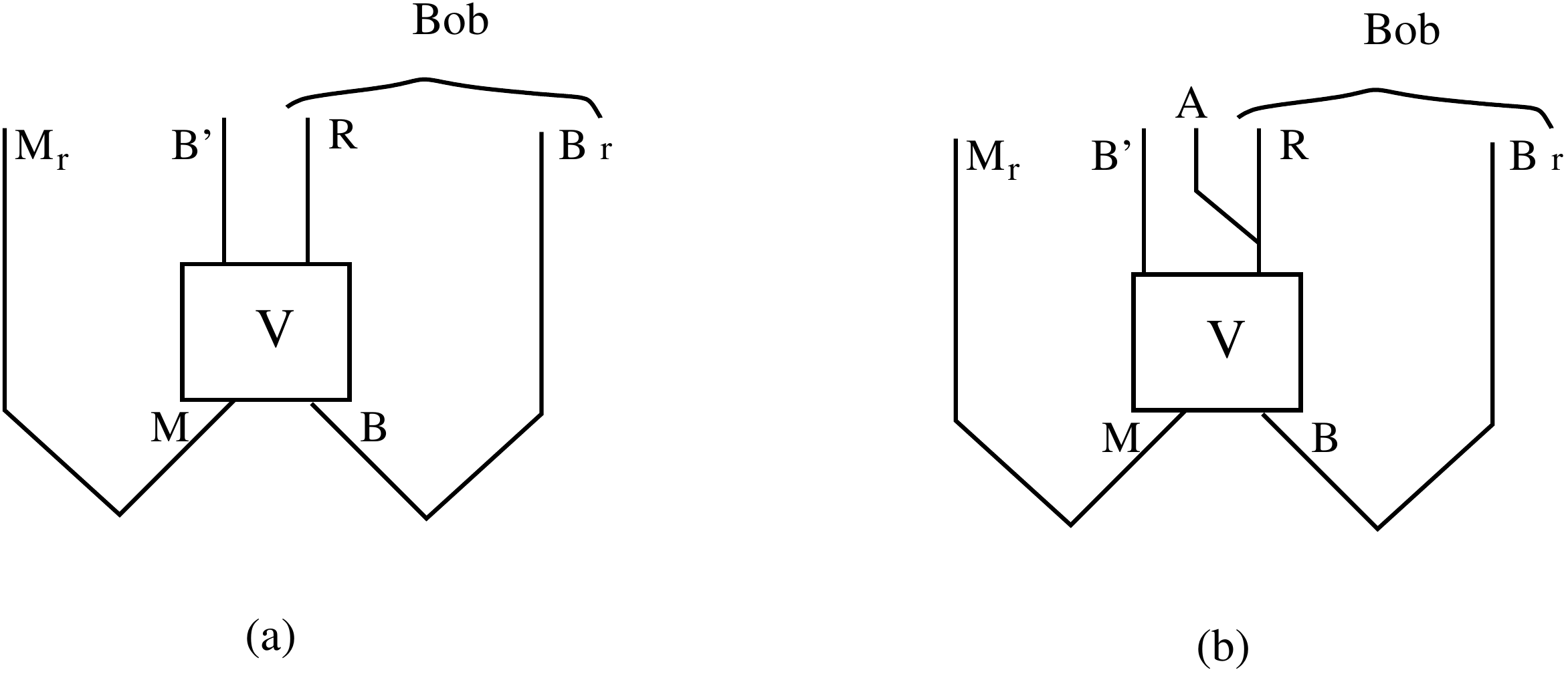}
\caption{Setup for the Hayden and Preskill discussion. We have a system $B$ maximally entangled with a reference system $B_r$ and also a 
message $M$ maximally entangled with a reference system $M_r$. A general unitary, $V$,  acts on $M\times B$. In (a) Bob has acess to the quantum information in 
 $R$. In (b) Bob has access to classical information in $R$. The apparatus $A$
  represents the system that implements the decoherence for the measurement.  }
\label{HPSetup}
\end{center}
\end{figure}

 Abstractly, we start from a system called $B$ that is maximally entangled with $B_r$ of size $|B|$. We have a message system of dimension $|M|$, also maximally entangled with 
 a system $M_r$ of the same dimension.  See figure \ref{HPSetup}. 
 We put them together to get a system of dimension $N = |B||M|$ and we apply a general unitary transformation $V$. The result of this transformation is then split into a 
 system $B'$ and $R$ of dimensions $|B'|$ and $|R|$. Bob has acess to $B_r$ and either quantum or classical information in $R$. 

 The task is to recover the message from the systems that Bob has acces to. This is equivalent to recovering a purification of the system $M_r$. A necessary and sufficient condition for this is that $M_r$ should be (to a good approximation) uncorrelated with whatever Bob doesn't have \cite{hayden2008decoupling,dupuis2010decoupling}. In the case where Bob gets full quantum control over $R$, we require $\|\rho^{M_r B'} - \rho^{M_r}\otimes \rho^{B'}\|_1\ll 1$. In the case where he only gets classical information about $R$, we require $\|\rho^{M_rB'A}-\rho^{M_r}\otimes \rho^{B'A}\|_1\ll 1$. In both cases the relevant norm is the 1-norm, which is the sum of the absolute values of the eigenvalues.

\subsection{Bob  accesses quantum information in the new radiation} 
This is simply a review of \cite{Hayden:2007cs}. The full pure state is 
\be
|\psi \rangle = { 1 \over \sqrt{N} }   V_{r I'}^{~~~m I} |m\rangle |I \rangle |I'\rangle |r \rangle ~,~~~~m\in M_r ~,~~~r\in R ~,~~~I \in B_r ~,~~~I' \in B' 
\ee
where repeated indices are summed. In such a state, it is eash to check that $\rho_{M_r}\otimes \rho_{B'}$ is maximally mixed. To ensure Bob can recover the message, we need to show that $\rho = \rho^{ M_r B' }$ is also close to maximally mixed in the 1-norm. As a first step, it is useful to compute $Tr[\rho^2]$. We have that 
\be \la{rhsq}
\rho_{m I'; n J'} = { 1 \over N } V_{r I'}^{~~~m I} (V^\dagger)_{n I}^{~~~  r J'}  ~,~~~~~~~Tr[\rho^2] = { 1 \over N^2}  V_{r I'}^{~~m I } (V^\dagger)_{n I }^{~~ r J'} V_{s J'}^{~~n L} (V^\dagger)_{m L}^{~~ s I' } 
\ee
We now average $tr[\rho^2]$ over the unitary group using 
%\bea
%  \langle U_{a}^{~\alpha} U_b^{\beta} (U^\dagger)_{\alpha'}^{~a'} (U^\dagger)_{\beta'}^{~b'} \rangle&=&
% { 1 \over N^2-1} \left( \delta^{a}_{a'} \delta^b_{b'} \delta^\alpha_{\alpha'} \delta^\beta_{\beta'} 
%+  \delta^{b}_{a'} \delta^a_{b'} \delta^\beta_{\alpha'} \delta^\alpha_{\beta'}  \right) 
%\cr
%&&
 %- { 1 \over N (N^2-1) } \left(  \delta^{a}_{a'} \delta^b_{b'} \delta^\beta_{\alpha'} \delta^\alpha_{\beta'} +
 %\delta^{b}_{a'} \delta^a_{b'} \delta^\alpha_{\alpha'} \delta^\beta_{\beta'}  \right)
 %\eea
 \bea
  \langle V_{a}^{~\bar c } V_b^{\bar d } (V^\dagger)_{\bar c'}^{~a'} (V^\dagger)_{\bar d'}^{~b'} \rangle&=&
 { 1 \over N^2-1} \left( \delta^{a}_{a'} \delta^b_{b'} \delta^{\bar c}_{\bar c'} \delta^{\bar d}_{\bar d'} 
+  \delta^{b}_{a'} \delta^a_{b'} \delta^{\bar d}_{\bar c'} \delta^{ \bar c }_{\bar d'}  \right) 
\cr
&&
 - { 1 \over N (N^2-1) } \left(  \delta^{a}_{a'} \delta^b_{b'} \delta^{\bar d}_{\bar c'} \delta^{ \bar c }_{\bar d'} +
 \delta^{b}_{a'} \delta^a_{b'} \delta^{\bar c}_{\bar c'} \delta^{\bar d}_{\bar d'}   \right)
 \eea
 Notice that each index $a$ of $\bar c$ in this formula becomes a pair of indices like $m L$ or $ r I'$ for our application. 
 Using this for \nref{rhsq} we get 
 \bea
 \langle Tr[\rho^2]\rangle &= &  { 1 \over N^2} \left[ 
{ 1 \over N^2 -1} \left( ( \delta^{mI}_{n I } \delta^{nL}_{m L  } )(\delta^{ r J'}_{r I'} \delta^{ s I'}_{s J'} )+ ( \delta^{mI}_{m L } \delta^{ nL}_{ nI})( \delta^{r J' }_{s J'} \delta^{s I'}_{r I' }) 
\right) - \cdots 
\right] 
\cr
&=& { 1 \over N^2} \left[ 
{ 1 \over N^2 -1} \left( |M||B|^2 \times |R|^2|B'| + |M|^2|B| \times  |R||B'|^2 \right) \right. 
\cr
& &- \left. { 1 \over N (N^2-1) } \left( |M||B|^2 \times  |R||B'|^2  +  |M|^2|B| \times  |R|^2|B'| \right) \right] 
\cr \la{AvUn}
\langle Tr[\rho^2]\rangle  & \sim & { 1 \over |B'| |M|  } \left[ 1 + { |M|^2 \over |R|^2 } - { 1 \over |R|^2 } \right] 
\eea
 where in the last step we assumed $N\gg 1$ and neglected the last term. Note that $D= |B'| |M|$ is the dimension of the space where the radiation lives. 
 We now use the Cauchy-Schwarz inequality to bound the average of the 1-norm in terms of the trace of $\rho^2$ (we suppress the $\langle ~~\rangle$ symbols in all terms):
 \be \la{NormOne}
 \|\rho - { {\bf 1 } \over D } \|_1^2  \equiv \left( \sum_{i=1}^D |\rho_i - { 1 \over D} | \right)^2 \leq  D  Tr[ ( \rho -{ 1 \over D} )^2]   =   D Tr[\rho^2] -1 = 
   { |M|^2 \over |R|^2 } - { 1 \over |R|^2 } 
 \ee
 where $\rho_i$ are the eigenvalues of $\rho$. Therefore if $|R|$ contains a few qubits more than $|M|$ then indeed Bob has a purification of $M_r$ and hence the message $M$.
 
 \subsection{Bob accesses only classical information in the new radiation } 
 
 This case is very similar to the previous one. Now we need to also take into account the measurement device which starts in the vacuum state
 and, after measuring the radiation, is in the $|r\rangle_A$ state. In other words, we get 
 $|r\rangle |0 \rangle_A  \to |r\rangle |r \rangle_A $. 
 We now want to compute the density matrix $\rho^{M B' A} $ which also includes the apparatus:
 \be \la{rhoAp}
 \rho^{MB'A}_{~~mI'r;nJ'  s}  = { 1 \over N } V_{ r I'}^{~~m I} (V^\dagger)_{n I }^{~~ r J'} \delta_{rs} ~,~~~~ 
 Tr[\rho^2] =  { 1 \over N^2 } V_{ r I'}^{~~m I} (V^\dagger)_{n I}^{~~  r J'} V_{r J'}^{~~n L} (V^\dagger)_{m L}^{~~  r I'} 
 \ee
 The difference from \nref{rhsq} is that,  since now $\rho$ has an $r$ index, in $Tr[\rho^2]$ we get only one sum over the $r$ indices instead of two sums. 
The average over the unitary group is similar to \nref{AvUn}, except some terms will miss factors of $|R|$. This gives 
 \be
  \langle Tr[\rho^2]\rangle \sim  { 1 \over |B'| |M| |R|  } \left[ 1 + { |M|^2 \over |R| } - { 1 \over |R|  } \right] 
 \ee
 where again we assumed $N\gg1$. Now  $D =  |B'| |M| |R|$. Again, as in \nref{NormOne},
 we find that 
 \be
  \langle \|\rho - { {\bf 1 } \over D } \|^2_1 \rangle =   { |M|^2 \over |R|  } - { 1 \over |R|  }.
 \ee
To make this small we need $|R|\gtrsim |M|^2$, which means that the number of classical bits should be slightly more than {\it twice} the number of qubits in the message $M$. Again we conclude that under these circumstances the qubits in $M_r$ should be purified by the system that Bob has access to.

 \section{Improved bound on information transfer}\label{app:inf}
In the main text, \nref{FiBound}, we used the uncertainty principle to bound the amount of information that could be sent through a null window of size $\Delta x^+$. Here we will give an improved version, using an argument from \cite{Blanco:2013lea}.
\begin{figure}[h]
\begin{center}
\includegraphics[width=.25\textwidth]{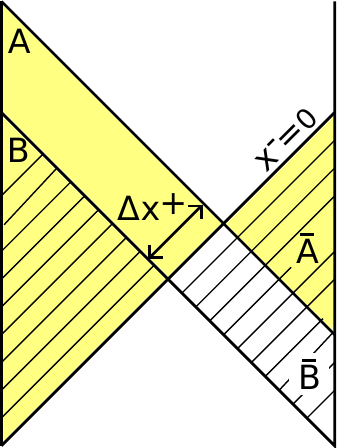}
\caption{The region $A$ is the yellow triangle on the left. The region $B$ is the shaded triangle. The regions $\bar{A}$ and $\bar{B}$ are their complements.}
\label{infpicture}
\end{center}
\end{figure}

We have drawn some spacetime regions in figure \ref{infpicture}. The idea of this configuration is that the region $B$ represents what is visible to the observer at the left boundary before we apply the double-trace deformation, and the region $A$ represents what is visible afterwards. We have assumed that the $OO$ operator insertion has a large boost relative
to the matter we are considering, so that its main effect is to produce a $\Delta X^+$ displacement, and we neglect the $\Delta X^-$ one.\footnote{
The final bound \nref{und} is still valid in the general case.} The regions $\bar{A},\bar{B}$ are simply the complements of $A$ and $B$. Notice that $B\subset A$ and $\bar{A}\subset \bar{B}$, so by monotonicity of the relative entropy,
\begin{align}
S(\rho_A|\sigma_A) \ge S(\rho_B|\sigma_B), \hspace{20pt} S(\rho_{\bar{B}}|\sigma_{\bar{B}}) \ge S(\rho_{\bar{A}}|\sigma_{\bar{A}}).
\end{align}
Here we take the state $\sigma$ to be the vacuum of the quantum fields in $AdS_2$, which is also the state appropriate for the thermofield double. We take $\rho$ to be a state in which there is some added matter propagating in $AdS_2$, which is entangled with a reference system $M_r$. Taking the sum of the two inequalities above, and writing the relative entropies in terms of the modular Hamiltonians $K$, we have
\be
S(A) - S(B) + S(\bar{B}) - S(\bar{A}) \le (K_A - K_{\bar{A}}) - (K_B - K_{\bar{B}}) = \hat{K}_A - \hat{K}_B.\label{comb}
\ee
The modular Hamiltonians for these regions are simply
\be
\hat{K}_A = -2\pi \int_{-\infty}^\infty dx^+ (x^+-\Delta x^+)T_{++} , \hspace{20pt} \hat{K}_B = -2\pi \int_{-\infty}^\infty dx^+ T_{++} x^+,
\ee
where $x^{\pm}$ are Kruskal coordinates, and the integral is over $x^-=0$. We conclude that $\hat{K}_A - \hat{K}_B = 2\pi \Delta x^+\int dx^+ T_{++} = 2\pi\Delta x^+(-P_+)$. 

We can also simplify the expression on the LHS of (\ref{comb}). First, we write
\be\label{mut}
I(A,M_r) - I(B,M_r) = S(A) + S(M_r) - S(AM_r) - S(B) - S(M_r) +S(BM_r).
\ee
Using purity of the entire system (including the system $M_r$ that purifies whatever matter we have included in the bulk) it follows that $S(AM_r) = S(\bar{A})$ and $S(BM_r) = S(\bar{B})$. As pointed out in \cite{Blanco:2013lea}, this turns (\ref{mut}) into precisely the LHS of (\ref{comb}). So we conclude
\be\label{und}
I(A,M_r) - I(B,M_r)\le 2\pi \Delta x^+(-p^{\text{total}}_+).
\ee
Now, recall that the role of $M_r$ was to purify some message system that we imagine adding from the right side. In this situation, $I(B,M_r) = 0$, and $I(A,M_r)$ is a measure of how much of the information of the message is contained in $A$ after applying the double trace operation. We can then plug in (\ref{xmshift}) and (\ref{ptotb}) to get a bound relating this information to $g$, and using (\ref{nequbits}) to the number of bits exchanged. It would be nice to have sharper versions of these last steps.

\mciteSetMidEndSepPunct{}{\ifmciteBstWouldAddEndPunct.\else\fi}{\relax}
\bibliographystyle{utphys}
\bibliography{TraversableDraft}{}

\end{document}